\documentclass[preprint,12pt]{elsarticle}

\usepackage{amsmath,amssymb}
\usepackage{graphicx}
\usepackage{geometry}
\usepackage{microtype}
\usepackage{booktabs}
\usepackage{siunitx}
\usepackage[version=4]{mhchem}
\usepackage[table]{xcolor}
\usepackage{makecell}
\usepackage{xurl}
\usepackage{hyperref}
\hypersetup{hidelinks}
\usepackage{subcaption}
\usepackage[section]{placeins}
\usepackage{cleveref}
\usepackage{lineno}

\usepackage{array}
\usepackage{multirow}

\makeatletter
\def\fps@figure{!tb}
\makeatother

\newcommand{\Dmtb}{D_{\mathrm{m}}^{{}^{149\mathrm{g}}\mathrm{Tb}}}

\definecolor{matApproved}{HTML}{1A9850}
\definecolor{matPhase23}{HTML}{66BD63}
\definecolor{matPhase2}{HTML}{93B058}
\definecolor{matPhase12}{HTML}{FDAE61}
\definecolor{matPhase1}{HTML}{F46D43}
\definecolor{matPreclin}{HTML}{D73027}

\journal{Applied Radiation and Isotopes}

\begin{document}

\begin{frontmatter}

\title{Scalable Terbium-149 Production from Highly Enriched Gadolinium-150 Targets}

\author[mf]{J.~F.~Parisi\corref{cor1}}
\cortext[cor1]{jason@marathonfusion.com}
\author[mf]{A.~Rutkowski}
\address[mf]{Marathon Fusion, 150 Mississippi Street, San Francisco, CA 94107, USA}

\begin{abstract}
We propose a two-stage production method to overcome existing supply-constraints for the alpha-emitter \(^{149\mathrm{g}}\)Tb, a promising candidate for Targeted Alpha Therapy (TAT) with no existing globally scalable production pathway. Although awaiting experimental measurement of the \(^{150}\)Gd(p,2n)\(^{149\mathrm{g}}\)Tb cross section, the proposed method could produce \(^{149\mathrm{g}}\)Tb at clinical scale and beyond on readily available proton cyclotrons, enabled by production of the extinct but long-lived isotope \(^{150}\)Gd, a pure alpha emitter with a 1.79 million year half-life. Stage one generates \(^{150}\)Gd feedstock by irradiating natural Eu or enriched \(^{151}\)Eu with $\gtrsim$10 MeV protons, neutrons or photons. Stage two produces \(^{149\mathrm{g}}\)Tb from fabricated \(^{150}\)Gd targets by driving the \(^{150}\)Gd(p,2n)\(^{149\mathrm{g}}\)Tb reaction with $\gtrsim$14 MeV protons, accessible on over 700 reported cyclotrons worldwide. While not yet experimentally measured, the most recent TALYS evaluation (TENDL-2025) predicts the \(^{150}\)Gd(p,2n)\(^{149\mathrm{g}}\)Tb cross section to peak at $\sim$\qty{457}{mb} at $\sim$\qty{18}{MeV} proton energy, predicting an admin-time radionuclidic purity of $\sim$98\% on a thin \(^{150}\)Gd target after \qty{4.1}{h} cooldown for $^{149\mathrm{g}}$Tb, with $^{149\mathrm{g}}$Tb and $^{150}$Tb together accounting for $\sim99.7\%$ of total Tb activity. \(^{149\mathrm{g}}\)Tb activity is $\sim$\qty{1200}{MBq/\uA} at the end of a two-hour 20 MeV proton cyclotron run in a thick \(^{150}\)Gd target ($\sim$\qty{4230}{MBq/\uA} at saturation), sufficient to deliver tens to thousands of $\sim$50 MBq doses per irradiation with modest proton beam current. Given that a \qty{50}{MBq} dose of \(^{149\mathrm{g}}\)Tb weighs only \qty{0.27}{ng}, a minimal quantity of \(^{150}\)Gd target material is required ($\sim$\qty{2}{\micro g}/dose), even when accounting for losses during extraction and target re-fabrication. Existing cyclotrons can be used to produce sufficient quantities of material for preclinical and clinical work prior to scaling, enabling progress on medical testing in parallel with efforts to scale isotope supply through other routes. We show conclusions are robust to administered dose sizes  more than an order of magnitude higher than the assumed 50 MBq/dose. Existing neutron, proton, and photon sources are of sufficient rate to generate tens to hundreds of mg of \(^{150}\)Gd per year, sufficient for multiple targets capable of high rate production of \(^{149}\)Tb. We expect future deuterium-tritium fusion fast-neutron sources to be capable of providing grams to kg of \(^{150}\)Gd per year. Fast fusion neutrons appear to offer the most scalable pathway for \(^{150}\)Gd production: even with a $^{149 \mathrm{g} }$Tb dose size of 1 GBq and 40 million doses/yr, we estimate this would require neutrons produced by only 6.8 megawatts of steady-state deuterium-tritium power to produce the required $^{150}$Gd, far below expected capacity in the next decade. Experimental validation of the \(^{150}\)Gd(p,2n)\(^{149\mathrm{g}}\)Tb cross section is the necessary next step and is under way. The route described here, if validated, would enable \(^{149\mathrm{g}}\)Tb supply at the scale needed to support clinical development of $^{149\mathrm{g}}$Tb-based TAT.
\end{abstract}

\end{frontmatter}

\section{Introduction}

\begin{figure}[!tb]
\centering
\includegraphics[width=0.7\columnwidth]{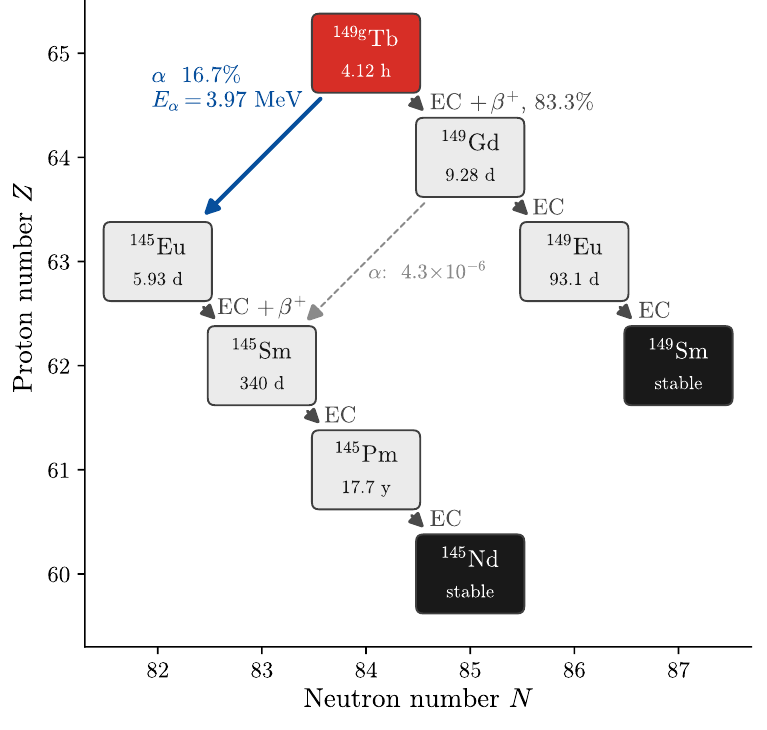}
\caption{Decay paths of \(^{149\mathrm{g}}\)Tb.}
\label{fig:Tb149g_decay}
\end{figure}

Targeted Alpha Therapy (TAT) exploits the short range (tens of micrometers in tissue, corresponding to only several human cell diameters) and high linear energy transfer of alpha particles to destroy individual cancer cells while sparing surrounding healthy tissue~\cite{kim2012overview,parker2013}. Among candidate alpha emitters, \(^{149\mathrm{g}}\)Tb occupies a unique position, being the only alpha emitter with a clinically useful half-life that is simultaneously suitable for positron-emission-tomography (PET) imaging. First discovered in 1950 by Rasmussen~\cite{rasmussen1950} via mass spectrographic identification of \qty{4.1}{\hour} alpha activity following \SI{150}{MeV} proton bombardment of gadolinium oxide at the Berkeley 184-inch cyclotron, \(^{149\mathrm{g}}\)Tb has been recognized for over three decades as a TAT candidate~\cite{allen1996alpha}, and while small animal preclinical trials have shown its imaging and therapeutic effectiveness, the isotope has never been produced in quantities sufficient for clinical use. Its \qty{4.1}{\hour} half-life is short enough to limit whole-body dose yet long enough for same-day clinical use. The \SI{16.7}{\percent} alpha branching fraction provides therapy, while the remaining \qty{83.3}{\percent} of decays proceed via positron emission and electron capture, also allowing PET imaging without a companion diagnostic isotope~\cite{muller2012}. Additionally, \(^{149\mathrm{g}}\)Tb has the lowest-energy alpha particle among TAT alpha emitters, and unlike some other TAT candidate isotopes has no additional alphas in the decay chain. These features  minimize damage to surrounding healthy cells while efficiently destroying tumor cells. In \Cref{fig:Tb149g_decay} we show the decay scheme of \(^{149\mathrm{g}}\)Tb. The isomer \(^{149\mathrm{m}}\)Tb decays via $\beta^+$/EC (21.0\%/79.0\%) to \(^{149}\)Gd and $\alpha$ (0.022\%) to \(^{145}\)Eu with half-life 4.2 mins, and is not useful for TAT. Therefore any production scheme must prioritize producing \(^{149\mathrm{g}}\)Tb, not \(^{149\mathrm{m}}\)Tb.

\begin{figure}[!tb]
\centering
\includegraphics[width=\columnwidth]{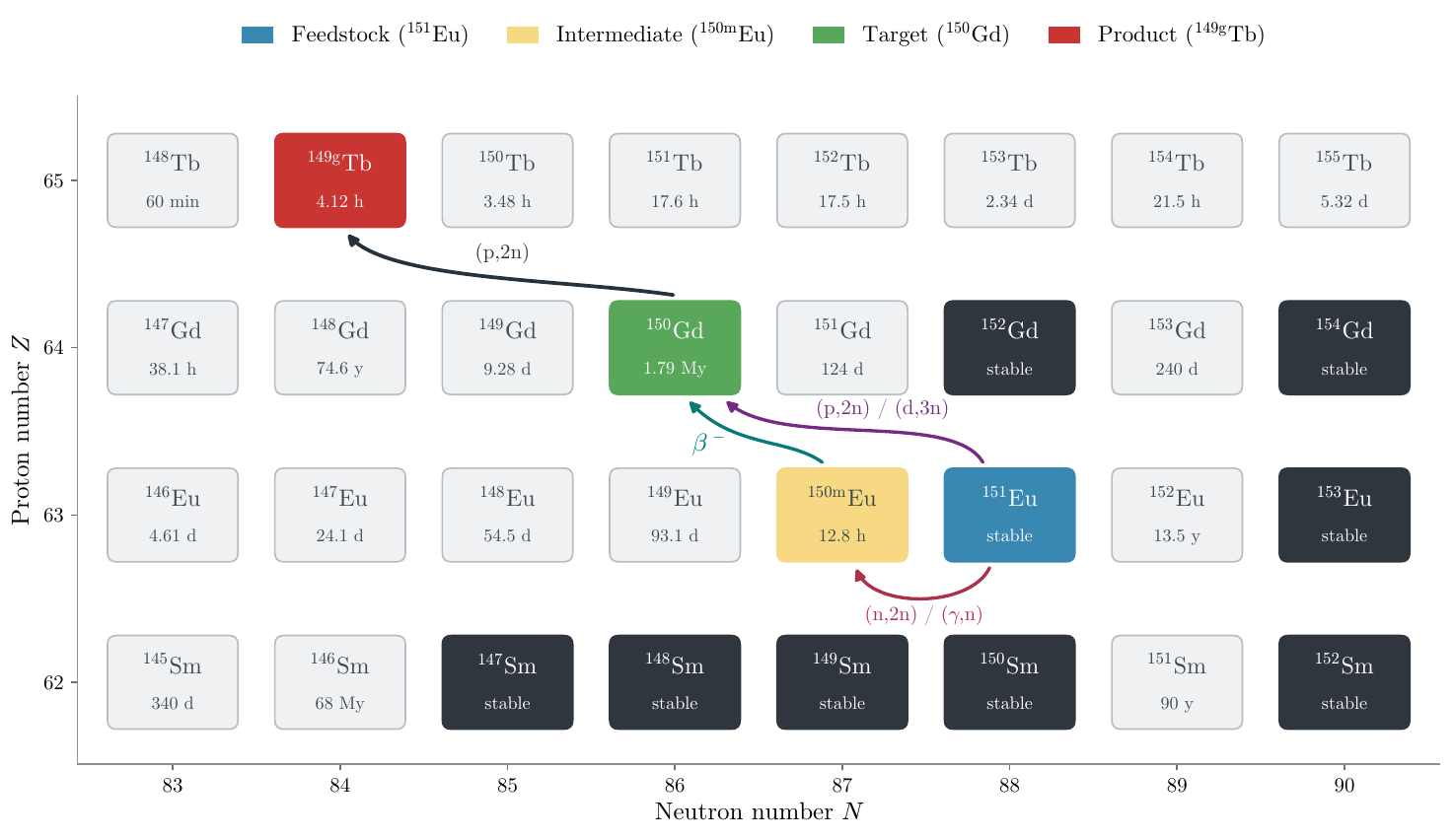}
\caption{Proposed \(^{149\mathrm{g}}\)Tb production scheme on the chart of nuclides. $^{151}$Eu (blue) is transmuted to $^{150}$Gd (green) via (p,2n), (d,3n), (n,2n), or ($\gamma$,n) reactions. The $^{150}$Gd is then irradiated with $\sim$14-25 MeV protons on a proton cyclotron, producing $^{149 \mathrm{g} }$Tb (red) via (p,2n).}
\label{fig:chart}
\end{figure}

\(^{149\mathrm{g}}\)Tb is part of the terbium radionuclide quadruplet \cite{muller2012}, comprising alpha therapy (\(^{149\mathrm{g}}\)Tb), PET (\(^{149\mathrm{g}}\)Tb, \(^{152}\)Tb), SPECT (\(^{155}\)Tb), and beta therapy (\(^{161}\)Tb). All four share effectively identical chemistry, so a radiopharmaceutical developed for one can translate to the others. Promising preclinical results for \(^{149\mathrm{g}}\)Tb have been demonstrated with rituximab \cite{beyer2004}, folate-based targeting agents~\cite{muller2014}, PSMA-targeting agents \cite{umbricht2019}, and somatostatin receptor targeting agents DOTATATE and DOTA-LM3 \cite{mapanao2025}. Another preclinical study \cite{muller2012} explored the use of multiple terbium isotopes capable of functioning as a theranostic across PET and SPECT imaging as well as beta minus and alpha decays, and a later study \cite{muller2017alphapet} specifically performed PET imaging alongside the alpha treatment using the positron emission from \(^{149}\)Tb. A separate in-vitro study directly compared the effectiveness of \(^{149}\)Tb and \(^{213}\)Bi on cell suspensions and cell pellets, including consideration of radiotoxicity differences \cite{miederer2003radiotoxicity}.

\begin{figure}[!tb]
\centering
\includegraphics[width=\columnwidth]{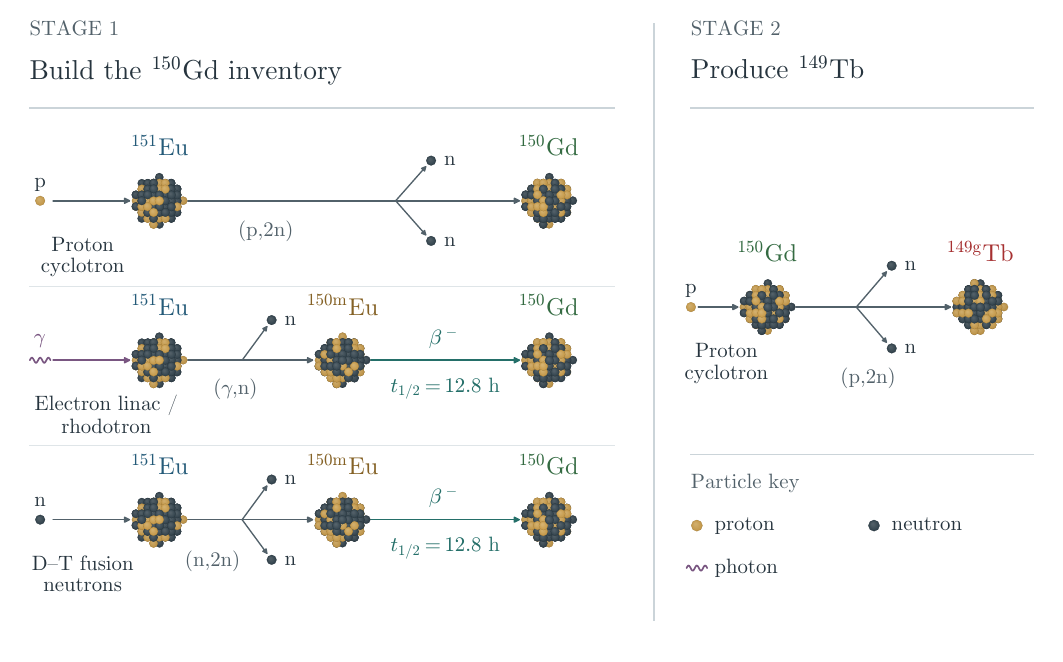}
\caption{Two-stage production scheme corresponding to \Cref{fig:chart}.}
\label{fig:schematic_production}
\end{figure}

In this article, we propose a scheme to resolve the TAT supply barrier by treating the extinct nuclide \(^{150}\)Gd as a near-stable feedstock, fabricated into targets and then irradiated in proton cyclotrons to make \(^{149\mathrm{g}}\)Tb on demand. The scheme is shown schematically on the chart of nuclides in \Cref{fig:chart}, and the reactions are shown in \Cref{fig:schematic_production}.

We structure this article as follows. Section~\ref{sec:existing} reviews existing \(^{149\mathrm{g}}\)Tb production techniques. Sections~\ref{sec:step1} and~\ref{sec:step2} describe production of the \(^{150}\)Gd feedstock and its conversion to \(^{149\mathrm{g}}\)Tb. Section~\ref{sec:production} connects these yields to material requirements, global supply, and the first preclinical doses. Section~\ref{sec:discussion} discusses the implications and experimental priorities. In order to keep the main text focused and readable, we have put much of the technical analysis in the appendices, including dose-size sensitivity, nuclear-data comparisons, target optimization, self-sufficient targets, and supporting derivations. We strongly encourage readers seeking a more detailed technical assessment to consult these appendices alongside the main text.

\section{Challenges with Existing \(^{149\mathrm{g}}\)Tb Production Techniques}\label{sec:existing}

\begin{table}[!htbp]
\centering
\scriptsize
\setlength{\tabcolsep}{4pt}
\renewcommand{\arraystretch}{1.15}
\caption{Leading alpha emitters for Targeted Alpha Therapy. Clinical maturity and active trial counts as of 9 September 2026~\cite{clinicaltrials_counts2026}. The $\#\alpha$ column counts alpha emissions per decay chain.}
\label{tab:tat_landscape}
\begin{tabular}{@{}l c c c l >{\centering\arraybackslash}m{3.2cm}@{}}
\toprule
\textbf{Isotope} & \textbf{$t_{1/2}$} & \textbf{$\#\alpha$} & \makecell{\textbf{Active trials}\\\textbf{(recruiting)}} & \textbf{$E_\alpha$ (MeV)} & \textbf{Clinical maturity} \\
\midrule
$^{223}$Ra & 11.4\,d & 4 & \,18 (7) & \makecell[l]{5.72 / 6.82 /\\ 7.39 / 6.62}
& \cellcolor{matApproved!15}Approved (Xofigo)\,\cite{xofigo_fda} \\
\midrule
$^{225}$Ac & 9.92\,d & 4 & 51 (36) & \makecell[l]{5.83 / 6.34 /\\ 7.07 / 8.38}
& \cellcolor{matPhase23!15} Phase III~\cite{action1_registry} \\
\midrule
$^{212}$Pb/$^{212}$Bi & 10.6\,h & 1 & 12 (9) & \makecell[l]{6.05 (36\%) /\\ 8.78 (64\%, via $^{212}$Po)}
& \cellcolor{matPhase2!15}Phase II\,($^{212}$Pb)~\cite{alphamedix_registry} \\
\midrule
$^{211}$At & 7.21\,h & 1 & 6 (5) & \makecell[l]{5.87 (42\%) /\\ 7.45 (58\%, via $^{211}$Po)}
& \cellcolor{matPhase12!15} Phase I/II~\cite{at211_registry} \\
\midrule
$^{213}$Bi & 45.6\,min & 1 & 0 & \makecell[l]{5.87 (2\%) /\\ 8.38 (98\%, via $^{213}$Po)}
& \cellcolor{matPhase12!15} Phase I/II completed~\cite{rosenblat2010bi,bi213_registry} \\
\midrule
$^{227}$Th & 18.7\,d & 5 & 0 & \makecell[l]{6.04 / 5.72 / 6.82 /\\ 7.39 / 6.62}
& \cellcolor{matPhase1!15} Phase I completed~\cite{th227_registry,th227_psma_registry} \\
\midrule
$^{149\mathrm{g}}$Tb & 4.12\,h & 0.17 & 0 & \makecell[l]{3.97\\ (BR 16.7\%)}
& \cellcolor{matPreclin!15}Pre-clinical\,\cite{vandermeulen2026ici} \\
\bottomrule
\end{tabular}
\end{table}

\begin{table}[!htbp]
\centering
\scriptsize
\caption{Nuclear properties of relevant europium, gadolinium, and terbium isotopes. Europium abundances: \cite{ciaaw_europium}.}
\label{tab:nuclides}
\begin{tabular}{llll}
\toprule
\textbf{Nuclide} & \textbf{$t_{1/2}$} & \textbf{Decay mode} & \textbf{Nat.\ abund.} \\
\midrule
$^{150\mathrm{m}}$Eu & \qty{12.8}{h}      &  $\beta^-$ (89\%), EC/$\beta^+$ (11\%)           & $-$ \\
$^{151}$Eu & 4.6 $\cdot 10^{18}$ yr      & $\alpha$           &  47.81\% \\
$^{152}$Eu & 13.5 yr      &  EC/$\beta^+$ (72\%), $\beta^-$ (28\%)           & $-$ \\
$^{152\mathrm{m1}}$Eu & \qty{9.3}{h}      &  $\beta^-$ (73\%), EC/$\beta^+$ (27\%)           & $-$ \\
$^{153}$Eu & stable      & $-$           &  52.19\% \\
\midrule
$^{148}$Gd & 86.9 yr      & $\alpha$          & $-$ \\
$^{149}$Gd & \qty{9.28}{d}       & EC ($\sim$100\%), $\alpha$ ($4.3 \cdot 10^{-4}$\%)                 & $-$ \\
$^{150}$Gd & $1.79 \cdot 10^6$ yr & $\alpha$ ($\sim$100\%)  & $-$ \\
$^{151}$Gd & \qty{124}{d}        & EC ($\sim$100\%), $\alpha$ ($1.1 \cdot 10^{-6}$\%)                 & $-$ \\
$^{152}$Gd & $1.09 \cdot 10^{14}$ yr & $\alpha$       & 0.20\% \\
$^{153}$Gd & \qty{240.7}{d}      & EC                 & $-$ \\
$^{154}$Gd & stable       & $-$                        & 2.18\% \\
\midrule
$^{148}$Tb & \qty{60}{min}       & EC/\(\beta^+\) ($\sim$100\%), \(\alpha\) (0.0094\%) & $-$ \\
$^{149 \mathrm{g} }$Tb & \qty{4.12}{h}     & EC/\(\beta^+\) (83.3\%), \(\alpha\) (16.7\%) & $-$ \\
$^{149 \mathrm{m} }$Tb & \qty{4.16}{min}     &  EC/\(\beta^+\) (99.98\%), \(\alpha\) (0.022\%) & $-$ \\
$^{150}$Tb & \qty{3.48}{h}      & EC/\(\beta^+\) & $-$ \\
$^{151}$Tb & \qty{17.6}{h}      & EC/\(\beta^+\) & $-$ \\
$^{152}$Tb & \qty{17.5}{h}      & EC/\(\beta^+\) & $-$ \\
\bottomrule
\end{tabular}
\end{table}

\(^{149\mathrm{g}}\)Tb has been produced at a handful of facilities worldwide (CERN-ISOLDE/MEDICIS~\cite{cern_medicis,favaretto2024}, iThemba LABS~\cite{steyn2014}, Kurchatov Institute~\cite{moiseeva2020}), but total global production delivered to end users across all sources is $\sim$\SI{1}{GBq\,yr^{-1}}~\cite{naskar2021}, orders of magnitude below what a single clinical trial would require. Supply has been explicitly identified as the primary obstacle to clinical adoption~\cite{naskar2021,favaretto2024}.

 A range of production methods have been pursued for $^{149\mathrm{g}}\mathrm{Tb}$ production, as summarized in \cite{moiseeva2024}.

The earliest recorded experimental work on \(^{149\mathrm{g}}\)Tb production \cite{alexander1963} used a range of heavy ions like \(^{11}\)B and \(^{12}\)C, in which the highest cross section was recorded for $^{142}\mathrm{Nd}(^{11}\mathrm{B},\mathrm{4n})^{149\mathrm{g}}\mathrm{Tb}$ at only \qty{58}{mb}.  Heavy ion production pathways suffer not just from small cross sections, but high required energies to access the reaction, and even more critically, the need for highly specialized  heavy ion accelerators with limited availability.  For these reasons, it is highly unlikely that pathways of this type will scale to meaningful production rates.

\begin{table}[!tb]
\centering
\caption{Main routes for $^{149\mathrm{g}}$Tb production in literature}
\label{tab:tb149_routes_yield_energy}
\renewcommand{\arraystretch}{1.25}
\resizebox{\textwidth}{!}{%
\begin{tabular}{
  m{2.8cm}
  m{2.8cm}
  m{3.4cm}
  m{5.2cm}
  m{2.2cm}
}
\hline
\textbf{Reaction} &
\textbf{Radionuclidic purity} &
\textbf{Max $\mathrm{^{149g}Tb}$ Production Cross Section [barn]} &
\textbf{Beam energy requirement (threshold, cross section peak) [MeV]} &
\textbf{Ref.} \\
\hline

$\mathrm{^{152}Gd(p,4n)}$ &
Low &
0.25
&
(30, 40)
&
\cite{steyn2014, formentocavaier2020} \\
\hline

$\mathrm{^{151}Eu}(^3\mathrm{He,5n})$ &
43\% &
0.75
&
(31,45)
&
\cite{moiseeva2020, zagryadskii2017} \\
\hline

$\mathrm{^{nat}Ta(p,x)}$ &
99.9\%\textsuperscript{*} &
0.022
&
(300,1100)
&
\cite{favaretto2024,verhoeven2020,winsberg1964} \\
\hline
$\mathrm{^{nat}Nd}(^{11}\mathrm{B,x})$ &
Low &
0.058 (max)
&
($\sim$ 40,55.2)
&
\cite{zaitseva2003,lahiri1999,sarkar1997,alexander1963} \\
\hline
$\mathrm{^{150}Gd(p,2n)}$ &
$\lesssim$ 98\% (modeled) &
0.457 (TENDL-2025)
&
(13.2, $\sim$ 18) (TENDL-2025)
&
[This work] \\
\hline
\end{tabular}%
}

\vspace{0.5em}
\raggedright
\footnotesize
\textsuperscript{*}Achieved after online isotope separation. The energy column lists the excitation-function (threshold, peak). The beam energies used in the cited experiments are higher (e.g.\ \qty{66}{MeV} for $^{152}$Gd(p,4n)~\cite{steyn2014} and \qty{70}{MeV} for $^{151}$Eu($^3$He,5n)~\cite{moiseeva2020}) so as to integrate the thick-target yield over the full slowing-down window.
\end{table}

One cyclotron route is \(^{152}\)Gd(p,4n)\(^{149\mathrm{g}}\)Tb, which requires $\sim$\SI{66}{MeV} protons and an enriched \(^{152}\)Gd target~\cite{steyn2014}. However, \(^{152}\)Gd has only \SI{0.20}{\percent} natural abundance, and the highest commercially available enrichment at present is $\sim$\SI{30}{\percent}~\cite{naskar2021}. The remaining $\sim$\SI{70}{\percent} of non-\(^{152}\)Gd isotopes undergo their own (p,xn) reactions, co-producing \(^{150}\)Tb (\SI{31}{mb}), \(^{151}\)Tb (\SI{96}{mb}), \(^{152}\)Tb (\SI{114}{mb}), and \(^{153}\)Tb (\SI{124}{mb}) that collectively contribute significantly more to activity than the \SI{250}{mb} \(^{149\mathrm{g}}\)Tb cross section~\cite{steyn2014,naskar2021}. Because all Tb isotopes share similar chemistry, chemical purification cannot remove these contaminants, and the only recourse is mass separation via calutrons, which is far too slow to extract high specific activity \(^{149\mathrm{g}}\)Tb~\cite{naskar2021}.

Several groups have demonstrated production of \(^{149\mathrm{g}}\)Tb from spallation of Ta targets by high energy ($\sim$\ \qty{1.4}{GeV}) protons \cite{favaretto2024, verhoeven2020, winsberg1964}. This pathway is currently pursued at CERN-ISOLDE, where $\sim$\SI{1000}{MBq} is collected in the target per day, but after transport to the Paul Scherrer Institute (one half-life lost), mass separation, and radiochemical purification, only $\sim$\SI{260}{MBq} remains~\cite{favaretto2024}. This facility is currently unavailable for routine \(^{149\mathrm{g}}\)Tb production. The \SI{4.1}{hr} half-life compounds all of these difficulties, as every hour of processing or transport costs $\sim$\SI{15}{\percent} of the product, requiring production, separation, transport, labeling, and injection to occur same day at a single site.

Two recent studies \cite{moiseeva2020,zagryadskii2017} have demonstrated a production route starting from \(^{151}\)Eu using \(^{3}\)He beams.
While this reaction has a reasonably high cross section and an abundant feedstock isotope, it relies on the use of rare high energy \(^{3}\)He beams that do not exist broadly.
It also suffers from the coproduction of many other terbium radioisotopes that reduce the radionuclidic purity of the product. The cross sections of these other reactions are larger.
Within some limits, the purity can be increased through use of higher energies and thinner targets, but this results in tradeoff on yield.

In this work, we propose a two-stage approach that could address existing \(^{149\mathrm{g}}\)Tb production challenges, simultaneously achieving high yield and radionuclidic purity using readily available cyclotrons and using commonly available feedstocks.

\section{Stage One: \(^{150}\)Gd Production}\label{sec:step1}

 The first stage of the two-stage \(^{149\mathrm{g}}\)Tb production scheme generates the feedstock \(^{150}\)Gd, which is extinct and therefore must be produced via transmutation. Three routes on a \(^{151}\)Eu target are most attractive for producing \(^{150}\)Gd: (p,2n) with a cyclotron, (n,2n) with a fast-neutron source, and (\(\gamma\),n) with a photon source such as a linear electron accelerator (LINAC). The proton routes use widely available accelerator infrastructure, the photon route is available on accelerators capable of producing photons with $\sim$15 MeV, and the neutron route is available to deuterium-tritium (D-T) neutron generators whose availability, rate, and flux are expected to grow significantly in coming years. We expect D-T neutrons will be the lowest-cost pathway to produce high-purity $^{150}$Gd in quantities sufficient to support clinical-scale use of $^{149}$Tb.

Several additional routes for \(^{150}\)Gd production (multi-step neutron reactions and other charged-particle beams) are conceptually possible-but far less efficient than the direct proton, photon, and neutron routes-and are described in \ref{sec:alternative_gd150}.

\subsection{Proton route: \(^{151}\)Eu(p,2n)\(^{150}\)Gd}

\begin{figure}[!tb]
\centering
\includegraphics[width=\textwidth]{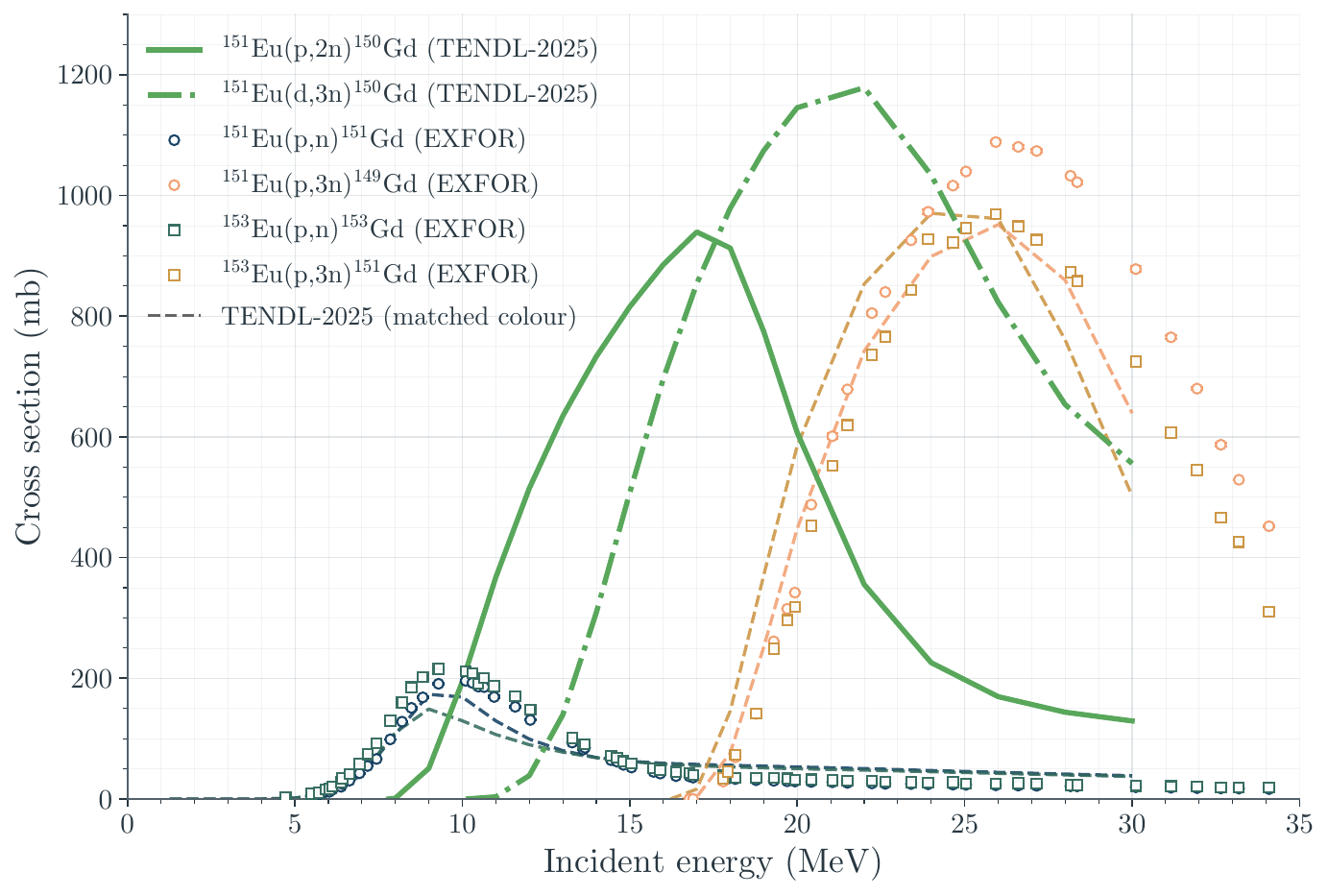}
\caption{Predicted cross sections for Stage-1 routes that produce \(^{150}\)Gd from a \(^{151}\)Eu target: \(^{151}\)Eu(p,2n)\(^{150}\)Gd (TENDL-2025) peaks near \SI{940}{mb} at \qty{17}{MeV}, and \(^{151}\)Eu(d,3n)\(^{150}\)Gd (TENDL-2025) peaks near \qty{1180}{mb} at \qty{22}{MeV}. Neither of these two feedstock-producing reactions has a published experimental measurement. The EXFOR points show measured neighboring channels for comparison.}
\label{fig:Eu151_p2n_new}
\end{figure}

\begin{figure}[!tb]
    \centering
    \begin{subfigure}[t]{0.49\textwidth}
    \centering
    \includegraphics[width=\textwidth]{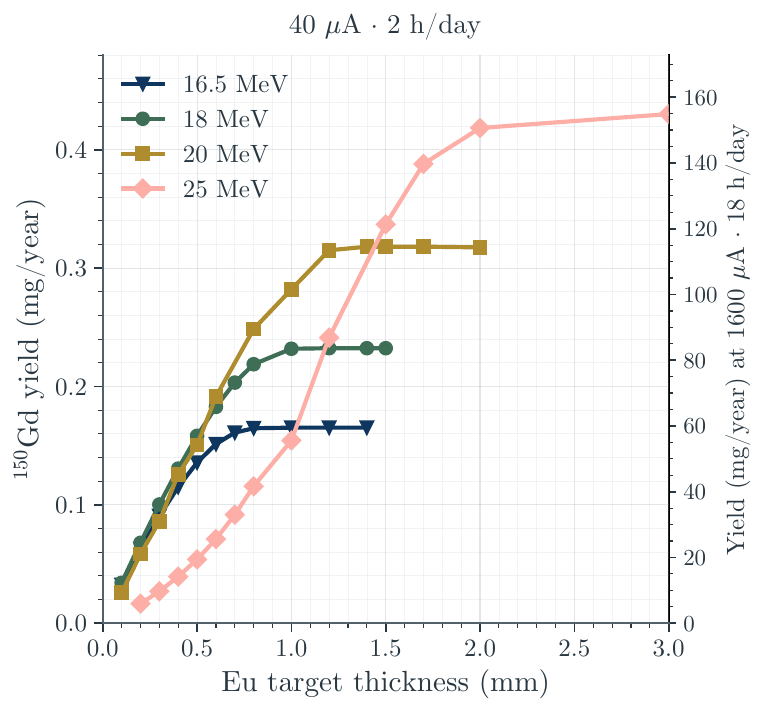}
    \caption{Yield}
    \end{subfigure}
    \hfill
    \begin{subfigure}[t]{0.49\textwidth}
    \centering
    \includegraphics[width=\textwidth]{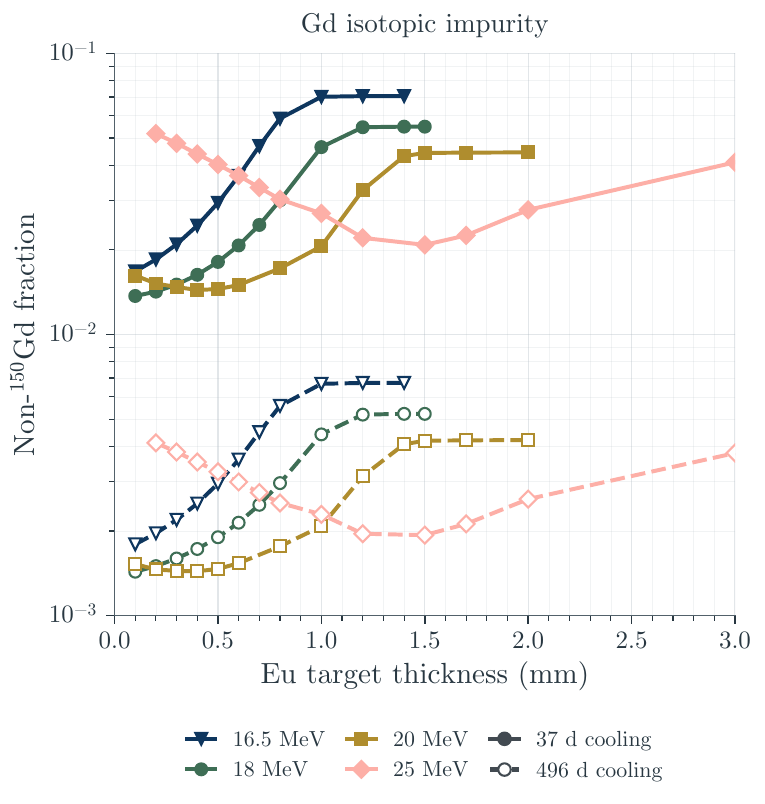}
    \caption{Isotopic impurity}
    \end{subfigure}
    \caption{(a) \(^{150}\)Gd yield versus \(^{151}\)Eu target thickness for one year of proton irradiation at 16.5, 18, 20, and 25 MeV computed with ISOTOPIA \cite{koning2025isotopia}. The left axis gives the yield for a \qty{40}{\uA} hospital-class cyclotron operating 2 hours per day. The right axis rescales the same curves to a \qty{1600}{\uA} TR-30-class machine operating 18 hours per day (a $\sim$360$\times$ factor in effective beam-on charge). (b) Non-\(^{150}\)Gd fraction of the Gd inventory after \qty{37}{d} cooldown ($\sim$4~half-lives of \(^{149}\)Gd, filled markers, solid) and after \qty{496}{d} (open markers, dashed). The fractions are identical for both beam regimes since impurity ratios depend only on cross sections and cooldown.}
    \label{fig:Eu151_thickness_cooled}
\end{figure}

One route bombards an enriched \(^{151}\)Eu target with protons at roughly 14-25 MeV,
\begin{equation}
    \ce{^{151}Eu(p,{2n})^{150}Gd}.
    \label{eq:Gd150_protons}
\end{equation}
The cross section is predicted to peak near \qty{1}{b} at \qty{17}{MeV} (\Cref{fig:Eu151_p2n_new}). No experimental measurement is published.  The reaction threshold near \qty{10}{MeV} is accessible on over 1200 reported proton cyclotrons (see \ref{sec:cyclotron_dep}). The (p,n) channel competes at lower energy and produces \(^{151}\)Gd, which decays back to \(^{151}\)Eu over \qty{124}{d}. While the reaction in \Cref{eq:Gd150_protons} has not been reported experimentally, both (p,n) and (p,3n) have been measured, shown in \Cref{fig:Eu151_p2n_new}. If deuteron beams are available, there is also a (d,3n) pathway available with a large predicted cross section, also shown in \Cref{fig:Eu151_p2n_new}.

This stage-1 feedstock calculation uses the beam currents typical of the machines that would run it (\qty{40}{\uA} hospital-class and \qty{1600}{\uA} TR-30-class), distinct from the \qty{100}{\uA} reference current adopted for the stage-2 $^{149\mathrm{g}}$Tb production estimates. We calculate the \(^{150}\)Gd yield for the (p,2n) pathway using the isotope production code ISOTOPIA \cite{koning2025isotopia}  and calculate isotopic purity in post-processing. Results are shown in \Cref{fig:Eu151_thickness_cooled}. \(^{150}\)Gd yield saturates once the target thickness reaches the proton stopping range (\qty{1.84}{mm} at \qty{20}{MeV}). \(^{150}\)Gd isotopic purity is shown after \qty{37}{d} cooldown ($\sim$4~half-lives of \(^{149}\)Gd, filled markers, solid) and after \qty{496}{d} (open markers, dashed). Higher beam energies produce more \(^{149}\)Gd contamination at end-of-irradiation (EOI) but cool to comparable purity within a year. Because \(^{149}\)Gd is much shorter lived (\qty{9.3}{d} half-life) than \(^{151}\)Gd (\qty{123.8}{d} half-life), increasing target thickness at higher beam energy tends to improve \(^{150}\)Gd isotopic purity.
Regardless of target thickness and beam energy, the \(^{150}\)Gd isotopic purity always exceeds 99\% after a two year cooldown. \Cref{fig:Eu151_thickness_cooled} shows \(^{150}\)Gd yields of $\sim$0.1-\qty{0.4}{mg} per year for a \qty{40}{\uA} proton beam operating for two hours per day. Much higher yields are possible on cyclotrons with higher beam current operating at higher duty factor such as the TR-30 \cite{duh2004current}, which can produce \qty{1600}{\uA} proton beams. We show the annual yield for a TR-30-class machine with 75\% beam uptime in \Cref{fig:Eu151_thickness_cooled}. On the order of \qty{100}{mg} of \(^{150}\)Gd can be produced annually.

\subsection{Photon route: \(^{151}\)Eu(\(\gamma\),n)\(^{150\mathrm{m}}\)Eu \(\to\) \(^{150}\)Gd}

\begin{figure}[!tb]
\centering
\includegraphics[width=0.85\textwidth]{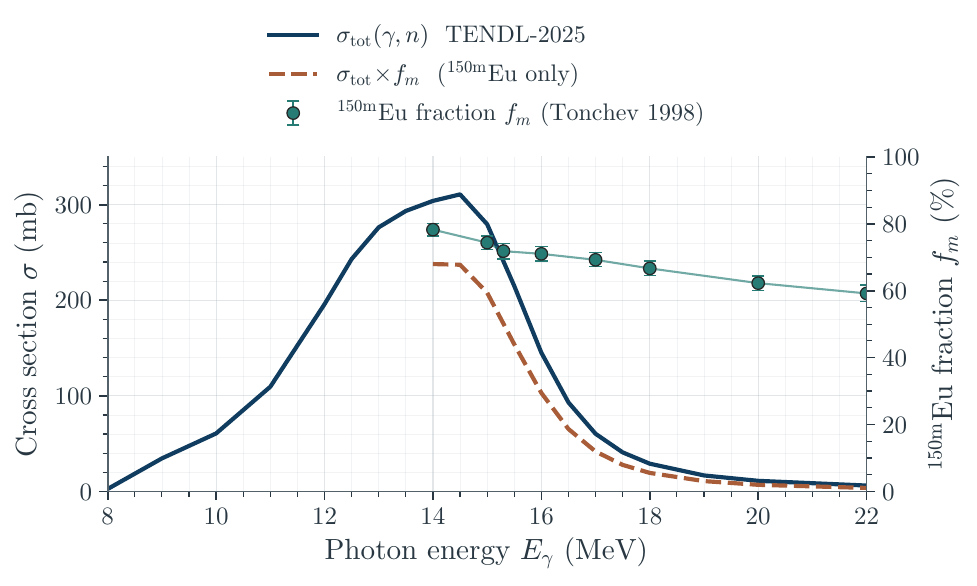}
\caption{TENDL-2025 \(^{151}\)Eu(\(\gamma\),n)\(^{150}\)Eu total cross section, peaking near \qty{310}{mb} at $\sim$\qty{14.5}{MeV}, with the measured \(^{150\mathrm{m}}\)Eu isomer fraction from Tonchev~\cite{tonchev1998deformation}.}
\label{fig:Eu151_gn}
\end{figure}

The second route uses bremsstrahlung photons that can be produced in a LINAC or Rhodotron,
\begin{equation}
    {}^{151}\mathrm{Eu}(\gamma,\mathrm{n}){}^{150 \mathrm{m}}\mathrm{Eu} \xrightarrow{\beta^-\ (12.8\,\mathrm{h},\,89\%)} {}^{150}\mathrm{Gd}.
    \label{eq:Eu151_gn}
\end{equation}
This technique was used in 1965 \cite{ogawa1965alpha} to produce \(^{150}\)Gd in very small quantities. The total \(^{151}\)Eu(\(\gamma\),n) cross section peaks near \qty{310}{mb} at \qty{14.5}{MeV} (TENDL-2025). Using the Tonchev~\cite{tonchev1998deformation} measured branching ratio into the \(^{150\mathrm{m}}\)Eu isomer, the effective cross section ultimately feeding \(^{150}\)Gd peaks near \qty{238}{mb} at \qty{14}{MeV}. The cross sections and isomer branching are shown in \Cref{fig:Eu151_gn}.

Other photonuclear medical isotope production processes are already being pursued, including for \(^{67}\)Cu. Recent work demonstrated production of about 300 GBq ($\sim$ \qty{10}{\micro g}) per six day batch in a \qty{40}{MeV} rhodotron \cite{hawkins2025}. As a rough, conservative heuristic estimate we can assume similar target design between these cases and compare estimated production rates based on the ratio of peak cross sections. Because the \(^{151}\)Eu(\(\gamma\),n)\(^{150\mathrm{m}}\)Eu reaction cross section is about 80 times larger, this corresponds to about \qty{1.92}{mg} of \(^{150}\)Gd produced per six day run, or about \qty{105}{mg} over a year of operation at 90$\%$ uptime.

\subsection{Fast-neutron route: \(^{151}\)Eu(n,2n)\(^{150\mathrm{m}}\)Eu \(\to\) \(^{150}\)Gd}

\begin{figure}[tb]
\centering
\includegraphics[width=0.85\textwidth]{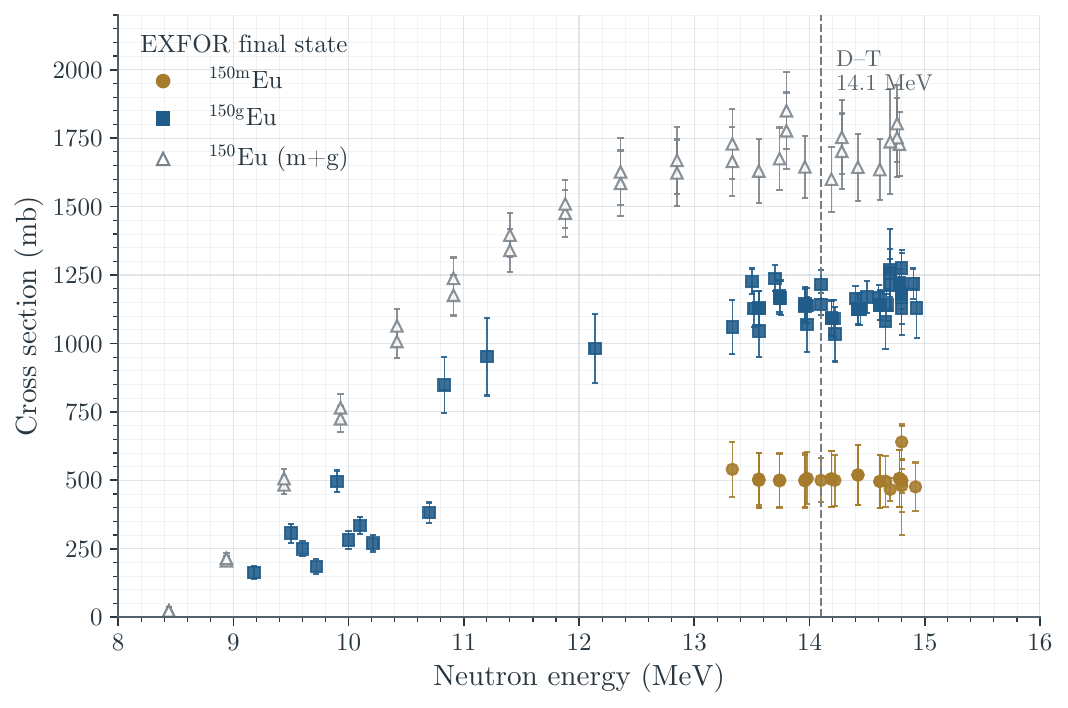}
\caption{EXFOR cross sections of (n,2n) reactions on \(^{151}\)Eu to produce the metastable and ground states of \(^{150}\)Eu. Selected measurements are shown.}
\label{fig:eu150_m_g_from_neutrons}
\end{figure}

The third primary route bombards \(^{151}\)Eu with fast neutrons (e.g., \qty{14.1}{MeV} D-T fusion neutrons),
\begin{equation}
    {}^{151}\mathrm{Eu}(\mathrm{n},\mathrm{2n}){}^{150 \mathrm{m}}\mathrm{Eu} \xrightarrow{\beta^-\ (12.8\,\mathrm{h},\,89\%)} {}^{150}\mathrm{Gd}.
\end{equation}
The (n,2n) cross section to the \qty{12.8}{h} \(^{150\mathrm{m}}\)Eu isomer is measured at $\approx\qty{0.5}{b}$ at \qty{14}{MeV}, shown in \Cref{fig:eu150_m_g_from_neutrons}.

 For the yield estimate, we solve the Bateman system for this chain with cross sections at \qty{14.1}{MeV} averaged across all available EXFOR measurements (\qty{495}{mb} for \(^{150\mathrm{m}}\)Eu and \qty{1125}{mb} for \(^{150\mathrm{g}}\)Eu \cite{nethaway1972cross,konno1993activation,qaim1996excitation,meadows1996measurement,filatenkov2016neutron,luo2018activation}). \Cref{tab:n_route_eu_inventory} shows the annual \(^{150}\)Gd production rate for ${}^\mathrm{nat}\mathrm{Eu}$ and enriched ${}^{151}\mathrm{Eu}$ blankets, along with the Eu blanket mass. Solving the Bateman equations, we estimate a \(^{150}\)Gd production rate of \qty{223}{\mg} per kilowatt year of deuterium-tritium fusion power (corresponding to a \num{3.6e14} neutrons/s source operating at 100\% uptime). Therefore a one megawatt D-T fusion neutron source could produce \qty{223}{g} of $^{150}$Gd per year and with a first-wall fast neutron flux of \qty{e14}{\cm^{-2} s^{-1}} would require $\sim$\qty{250}{kg} of enriched \ce{^151Eu} (wall area $A = S/\phi \approx \qty{3600}{\cm\squared}$ at the areal density $\sigma_{\mathrm{Eu}}^{\mathrm{areal}} \approx \qty{70}{\g/\cm\squared}$ for $50\%$ absorption).
 Using natural europium feedstock would produce roughly \qty{107}{g} of \ce{^150Gd} per year under the same assumptions, accompanied by other Gd isotopes. We examine their relative abundances below.
Fusion neutrons have been proposed before to transmute materials into isotopes whose value per neutron exceeds the electricity value \cite{rutkowski2025scalable,parisi2025isotope,parisi2026neutronvalue,Parisi2026BetaBattery,Parisi2026FusionBattery}, sometimes by many orders of magnitude ~\cite{engholm1986radioisotope,parisi2025j,evitts2025theoretical,parisi2026scalable}.
Producing $^{150}$Gd via fast-neutron irradiation of europium is not qualitatively different to the other suggested schemes, although it has a significant advantage compared with radioisotopes in that the $^{150}$Gd is long-lived, and therefore does not need to be extracted immediately after being produced.

\begin{table}[!tb]
\centering
\caption{D-T fast-neutron route to $^{150}$Gd.  Annual yield and Eu blanket inventory required for $50\%$ of the incident $14.1$~MeV neutrons to undergo an inelastic reaction. ``Enriched'' uses $100\%$~$^{151}$Eu in the blanket. ``Natural'' uses $^{\mathrm{nat}}$Eu ($47.8\%$~$^{151}$Eu).  The blanket flux $\phi$ at the front face sets the required wall area $A = S/\phi$ and therefore the blanket mass.}
\label{tab:n_route_eu_inventory}
\small
\renewcommand{\arraystretch}{1.35}
\setlength{\tabcolsep}{3pt}
\begin{tabular*}{\textwidth}{@{\extracolsep{\fill}}c c c c c c c@{}}
\toprule
\multirow{3}{*}{\makecell{$S$ \\ (n/s)}} &
\multirow{3}{*}{\makecell{D-T \\ power}} &
\multicolumn{2}{c}{$^{150}$Gd yield (per year)} &
\multicolumn{3}{c}{Natural-Eu blanket mass} \\
\cmidrule(lr){3-4} \cmidrule(lr){5-7}
& & \multirow{2}{*}{enriched $^{151}$Eu} & \multirow{2}{*}{$^{\mathrm{nat}}$Eu} &
\multicolumn{3}{c}{at flux $\phi$ (n cm$^{-2}$ s$^{-1}$)} \\
& & & & $10^{10}$ & $10^{13}$ & $10^{16}$ \\
\midrule
$10^{14}$ & \qty{282}{\watt} & \qty{63}{\milli\gram} & \qty{30}{\milli\gram} & \qty{700}{\kg} & \qty{700}{\g} & \qty{700}{\milli\gram} \\
$10^{17}$ & \qty{282}{\kW} & \qty{63}{\g} & \qty{30}{\g} & \qty{700}{t} & \qty{700}{\kg} & \qty{700}{\g} \\
$10^{20}$ & \qty{282}{\MW} & \qty{63}{\kg} & \qty{30}{\kg} & \qty{700}{kt} & \qty{700}{t} & \qty{700}{\kg} \\
\bottomrule
\end{tabular*}
\end{table}

\subsubsection{Isotopic purity of neutron-produced gadolinium}
\label{ssec:neutron_gd_purity}

\begin{figure}[!tb]
\centering
\includegraphics[width=\textwidth]{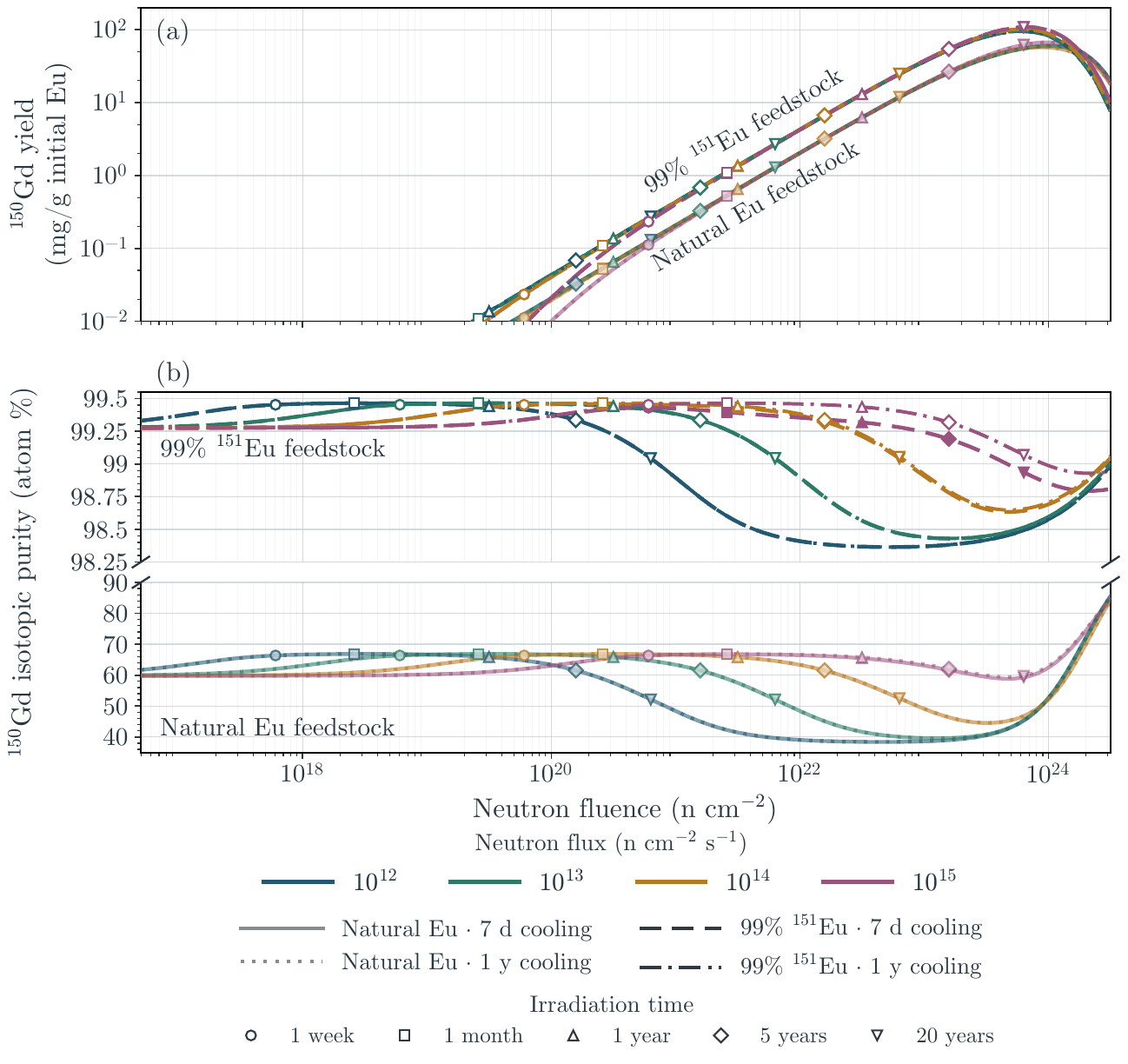}
\caption{$^{150}$Gd production from natural Eu and $99\%$-enriched $^{151}$Eu under uniform irradiation of \qty{14.1}{MeV} neutrons: (a) recovered yield per gram of initial Eu and (b) isotopic atom fraction among Gd isotopes. Gd is separated immediately at end of irradiation and cooled for seven days or one year. The uppermost plotted fluence corresponds to 100 years at \qty{e15}{\cm^{-2} s^{-1}} and 100,000 years at \qty{e12}{\cm^{-2} s^{-1}}.}
\label{fig:neutron_gd_purity}
\end{figure}

We now calculate $^{150}$Gd purity under fusion neutron irradiation. We show the recovered $^{150}$Gd yield and isotopic purity versus neutron fluence in \Cref{fig:neutron_gd_purity}, comparing natural Eu with $99\%$-enriched $^{151}$Eu at four neutron fluxes. Here purity is the atom fraction $x_{150}^{\rm Gd}=N(^{150}\mathrm{Gd})/\sum_A N(^{A}\mathrm{Gd})$ among all Gd isotopes. We solve OpenMC~\cite{romano2015openmc} with isomer cross sections, including the experimental $^{151}$Eu(n,2n) cross sections above and using TENDL-2025 for competing neutron reactions~\cite{tendl2025library}.

In both cooling cases, we separate Gd immediately at the end of irradiation and cool the isolated Gd fraction for seven days or one year. After one year of irradiation at \qty{e14}{\cm^{-2} s^{-1}} and seven days of cooling, the calculated $^{150}$Gd atom fractions are $65.8\%$ for natural Eu and $99.43\%$ for $99\%$-enriched $^{151}$Eu. The main impurity is $^{152}$Gd, fed during irradiation by $^{153}$Eu(n,2n) and by neutron capture followed by europium decay. Longer cooling allows $^{149}$Gd, $^{151}$Gd, and $^{153}$Gd to decay while retaining essentially all $^{150}$Gd. At large fluence $\sim 10^{24}$n cm$^{-2}$, further neutron reactions can significantly reduce the recovered $^{150}$Gd inventory even where its isotopic fraction increases; this is mainly because of $^{150}$Gd(n,$\gamma$) reactions. The reference result for $99\%$-enriched $^{151}$Eu exceeds the $99\%$ $^{150}$Gd feedstock purity assumed in \Cref{tab:practical}.

\section{Stage Two: \(^{150}\)Gd(p,2n)\(^{149\mathrm{g}}\)Tb}\label{sec:step2}

The second stage of our proposed scheme produces \(^{149\mathrm{g}}\)Tb. We propose irradiating a \(^{150}\)Gd target with \qtyrange{14}{25}{MeV} protons to drive the reaction
\begin{equation}
^{150}\text{Gd}\,\text{(p,2n)}\,^{149\mathrm{g}}\text{Tb}\,.
\end{equation}
The $\gtrsim$\qty{14}{MeV} beam energy can be supplied by cyclotrons at over 700 facilities~\cite{iaea_akp_2026}.
\(^{150}\)Gd is a pure alpha emitter~\cite{KHAZOV2016163}, which may facilitate easier handling.

\begin{figure}[!tb]
\centering
\includegraphics[width=0.8\textwidth]{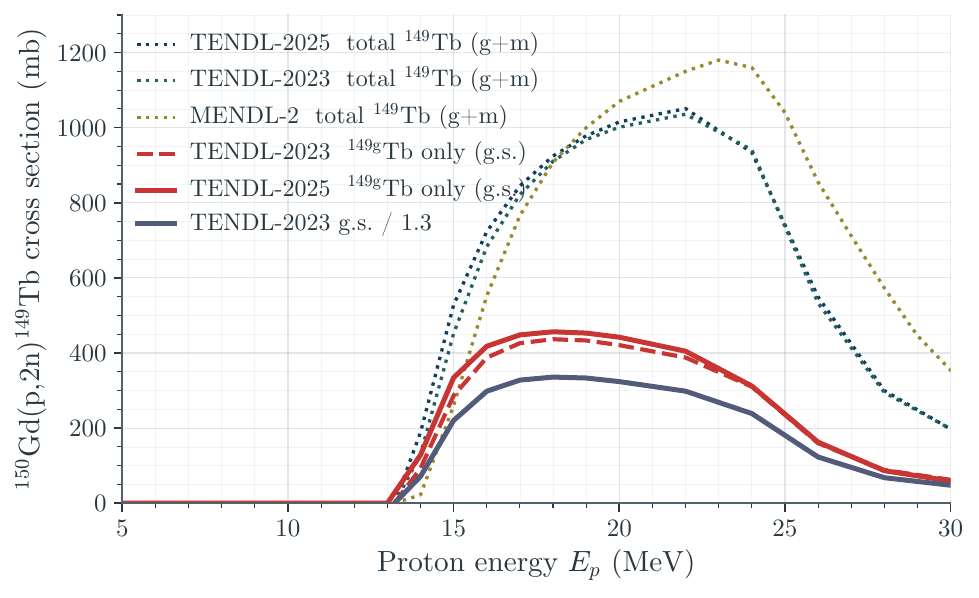}
\caption{\(^{150}\)Gd(p,2n) cross section predictions to populate both the ground and metastable state, and the ground state separately. The total \(^{149}\)Tb (g+m) is shown from MENDL-2, TENDL-2023, and TENDL-2025. The ground-state-only \(^{149\mathrm{g}}\)Tb is shown from TENDL-2023 (dashed) and TENDL-2025 (solid). The additional TENDL-2023 ground-state curve divided by 1.3 illustrates the neighboring-reaction comparison in \ref{sec:yield_validation}. TENDL-2025 modestly increases the \(^{149\mathrm{g}}\)Tb ground-state peak ($\sim$\qty{457}{mb} vs $\sim$\qty{437}{mb} in TENDL-2023 at \qty{18}{MeV}).}
\label{fig:gd150_p2n_XS}
\end{figure}

No experimental measurements of \(^{150}\)Gd(p,2n)\(^{149\mathrm{g}}\)Tb exist. We therefore rely on TENDL evaluations, referenced to data on the closest measured analog channels. In \Cref{fig:gd150_p2n_XS} we plot the TENDL-predicted \(^{150}\)Gd(p,2n) cross sections. Comparisons with six measured neighboring (p,xn) channels give TENDL-to-data peak ratios of \numrange{0.84}{1.22}. A separate TALYS model comparison with three measured Gd(p,2n)Tb channels overpredicts the data by an average factor of \(1.30\pm0.23\). These comparisons support the scale of the prediction but do not replace a measurement of \(^{150}\)Gd(p,2n)\(^{149\mathrm{g}}\)Tb. We use TENDL-2025 for the production estimates below. The comparisons and saturation-yield derivation are given in \ref{sec:yield_validation}.

\subsection{\texorpdfstring{\(^{149\mathrm{g}}\)}{149g}Tb Production Rates}\label{ssec:tb_production_rates}

We now use the ISOTOPIA code \cite{koning2025isotopia} with the TENDL-2025 nuclear data library to calculate \(^{149\mathrm{g}}\)Tb production rates from proton irradiation. We used a TENDL-2025-formatted ISOTOPIA library for the \(^{150}\)Gd + p target. The \(^{150}\)Gd(p,n), \(^{150}\)Gd(p,2n), and \(^{150}\)Gd(p,3n) cross sections are shown in \ref{sec:cross_section_isomer}.

 We use two activity fractions at administration: $P_{149}=A_{149\mathrm{g}}/A_{\rm Tb,total}$ and $P_{149+150}=(A_{149\mathrm{g}}+A_{150})/A_{\rm Tb,total}$. Here $A_{149\mathrm{g}}$ is ground-state Tb-149 activity, $A_{150}$ includes all surviving Tb-150 states, and $A_{\rm Tb,total}$ is total Tb activity. The combined fraction describes composition, and its clinical acceptability remains application dependent. Tb product recovery $\eta_{\rm Tb}$ is the fraction retained through separation, labeling, and dispensing, excluding radioactive decay. Unless stated otherwise, $\eta_{\rm Tb}=1$ and a dose means \qty{50}{MBq} of administered \(^{149\mathrm{g}}\)Tb. \ref{sec:estimate_Tb149g_dose} and~\ref{sec:dose_sensitivity} examine this dose assumption.

\begin{figure}[!tb]
\centering
\includegraphics[width=\textwidth]{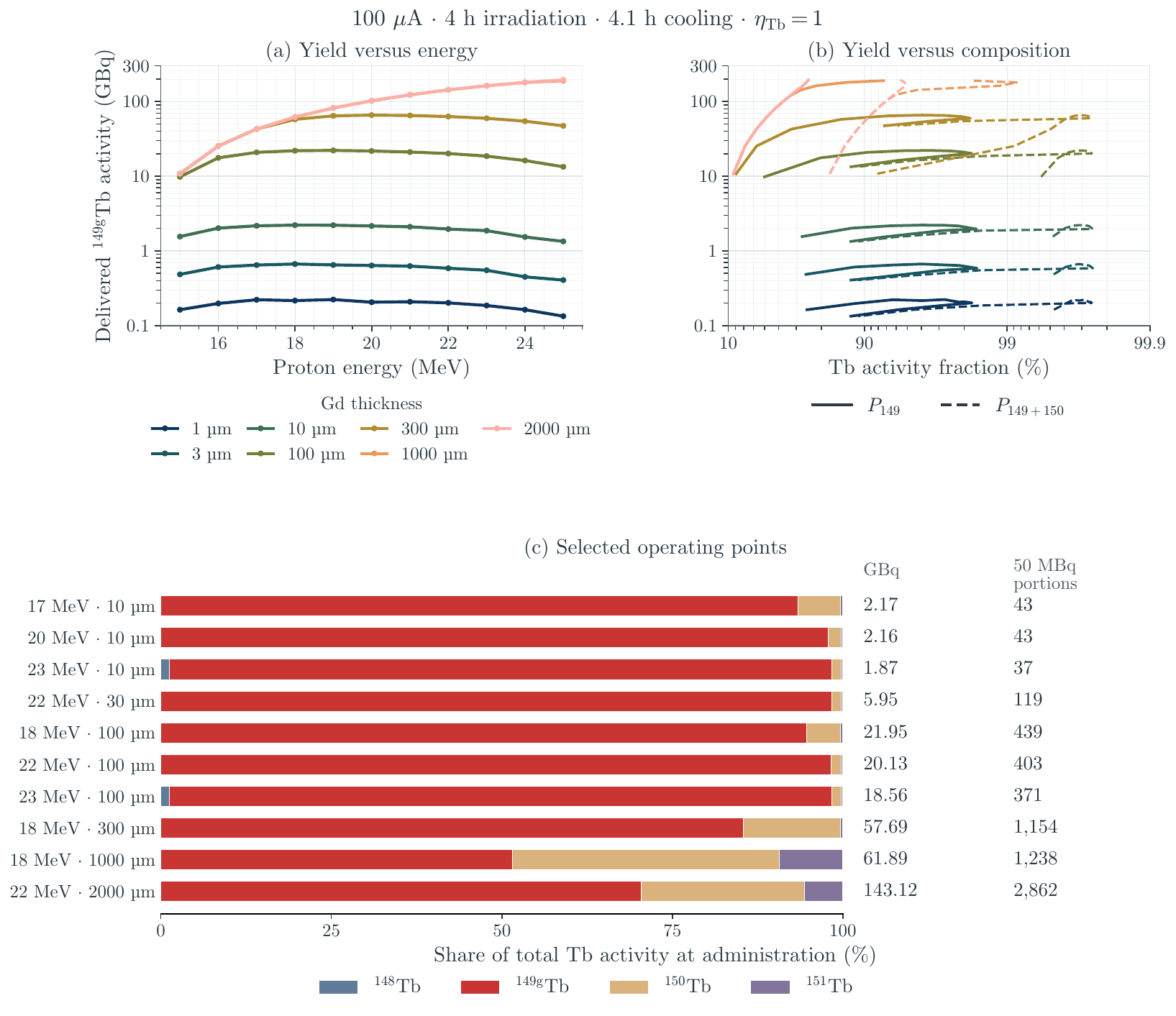}
\caption{ ISOTOPIA-based scan of $^{150}$Gd(p,X) on a isotopically pure $^{150}$Gd target at \qty{100}{\uA}, four-hour irradiation, 4.1-hour cooling, and $\eta_{\rm Tb}=1$. (a) Delivered $^{149\mathrm{g}}$Tb activity versus proton energy, colored by target thickness. (b) Delivered activity versus $P_{149}$ (solid) and $P_{149+150}$ (dashed). Line colors retain the same thickness key. (c) Tb activity composition at selected operating points, with delivered GBq and equivalent \qty{50}{MBq} administrations listed beside each bar. Short-lived trace Tb isomers are omitted.}
\label{fig:gd150_50uA_summary}
\end{figure}

In \Cref{fig:gd150_50uA_summary}, we show the \(^{149\mathrm{g}}\)Tb yield and terbium impurity fraction for a range of \(^{150}\)Gd target thickness and beam energies for a \qty{100}{\uA} proton beam. \Cref{fig:gd150_50uA_summary}(a) shows that for a given target thickness there is a yield-maximizing beam energy. \Cref{fig:gd150_50uA_summary}(b) shows the tradeoff between yield and terbium impurity activity fraction. The solid curves show $P_{149}$ and the dashed curves show $P_{149+150}$. The latter includes the co-produced positron emitter \(^{150}\)Tb, whose clinical suitability must be assessed for each application. The radionuclidic impurity is shown after a \qty{4.1}{\hour} cooldown.

In \Cref{fig:gd150_50uA_summary}(c) we plot the radioisotopic composition, ignoring the very short-lived terbium isomers. A wide range of operating points is available depending on the choice of beam energy and target thickness. A thin foil ($t \lesssim$ \qty{10}{\um}) at \qtyrange{20}{23}{MeV} barely loses beam energy, so the path-averaged cross section sits near the \(^{150}\)Gd(p,2n)\(^{149\mathrm{g}}\)Tb peak ($\sim$\qty{440}{mb}) and well above the \(^{150}\)Gd(p,n)\(^{150\mathrm{g}}\)Tb minimum ($\sim$\qty{6}{mb}), giving a predicted admin-time radionuclidic purity of $\sim$98\%  for $P_{149}$ and $\sim$99.7\% for $P_{149+150}$, at the cost of small yield per irradiation ($\sim$43 doses per 4-hour irradiation at \qty{100}{\uA}). At the other extreme, a thick (\qtyrange{1}{2}{mm}, full-stop) target at \qtyrange{20}{25}{MeV} slows the beam through the \(^{150}\)Gd(p,n) cross section peak near \qty{13}{MeV} and produces a large \(^{150}\)Tb activity ($\sim$24-39\% of total Tb at admin time), maximizing \(^{149\mathrm{g}}\)Tb yield ($\sim$1,200-2,900 doses per 4-hour irradiation at \qty{100}{\uA}). The corresponding activity fractions are $P_{149}\sim52$-70\% and $P_{149+150}\sim90$-95\%.

An important caveat is that these results for yield and impurity fraction all rest on TENDL-2025 (TALYS-2.1) predicted cross sections. To our knowledge, no cross sections for nuclear reactions on \ce{^150Gd} have ever been measured. Therefore an important part of developing this production scheme for \(^{149\mathrm{g}}\)Tb is not just measuring the $^{150}$Gd(p,2n) cross section, but all $^{150}$Gd(p,X) channels.

For a full-stop target at \qty{20}{MeV}, the predicted saturation activity per beam current is \qty{4230}{MBq/\uA}. This is about 18 times the reported \(^{151}\)Eu(\(^{3}\)He,5n) saturation yield per unit current at \qty{70}{MeV}~\cite{moiseeva2020}, while requiring a proton energy available at 113 reported cyclotrons. The full energy dependence and comparison with higher-energy direct routes are in \ref{sec:yield_validation}.

\section{Production and Material Estimates}\label{sec:production}

\begin{table}[!htbp]
\centering
\small
\caption{Production scheme parameters used in \Cref{sec:production}.}
\label{tab:practical}
\centering
\begin{tabular}{|p{0.62\textwidth}|c|}
\toprule
\midrule
$^{149\mathrm{g}}$Tb delivered dose & 50 MBq, \qty{0.27}{ng}  \\
Time to administration following production & \qty{4.1}{h}  \\
\ce{^150Gd} isotopic purity & 99\%at  \\
$^{151}$Eu isotopic purity & 97\%at  \\
$\eta_\mathrm{rec}$, ${}^{150}$Gd recovery efficiency & 99\%  \\
$T_\mathrm{irr}$, irradiation time & \qty{4}{h}  \\
proton beam energy & \qty{18}{MeV} \\
proton beam current & \qty{100}{\uA} \\
$\phi$, proton flux & $6.2 \cdot 10^{14}$ cm$^{-2}$ s$^{-1}$ \\
proton current density (at $E_\mathrm{ref} = \qty{18}{MeV}$, $J(E) = J_\mathrm{ref}\,E_\mathrm{ref}/E$ at fixed \qty{1.8}{\kilo\watt\per\cm^2} heat flux) & \qty{100}{\uA/\cm^2} \\
$m_{\mathrm{dose}}^{{}^{150}\mathrm{Gd}}$, ${}^{150}$Gd mass per delivered ${}^{149 \mathrm{g} }$Tb dose & \qty{1.9}{\ug} \\
$n_{\mathrm{irr}}$ irradiations per year  & 100 \\
\bottomrule
\end{tabular}
\end{table}

In this section we discuss practical considerations of production and material requirement estimates.

We start from the simple assumption that a dose is \qty{50}{MBq} and it requires one $^{149\mathrm{g}}$Tb half-life to extract and process $^{149\mathrm{g}}$Tb ready for chelation.
This means that we need to produce 100 MBq of $^{149\mathrm{g}}$Tb in the \ce{^150Gd} target to produce one deliverable dose of 50 MBq. We will assume the use of 99\% enriched \ce{^150Gd} feedstock and 97\% enriched $^{151}$Eu feedstock. These assumptions are listed in \Cref{tab:practical}.

We compare $^{149\mathrm{g}}$Tb yield per 4-hour irradiation per mg of \ce{^150Gd} target material (which we wish to maximize), and the activity fractions $P_{149}$ and $P_{149+150}$ defined in Section~\ref{ssec:tb_production_rates}.

With these parameters, we can calculate the required mass of \ce{^150Gd} per administered dose $m_{\mathrm{dose}}^{{}^{150}\mathrm{Gd}}$. \ref{sec:feedstock_balance} derives the feedstock balance. Without breeding in the target backing, the recurring make-up per dose is
\begin{equation}
\begin{aligned}
&
m_{\mathrm{dose}}^{{}^{150}\mathrm{Gd}}
& = \left[\frac{1-\eta_{\mathrm{rec}}}{f_{\mathrm{burn}}} + \eta_{\mathrm{rec}}\right] \frac{\sigma_{\mathrm{destr}}}{\sigma_\mathrm{p,2n}} \cdot \frac{T_{\mathrm{irr}}\,D_{\mathrm{MBq}}}{\eta_{\rm Tb}\,f_{\mathrm{sat}} f_{\mathrm{cool}}}\cdot \frac{M_{{}^{150}\mathrm{Gd}}}{N_{\mathrm{A}}},
\label{eq:per-dose_main}
\end{aligned}
\end{equation}
where $\eta_{\mathrm{rec}}$ is the \ce{^150Gd} recovery efficiency after each extraction from a cyclotron run, $\sigma_{\mathrm{destr}}$ is  the effective cross section for net removal of $^{150}$Gd from the recoverable feedstock, including production of $^{149\mathrm{g}}$Tb, with credit for $^{150}$Tb decay back to $^{150}$Gd limited to material returned to the feedstock, $\sigma_\mathrm{p,2n}$ is the $^{150}$Gd(p,2n)$^{149 \mathrm{g} }$Tb cross section, $T_\mathrm{irr}$ is the irradiation time, $D_\mathrm{MBq}$ is the activity per dose, $f_\mathrm{sat} $ and $f_\mathrm{cool}$ are the $^{149 \mathrm{g} }$Tb saturation fraction and cooldown fraction, $M_{{}^{150}\mathrm{Gd}}$ is the ${}^{150}$Gd molar mass, and $N_\mathrm{A}$ is Avogadro's constant. The burn fraction $f_{\rm burn}$ is the fraction of the initial $^{150}$Gd inventory lost to nuclear reactions during one irradiation, after tracking material that decays back into recoverable feedstock. With $\lambda_g=\ln2/T_{1/2}$ for $^{149\mathrm{g}}$Tb, $f_{\rm sat}=1-e^{-\lambda_gT_{\rm irr}}$ and $f_{\rm cool}=e^{-\lambda_gT_{\rm cool}}$, where $T_{\rm cool}$ is the time from end of irradiation to administration. In this expression the units are seconds for $T_{\rm irr}$ and Bq for $D_{\rm MBq}$.

Gd recovery $\eta_{\rm rec}$ and Tb product recovery $\eta_{\rm Tb}$ are separate efficiencies. The latter is the fraction of Tb retained through separation, labeling, and dispensing, excluding radioactive decay already included in $f_{\rm cool}$. If $D_{\rm run}$ denotes administered doses per irradiation, then for a fixed irradiated target $D_{\rm run}(\eta_{\rm Tb})=\eta_{\rm Tb}D_{\rm run}(1)$ and $m_{\rm dose}(\eta_{\rm Tb})=m_{\rm dose}(1)/\eta_{\rm Tb}$. At the reference conditions, the make-up is approximately \qty{1.9}{\micro\gram} per dose. Reducing $\eta_{\rm Tb}$ to 0.8 raises this to \qty{2.38}{\micro\gram} and reduces the dose count by 20\%. Recovery and current-density scans are given in \ref{sec:recovery_sensitivity}. The same recovery is applied to all co-separated Tb isotopes, so their activity fractions are unchanged.

\subsection{Target choice and throughput}\label{ssec:target_choice}

For scarce feedstock another useful metric is $\Dmtb$, the number of administered doses per milligram of initially irradiated $^{150}$Gd. This differs from $m_{\rm dose}^{{}^{150}\mathrm{Gd}}$, the recurring make-up consumed per dose after Gd recovery. $\Dmtb$ is useful because it measures the minimal viable feedstock inventory to produce a desired number of administered doses. At 99\% recovery of $^{150}$Gd feedstock, chemistry losses dominate, giving approximately $\Dmtb=10/m_{\rm dose}$ when $\Dmtb$ is in doses/mg and $m_{\rm dose}$ in $\mu$g/dose.

\begin{figure}[!tb]
    \centering
    \includegraphics[width=\textwidth]{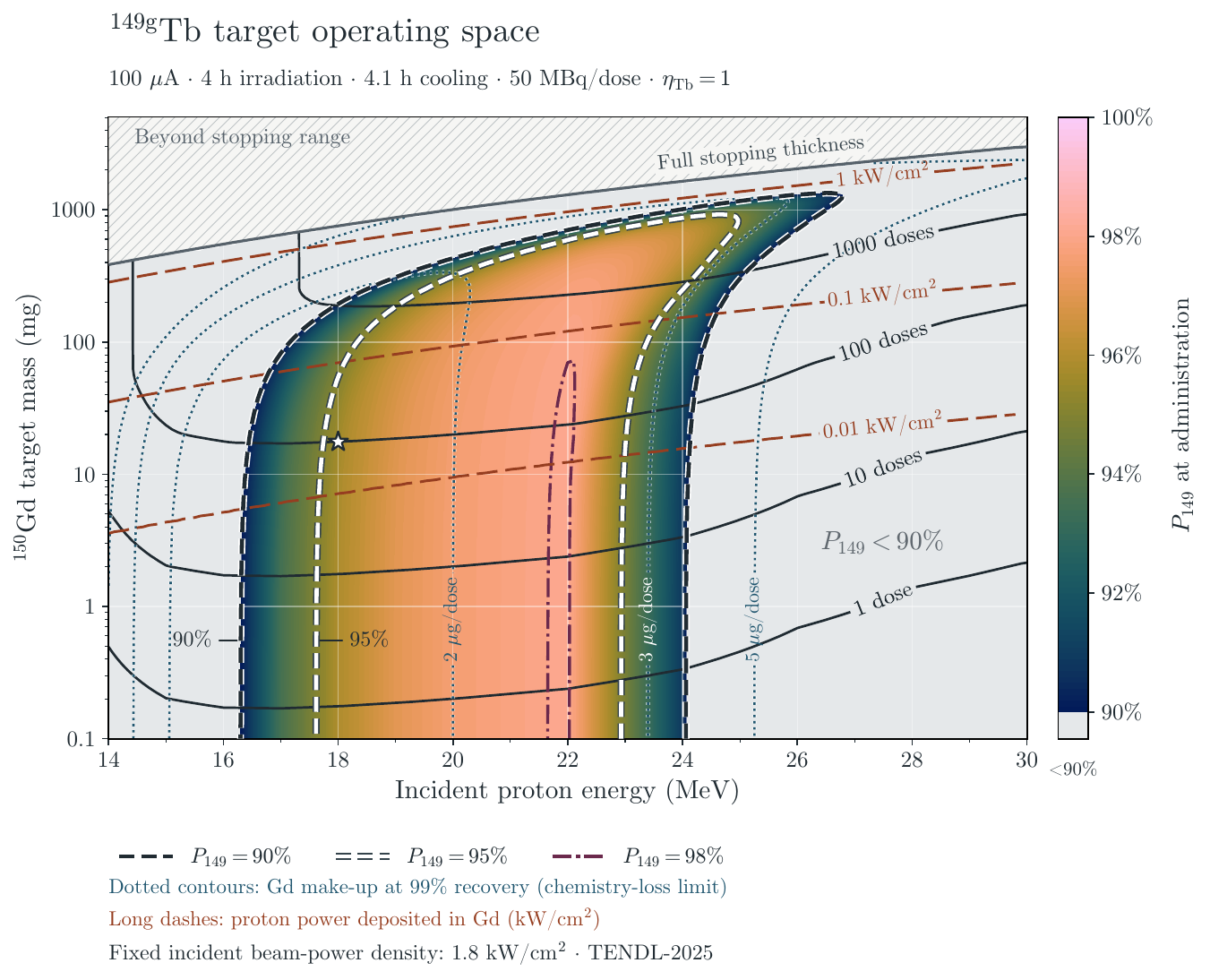}
    \caption{ $^{150}$Gd target operating space for $^{149\mathrm{g}}$Tb production. The star marks the \qty{18}{MeV}, \qty{17.6}{mg} reference target (100 doses, $P_{149}=95.6\%$). Conditions: \qty{100}{\uA}, \qty{4}{\hour} irradiation, \qty{4.1}{\hour} cooling, \qty{50}{MBq} per administration, $\eta_{\rm Tb}=1$, 99\% $^{150}$Gd enrichment, and TENDL-2025. A fixed incident beam-power density of \qty{1.8}{\kW/\cm^2} determines the beam area at each energy. }
    \label{fig:dose_throughput}
\end{figure}

We plot feedstock efficiency against total dose count in \mbox{\Cref{fig:dose_throughput}}. Each curve varies the Gd thickness at a fixed incident proton energy, so the exit energy also varies. These curves show the available operating choices, rather than a single optimized thickness or exit energy. Thin targets give nearly constant doses per milligram, while further thickening eventually reduces feedstock efficiency.

We show the target operating space in \Cref{fig:dose_throughput}, using incident proton energy and Gd target mass as the two design variables. Dose contours and purity colors show the production trade-off directly, and dotted contours give the $^{150}$Gd make-up mass per dose. Long brown dashes show the proton power deposited per unit area in the Gd. At each energy, increasing mass increases target thickness and reduces the proton exit energy.

Under a fixed \qty{1.8}{\kW/\cm^2} incident beam-power density, \qtyrange{16}{20}{MeV} protons give about 5-6 doses per mg in a four-hour irradiation. At \qty{18}{MeV} and \qty{100}{\uA}, a \qty{1.76}{mg} target supplies ten doses with $P_{149}\simeq95.9\%$ and $P_{149+150}\simeq99.4\%$. Increasing the target mass to \qty{17.6}{mg} supplies 100 doses with nearly the same composition. A longer discussion of tradeoffs is in \ref{sec:target_optimization}.

\subsection{Global supply requirements}\label{ssec:global_supply}

\begin{table}[!tb]
\centering
\caption{Summary of the two-stage production chain.}
\label{tab:summary}
\footnotesize
\begin{tabular}{p{6.3cm}p{6.9cm}}
\toprule
\multicolumn{2}{l}{\textbf{Stage 1: $^{150}$Gd feedstock from $^{151}$Eu}} \\
\midrule
protons: $^{151}$Eu(p,2n)$^{150}$Gd & $\sim$85 mg\,yr$^{-1}$ at \qty{1600}{\uA}, \qty{18}{MeV}, 75\% uptime \\
photons: $^{151}$Eu($\gamma$,n)$^{150\mathrm{m}}$Eu \;($\to{}^{150}$Gd) & $\sim$\qty{100}{\mg.yr^{-1}} at 40-85 kW e$^{-}$ Rhodotron ($E_e\!\gtrsim\!14$\,MeV), 90\% uptime \\
neutrons: $^{151}$Eu(n,2n)$^{150\mathrm{m}}$Eu \;($\to{}^{150}$Gd) & $\sim$180 mg\,yr$^{-1}$ per kilowatt of D-T fusion power, 80\% uptime \\
\midrule
\multicolumn{2}{l}{\textbf{Source required for global TAT supply ($\sim$4M doses\,yr$^{-1}$)}} \\
\midrule
Replacement rate & $\sim$7.5 g\,yr$^{-1}$ \(^{150}\)Gd (assuming $m_\mathrm{dose}=$1.9 $\mu$g) \\
\quad via protons & $\sim$140 mA at 18 MeV, 75\% uptime \\
\quad via photons & $\sim$3.0-6.4\,MW e$^{-}$, 90\% uptime \\
\quad via D-T neutrons & $\sim$\qty{42}{\kW}, 80\% uptime \\
\midrule
\multicolumn{2}{l}{\textbf{Stage 2: $^{149\mathrm{g}}$Tb in cyclotron}} \\
\midrule
Target & Thin $^{150}$Gd-containing material \\
Beam & $\gtrsim$14 MeV protons \\
Assumed $^{149\mathrm{g}}$Tb therapeutic dose & $\sim$\qty{50}{MBq} ($\sim$0.25\,ng) \\
Radionuclidic purity (admin time) &  $P_{149+150}\sim90$-99.7\%, $P_{149}\sim70$-98\% (thin to full-stop targets) \\
Specific activity & near carrier-free, $\sim$$2\times10^{5}$ TBq\,g$^{-1}$ ($\approx$ theoretical $^{149\mathrm{g}}$Tb) \\
\bottomrule
\end{tabular}
\end{table}

The total quantity of \(^{150}\)Gd needed to serve the global Targeted Alpha Therapy (TAT) patient population scales linearly with $m_\mathrm{dose}$ and the number of doses per year.
There are an estimated $\sim$20~million new cancer cases per year worldwide~\cite{globocan2022} and $\sim$10~million cancer deaths per year, of which $\sim$90\% may be attributable to metastatic cancer~\cite{dillekaas201990}. Therefore even modest penetration of radionuclide therapy into the metastatic population implies a large number of patients who could benefit from TAT.
 Taking an illustrative addressable population of 10 million patients per year, treating 10\% with four \qty{50}{MBq} cycles each requires 4 million doses per year.
At the reference conditions of \Cref{tab:practical}, $m_\mathrm{dose}= \qty{1.9}{\ug}$ of \(^{150}\)Gd is required per dose, corresponding to \qty{7.5}{\g} of \(^{150}\)Gd used per year.
A facility with 100 ${}^{150}$Gd irradiations per year and 100 therapeutic doses per irradiation could provide 10,000 treatments per year, and would require a supply of $\sim$\qty{19}{\mg} of ${}^{150}$Gd per year (assuming $m_\mathrm{dose}=\qty{1.9}{\ug}$) and an initial $^{150}$Gd target mass of $\sim$18 mg. Four hundred such facilities would be needed to provide 4 million doses of \(^{149\mathrm{g}}\)Tb per year. The \qty{7.5}{\g} of \(^{150}\)Gd per year requirement can fall significantly with improvements in $^{150}$Gd recovery efficiency.  We can consider the feasibility of scaling each of the three $^{150}$Gd production schemes identified earlier in this work:

\paragraph{Proton Sources} A \qty{1.6}{mA} \qty{18}{MeV} proton source with 75\% uptime over a year produces $\sim$\qty{85}{mg} of \(^{150}\)Gd per year (\Cref{fig:Eu151_thickness_cooled}), enough for $\sim$ 45,000 doses or $\sim$ 11,000 patients. While this likely would be far more than enough doses to support \(^{149\mathrm{g}}\)Tb  clinical trials, it would not suffice for scaling to millions of doses per year at $m_\mathrm{dose}= \qty{1.9}{\micro\gram}$.

\paragraph{Photon Sources} An IBA rhodotron of the type used at Northstar ($\sim$\qtyrange{40}{85}{\kW} electron beam power) ~\cite{hawkins2025} produces $\sim$\qty{100}{\mg} of \(^{150}\)Gd per year, enough for $\sim$ 53,000 doses. Similar to the proton source, this would be more than enough for clinical trials for $\sim$ 13,000 patients per year, but insufficient for millions of doses per year at $m_\mathrm{dose}=\qty{1.9}{\ug}$.

\paragraph{Neutron Sources}
We estimate D-T neutron sources can produce \qty{223}{\mg} per kilowatt year of fusion power. Assuming $\sim$80\% uptime, \qty{7.5}{\g} of \(^{150}\)Gd per year corresponds to $\sim$\qty{42}{\kW} of fusion capacity.
While such power fusion sources are unavailable in 2026, with only several hundred watts available commercially, we expect that this capacity will be available within five years.
Assuming a FLARE-like deuteron beam tritium gas target \cite{shine_neutron_imaging} supplies \qty{100}{W} of steady state D-T power, $\sim$\qty{20}{\mg} of \(^{150}\)Gd could be produced per year - although given the relatively low neutron flux, it would require significant quantities of enriched ${}^{151}$Eu.
It would be more economic to run with ${}^\mathrm{nat}$Eu, yielding $\sim$\qty{10}{mg} of \(^{150}\)Gd per year, although isotope enrichment may be necessary due to the co-produced \(^{152}\)Gd with a natural europium target.
In the medium- to long-term, if $m_\mathrm{dose}=\qty{1.9}{\micro\gram}$, we expect that the fusion neutron industry will have capacity to support millions of \(^{149\mathrm{g}}\)Tb doses if/when \(^{149\mathrm{g}}\)Tb receives FDA approval.

Over time, we also expect that $m_\mathrm{dose}$ will decrease as extraction and recovery processes become more efficient, and as the maximum-achievable proton beam flux increases. This reduces the quantity of \(^{150}\)Gd required from external sources. An enriched \(^{151}\)Eu target backing can replenish some of the consumed feedstock, as analyzed in \ref{sec:self_sufficient}.

These supply estimates assume \qty{50}{MBq} per administration. At \qty{1}{GBq}, dose counts fall by a factor of 20 and make-up per dose rises by the same factor. Four million such administrations require about \qty{150}{g} of $^{150}$Gd annually, corresponding to roughly \qty{0.85}{MW} of D-T fusion power at 80\% uptime. \ref{sec:dose_sensitivity} gives the full sensitivity analysis for irradiation and feedstock requirements as a function of dose size.

\begin{figure}[!tb]
\centering
\includegraphics[width=\textwidth]{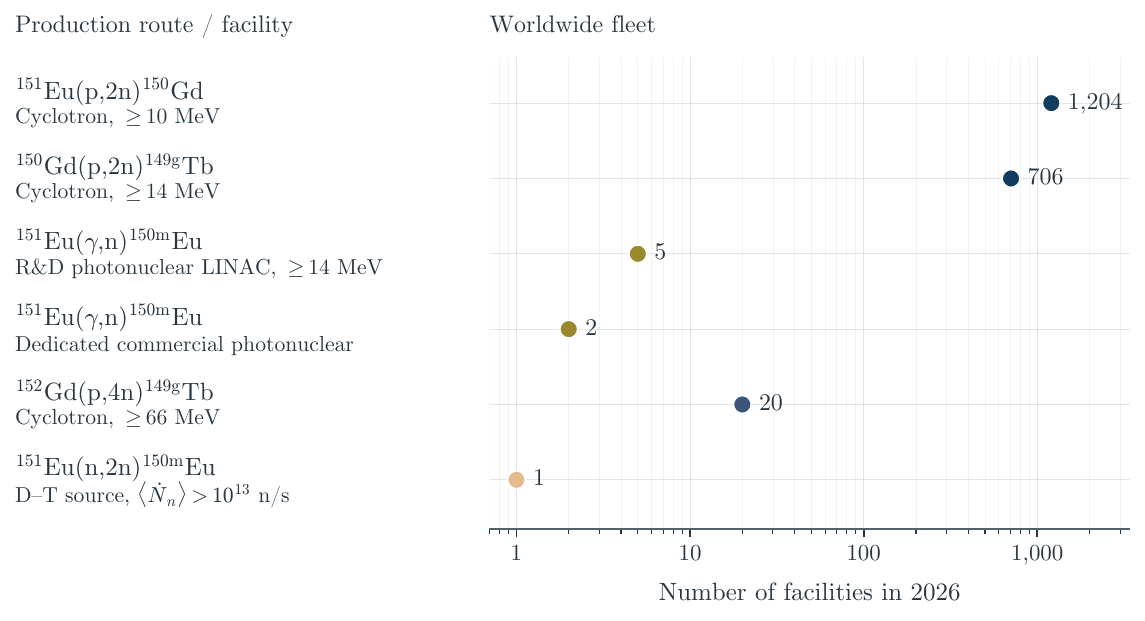}
\caption{Estimated number of facilities worldwide capable of \(^{150}\)Gd or \(^{149\mathrm{g}}\)Tb production at present (2026).}
\label{fig:production_facilities}
\end{figure}

In \Cref{fig:production_facilities} we plot the reported number of facilities worldwide capable of \(^{150}\)Gd or \(^{149\mathrm{g}}\)Tb production at present (2026).

\subsection{\texorpdfstring{Time to a preclinical dose}{Time to a preclinical dose}}\label{ssec:preclinical_ramp}

In this section, we briefly discuss how quickly a preclinical animal dose of \mbox{\(^{149\mathrm{g}}\)}Tb could be produced. The first supply milestone is enough \(^{149\mathrm{g}}\)Tb for animal studies, which typically use \qtyrange{2}{6}{MBq} per animal. Beyer et al.\ \cite{beyer2004} administered \qty{5.5}{MBq} of \(^{149}\)Tb-rituximab per mouse. M\"{u}ller et al.\ \cite{muller2014} used \qty{2.2}{MBq} and \qty{3.0}{MBq} of the \(^{149}\)Tb-folate conjugate cm09. Umbricht et al.\ \cite{umbricht2019} used \qty{5}{MBq} of \(^{149}\)Tb-PSMA-617 for PET/CT and either \qty{6}{MBq} once or \qty{3}{MBq} twice for therapy. Mapanao et al.\ \cite{mapanao2025} gave one or two \qty{5}{MBq} administrations of [\(^{149}\)Tb]Tb-DOTATATE and [\(^{149}\)Tb]Tb-DOTA-LM3. We therefore take \qty{5}{MBq} as a representative preclinical dose, one tenth of the reference human dose. The smaller feedstock requirement allows animal studies to begin with \(^{150}\)Gd bred on the same cyclotron that produces the \(^{149\mathrm{g}}\)Tb.

\begin{figure}[!tb]
\centering
\includegraphics[width=\textwidth]{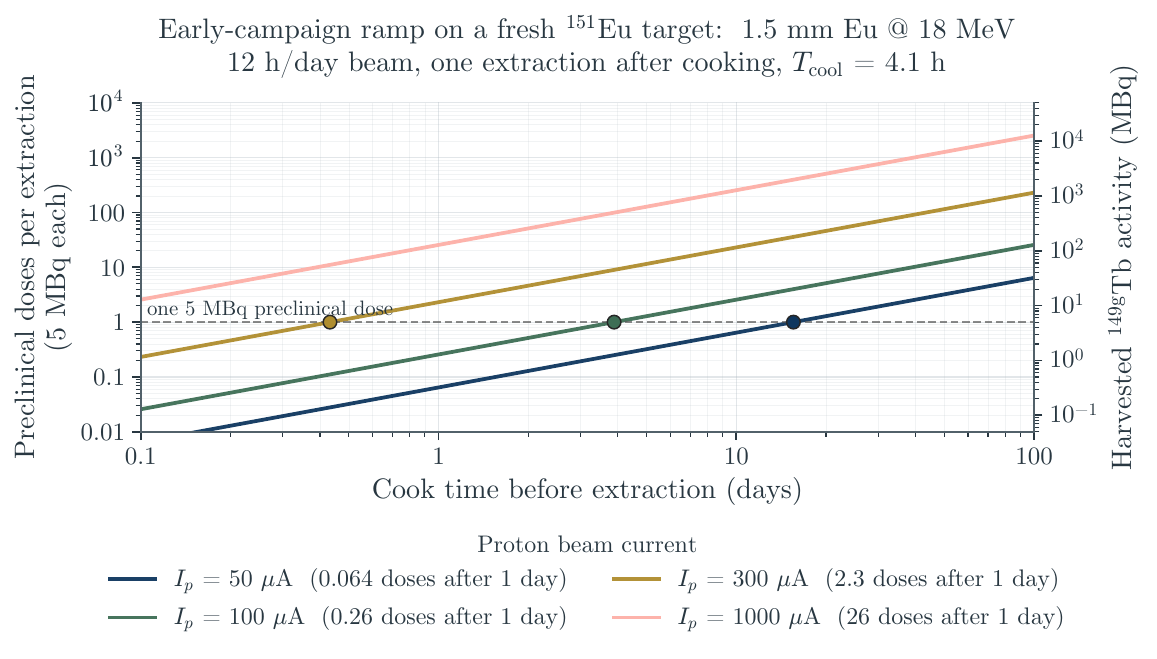}
\caption{ Time to a \qty{5}{MBq} preclinical \(^{149\mathrm{g}}\)Tb dose from a fresh \qty{1.5}{mm} enriched \(^{151}\)Eu target at four proton beam currents. Irradiation is at \qty{18}{MeV} for \qty{12}{h} per day, with one extraction after cook time $t$ and \qty{4.1}{h} cooling before administration. Curves use a duty-averaged, quasi-steady approximation for Tb activity. The dashed line marks one dose, and filled circles mark the threshold crossings. The right axis shows harvested \(^{149\mathrm{g}}\)Tb activity.}
\label{fig:preclinical_ramp}
\end{figure}

We plot the preclinical dose yield from a single extraction versus irradiation time in \Cref{fig:preclinical_ramp}. In this ``cook and harvest'' approach, irradiation first builds $^{150}$Gd within the Eu target and then converts part of it to $^{149\mathrm{g}}$Tb. We solve the coupled \(^{151}\)Eu \(\to\) \(^{150}\)Gd \(\to\) \(^{149\mathrm{g}}\)Tb system described in \ref{ssec:cook_harvest} for a fresh \qty{1.5}{mm} enriched \(^{151}\)Eu target irradiated at \qty{18}{MeV} for \qty{12}{h} per day. Here, extraction occurs only once, after a cook time $t$. Gd recovery losses therefore do not affect the first Tb yield, since they arise when Gd is recovered for later runs. Tb product recovery remains $\eta_{\rm Tb}=1$. We vary the proton beam current to determine how quickly the target reaches the required activity.

At \qty{300}{\uA}, one day of irradiation yields approximately \qty{11}{MBq} at administration, enough for two preclinical doses. Four days gives ten doses, and one month about 70. At \qty{100}{\uA}, typical of a hospital PET cyclotron, the first dose is available after about four days, and a month gives about eight doses. At \qty{1}{mA}, one day gives about 26 doses and a month gives several hundred. The yield depends approximately quadratically on current because increasing the current both builds the \(^{150}\)Gd inventory faster and converts it into \(^{149\mathrm{g}}\)Tb faster (\ref{ssec:cook_harvest}).

These estimates assume no \(^{150}\)Gd is initially present. However, a laboratory with an existing inventory could begin sooner. A \qty{5}{MBq} administration contains about \qty{26}{pg} of \(^{149\mathrm{g}}\)Tb and requires about \qty{18}{\micro g} of initially irradiated \(^{150}\)Gd under the thin-target reference conditions: \qty{18}{MeV}, \qty{100}{\uA/\cm^2}, four-hour irradiation, 4.1-hour cooling, and unit Tb recovery. The recurring make-up is much smaller, about \qty{0.2}{\micro g} per dose at 99\% Gd recovery (\ref{sec:gd_150_per_dose}). Tens of micrograms of \(^{150}\)Gd would therefore allow a laboratory to bypass the cook stage and begin preclinical production. Animal studies could proceed while \(^{150}\)Gd supply is scaled up (\Cref{sec:step1}).

\section{Discussion}\label{sec:discussion}

We have described a new production scheme for clinically-relevant quantities of alpha emitter \(^{149\mathrm{g}}\)Tb via the \(^{150}\)Gd(p,2n) reaction.
This is enabled by production of the feedstock \(^{150}\)Gd, an isotope that is currently extinct. \(^{150}\)Gd has a 1.79 million year half-life and therefore can be stockpiled and easily distributed. Irradiation of \(^{150}\)Gd with $\gtrsim$\qty{14}{MeV} protons is predicted to produce significant quantities of \(^{149\mathrm{g}}\)Tb due to its predicted cross section of hundreds of millibarns.

The next step is to measure the \(^{150}\)Gd(p,2n) cross section, along with the other proton-driven channels on \(^{150}\)Gd. If upcoming measurements confirm the TENDL-2025 \(^{150}\)Gd(p,2n)\(^{149\mathrm{g}}\)Tb prediction (and the comparably suppressed \(^{150}\)Gd(p,n)\(^{150\mathrm{g}}\)Tb channel), this route would scale to clinical supply on existing cyclotrons. Clinical development of $^{149\mathrm{g}}$Tb-based TAT has historically been held back by isotope supply. A globally scalable supply route addresses this constraint.

The feedstock \(^{150}\)Gd can be made via proton, photon, or neutron pathways on \(^{151}\)Eu. Long-term, we expect the fast-neutron pathway to be the most scalable, with production rates of approximately 220 grams per megawatt year of deuterium-tritium fusion power, more than enough to supply the global demand of \(^{150}\)Gd with a single near-term neutron source.

There are many other considerations that require further investigation. These include the impurity effects of \(^{152}\)Gd in \(^{150}\)Gd targets, handling of alpha emitters in targets such as \(^{148-152}\)Gd, the composition and design of the gadolinium and europium targets (such as metallic foils versus electroplated metal oxides), the optimal proton beam energy for minimizing terbium impurities and other elements such as \(^{148}\)Gd from \(^{148}\)Tb decay, the tolerability of co-produced \(^{150\mathrm{g}}\)Tb that has a similar half-life to \(^{149\mathrm{g}}\)Tb, and the clinical effectiveness of \(^{149\mathrm{g}}\)Tb.

If the predicted cross sections are confirmed, we expect this scheme to be among the most, if not the most scalable of any current alpha-emitter route for TAT, with margin to remain attractive even if the measured cross sections are  lower by a factor of several than the TENDL predictions. Three factors drive this scalability. First, the starting feedstock is inexpensive and abundant europium. Second, at least three independent paths produce ${}^{150}$Gd from europium (proton, photon, fast-neutron). Lastly, the second-stage \(^{150}\)Gd(p,2n)\(^{149\mathrm{g}}\)Tb step runs on widely available proton cyclotrons. Additional potential benefits include production of ${}^{149}$Tb on-demand, as opposed to generator pathways with deterministic decay losses, as well as the potential availability of theranostic pairs ${}^{152}$Tb and ${}^{155}$Tb.

\section*{Acknowledgments}

We are grateful for conversations with J. W. Engle, L. A. Bernstein, R. Schibli, H. VanBrocklin, N. P. van der Meulen, and J. S. Wexler, and to J. A. Schwartz for reviewing the paper.

\section*{AI-Usage Statement}

We made extensive use of Claude Opus and Fable in preparing this paper, primarily in automating large numbers of OpenMC and ISOTOPIA simulations and subsequent data visualization.

\section*{Data Availability and Reproducibility Statement}

Upon publication, we intend to make all scripts producing the results in this paper available on a public, permanent data repository.

\clearpage
\appendix

\section{Required therapeutic activity per cycle} \label{sec:estimate_Tb149g_dose}

In this appendix, we estimate the \(^{149\mathrm{g}}\)Tb per-cycle activity \(A_{^{149\mathrm{g}}\mathrm{Tb}}\) by analogy to the clinical dose of [\(^{212}\)Pb]Pb-TCMC-trastuzumab (\qtyrange{200}{500}{kBq/\kg}, or \qtyrange{14}{35}{MBq} for a \qty{70}{kg} patient \cite{poty2018alpha,meredith2014}). We emphasize that this is an initial estimate, and that detailed analysis and clinical trials are required to determine required dose size.

Cell killing in TAT can be approximated by a single-hit survival law \(S \approx e^{-A/A_0}\), where \(S\) is the surviving fraction of target cells, \(A\) is administered activity, and \(A_0\) is the activity reducing survival to \(1/e\) \cite{mird22}. The cytotoxic potency \(\mathcal{P} \equiv 1/A_0\) is heuristically,
\begin{equation}
\mathcal{P} \propto \tau_{\rm eff}\, f_\alpha \cdot g \cdot \mathrm{RBE}(L),
\label{eq:potency}
\end{equation}
 where the target time-integrated activity coefficient is \[\tau_{\rm eff}=\int_0^\infty A_{\rm target}(t)\,dt/A_{\rm admin},\] and \(f_\alpha\) is the alpha branching ratio per parent decay, \(g\) is the geometric efficiency (energy deposited in the target cell per emitted alpha), and \(\mathrm{RBE}(L)\) is the linear energy transfer (LET)-dependent relative biological effectiveness (RBE) for cell killing for linear energy transfer $L$ \cite{barendsen1968}. The expression in \Cref{eq:potency} is approximate since \(g\) and \(\mathrm{RBE}(L)\) are related through alpha energy, and  the proportionality constant absorbs cell-scale conversion factors. Uptake, target retention, and biological clearance are represented by $\tau_{\rm eff}$.
We assume this constant is matched between the two radiopharmaceuticals, valid for fast clearing peptides where physical half-life differences (\(^{149\mathrm{g}}\)Tb \qty{4.12}{\hour} vs \ce{^212Pb} \qty{10.6}{\hour}) do not substantially alter integrated decays at the target.

Matching potency between the two isotopes gives the required \(^{149\mathrm{g}}\)Tb activity,
\begin{equation}
\frac{A_{^{149\mathrm{g}}\mathrm{Tb}}}{A_{^{212}\mathrm{Pb}}} =  \frac{A_0^{149\mathrm{g}}}{A_0^{212}} = \frac{f_\alpha^{212}}{f_\alpha^{149\mathrm{g}}} \cdot \frac{\mathrm{RBE}(L_{^{212}\mathrm{Pb}})}{\mathrm{RBE}(L_{^{149\mathrm{g}}\mathrm{Tb}})}\cdot \frac{g_{^{212}\mathrm{Pb}}}{g_{^{149\mathrm{g}}\mathrm{Tb}}}\,\frac{\tau_{\rm eff}^{212}}{\tau_{\rm eff}^{149\mathrm{g}}}.
\label{eq:dosetranslation}
\end{equation}
Our goal is to calculate $A_{^{149\mathrm{g}}\mathrm{Tb}}$. The \ce{^212Pb} chain has one alpha per decay (\(f_\alpha^{212}=1.0\)) via \(^{212}\)Bi (35.9\%, \qty{6.05}{MeV}) and \(^{212}\)Po (64.1\%, 8.78 MeV), whereas \(^{149\mathrm{g}}\)Tb has one alpha in 16.7\% of decays at \qty{3.97}{MeV} (\(f_\alpha^{149\mathrm{g}}=0.167\)).
The \ce{^212Pb} chain alphas have track-averaged LETs of \qty{97}{keV/\um} (\ce{^212Bi}) and \qty{76}{keV/\um} (\ce{^212Po}) from  NIST ASTAR stopping powers \cite{nist_pstar}, giving a branching-weighted \(L_{^{212}\mathrm{Pb}}=\qty{83}{keV/\um}\).
\(^{149\mathrm{g}}\)Tb delivers \(L_{149\mathrm{g}}=\qty{143}{keV/\um}\) \cite{mird22}, so \(\mathrm{RBE}(L_{^{212}\mathrm{Pb}})\approx 4.8\) and \(\mathrm{RBE}(L_{149\mathrm{g}})\approx 6.8\).\footnote{We approximate the Barendsen 1968 V79 cell-killing RBE-vs-LET curve \cite{barendsen1968} by piecewise-linear interpolation through the points $(L, \mathrm{RBE}) = (50, 3.5),\,(100, 5.5),\,(150, 7.0)$ in keV/\(\mu\)m. This gives \(\mathrm{RBE}(L) = 3.5 + 0.04\,(L-50)\) for \(L \in [50,100]\) and \(\mathrm{RBE}(L) = 5.5 + 0.03\,(L-100)\) for \(L \in [100,150]\), giving \(\mathrm{RBE}(83)\approx 4.8\) and \(\mathrm{RBE}(143)\approx 6.8\). Both LETs are below the \(\sim\!150\)~keV/\(\mu\)m peak.}

The \qty{28}{\um} \(^{149\mathrm{g}}\)Tb range, versus \qtyrange{50}{85}{\um} for the \ce{^212Pb} chain, confines the alpha deposition to 2-4 cell diameters. Cellular dosimetry \cite{charlton1994} and surface-labeled cytotoxicity assays \cite{beyer2004} report  a per-alpha advantage of 3.7 of \(^{149\mathrm{g}}\)Tb over \(^{213}\)Bi \cite{mird22} which, divided by the LET-RBE ratio of 1.42 gives a geometric factor \(g_{^{149\mathrm{g}}\mathrm{Tb}}/g_{^{212}\mathrm{Pb}}\approx 2.6\) for surface-labeled cells.
 A more rigorous treatment using cellular S-values would refine this estimate. The following numerical example assumes $\tau_{\rm eff}^{212}=\tau_{\rm eff}^{149\mathrm{g}}$. The \qty{50}{MBq} reference remains a production-planning scenario rather than an established therapeutic activity. Substituting into Eq.~\ref{eq:dosetranslation} with the upper bound \(A_{^{212}\mathrm{Pb}}=\qty{35}{MBq}\) gives
\begin{equation}
A_{^{149\mathrm{g}}\mathrm{Tb}} = \,(35\,\mathrm{MBq}) \cdot \frac{1.0}{0.167} \cdot \frac{4.8}{6.8} \cdot \frac{1}{2.6} \approx \qty{57}{MBq}.
\end{equation}
We use \qty{50}{MBq} per cycle (\qty{0.27}{ng} of \(^{149\mathrm{g}}\)Tb), the upper bound dose considered in \cite{moiseeva2020}. For comparison, \cite{moiseeva2024} considers a much larger \qtyrange{10}{55}{MBq/\kg} dose range, equivalent to \qtyrange{0.7}{3.85}{GBq} for a \qty{70}{\kg} patient, more than an order of magnitude above our estimate. We retain the \qty{50}{MBq} figure for production-scale calculations throughout the paper, and perform a sensitivity scan to dose size in \ref{sec:dose_sensitivity}.
The intravenous route used here, against the intraperitoneal reference for \ce{^212Pb} \cite{meredith2014}, pushes this higher so bulky solid tumors with \(g_{149\mathrm{g}}/g_{^{212}\mathrm{Pb}}\to 1\) would require \(\sim\)\qtyrange{60}{150}{MBq}.
These are all simple estimates for the purposes of estimating dose size for the purposes of the main discussion in this paper.

\section{Dose Size Sensitivity}

\label{sec:dose_sensitivity}

In this appendix, we show how production metrics change as the assumed dose size $D_{\mathrm{MBq}}$ varies. Throughout this paper we adopt a reference administered activity of \qty{50}{MBq} per $^{149\mathrm{g}}$Tb dose, justified on potency-matching grounds in \ref{sec:estimate_Tb149g_dose}. The true therapeutic dose is not yet established:  estimates in the literature range from $\sim$\qty{10}{MBq} to $\sim$\qty{1}{GBq}, with \cite{moiseeva2024} considering \qtyrange{0.7}{3.85}{GBq} for a \qty{70}{\kg} patient, more than an order of magnitude above our \qty{50}{MBq} estimate.

For a fixed target and beam the delivered $^{149\mathrm{g}}$Tb activity per irradiation is set by the physics of \Cref{sec:step2} and does not depend on how it is divided among patients, so the number of administrable doses per irradiation is $A^{\mathrm{admin}}_{^{149\mathrm{g}}\mathrm{Tb}} / D_{\mathrm{MBq}}$ and scales as $1/D_{\mathrm{MBq}}$. Here $A^{\mathrm{admin}}_{^{149\mathrm{g}}\mathrm{Tb}}$ is the total recovered activity at administration, expressed in the same units as $D_{\rm MBq}$. The per-dose $^{150}$Gd feedstock requirement of \Cref{eq:per-dose_main} scales linearly with $D_{\mathrm{MBq}}$.

\begin{figure}[!tb]
    \centering
    \begin{subfigure}[t]{0.64\textwidth}
    \centering
    \includegraphics[width=\textwidth]{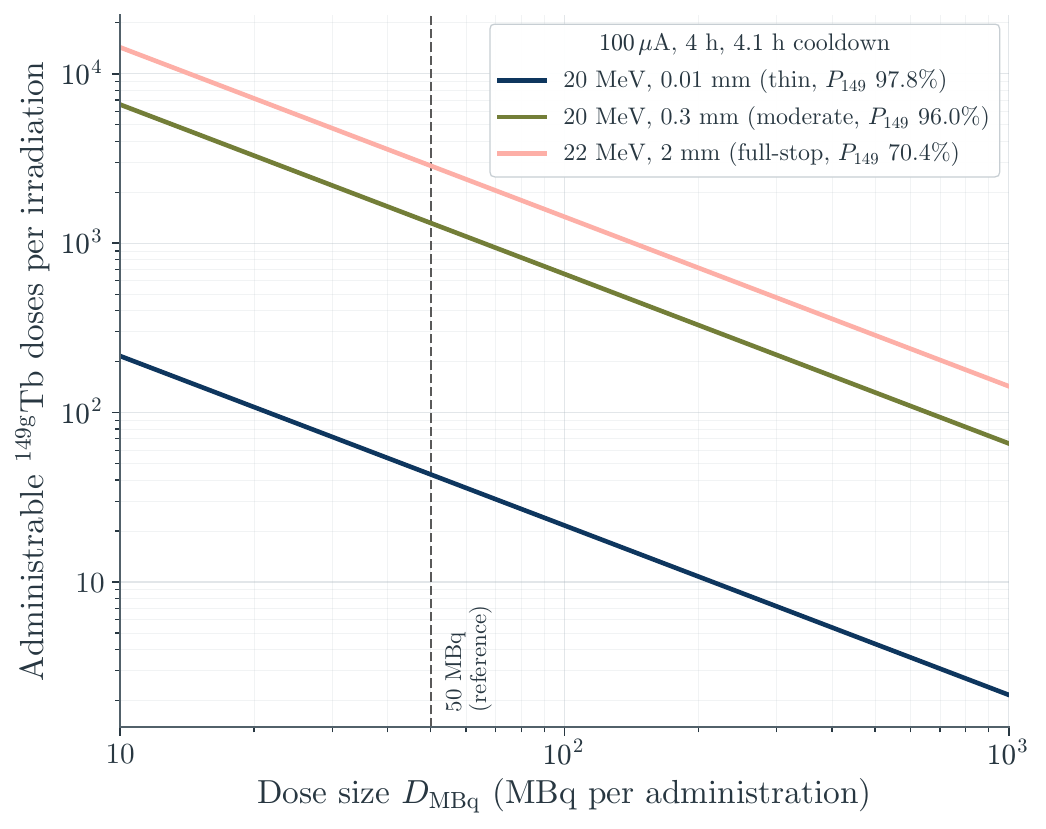}
    \caption{}
    \end{subfigure}
    \hfill
    \begin{subfigure}[t]{0.64\textwidth}
    \centering
    \includegraphics[width=\textwidth]{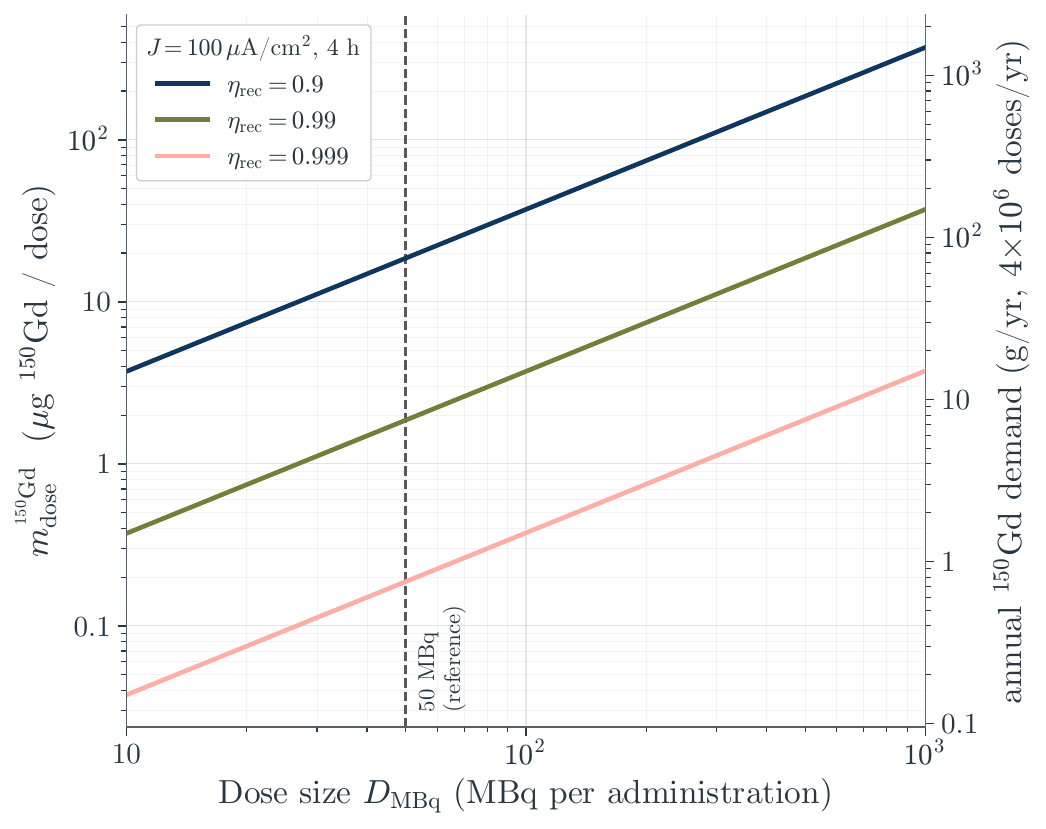}
    \caption{}
    \end{subfigure}
    \caption{Sensitivity of the production metrics to the assumed per-administration dose size $D_{\mathrm{MBq}}$, scanned from \qty{10}{MBq} to \qty{1}{GBq}. (a) Administrable $^{149\mathrm{g}}$Tb doses per irradiation for three representative (not optimized) operating points from the ISOTOPIA \qty{100}{\uA}, \qty{4}{\hour} scan (admin time, after a \qty{4.1}{\hour} cooldown),  covering the thin high-purity to full-stop high-yield range. (b) $^{150}$Gd make-up per dose $m_{\mathrm{dose}}^{{}^{150}\mathrm{Gd}}$ from \Cref{eq:per-dose_main} for three chemistry-recovery efficiencies. The right y-axis converts to annual global $^{150}$Gd demand at the $\sim$4~million doses/year estimated in \Cref{sec:production}.}
    \label{fig:dose_size_sensitivity}
\end{figure}

We show the  number of administrable doses per irradiation in \Cref{fig:dose_size_sensitivity}(a) for three representative operating points,  chosen to cover the yield range, from a thin high-purity target to a full-stop high-yield target. Dose size only partitions the delivered $^{149\mathrm{g}}$Tb activity into administrable portions. It does not change the produced material, so the radionuclidic purity and specific activity are fixed along each curve and only differ between the three operating points. The thin (\qty{20}{MeV}, \qty{10}{\um}) and moderate (\qty{20}{MeV}, \qty{300}{\um}) points have admin-time $P_{149}$ near 96-98\% (counting $^{150}$Tb as an impurity, after a \qty{4.1}{\hour} cooldown) at essentially carrier-free specific activity, while the full-stop point (\qty{22}{MeV}, \qty{2}{mm}) trades purity for yield, dropping to $\sim$70\% $P_{149}$ as the beam slows through the $^{150}$Gd(p,n)$^{150}$Tb peak.

At the reference \qty{50}{MBq} dose, the three operating points deliver $\sim$43 doses (thin, $P_{149}$ 97.8\%), $\sim$1,300 doses (moderate, $P_{149}$ 96.0\%), and $\sim$2,900 doses (full-stop, $P_{149}$ 70.4\%) per \qty{4}{\hour} irradiation. Moving to the high end of the estimated range, \qty{1}{GBq} per administration, reduces these to $\sim$2, $\sim$66, and $\sim$143 doses respectively. Even at \qty{1}{GBq} per dose the full-stop target supplies over a hundred doses per run, so our qualitative conclusion that a single cyclotron irradiation covers a clinically meaningful number of doses still holds even at the upper end of dosage.

\begin{figure}[!tb]
    \centering
    \includegraphics[width=0.8\textwidth]{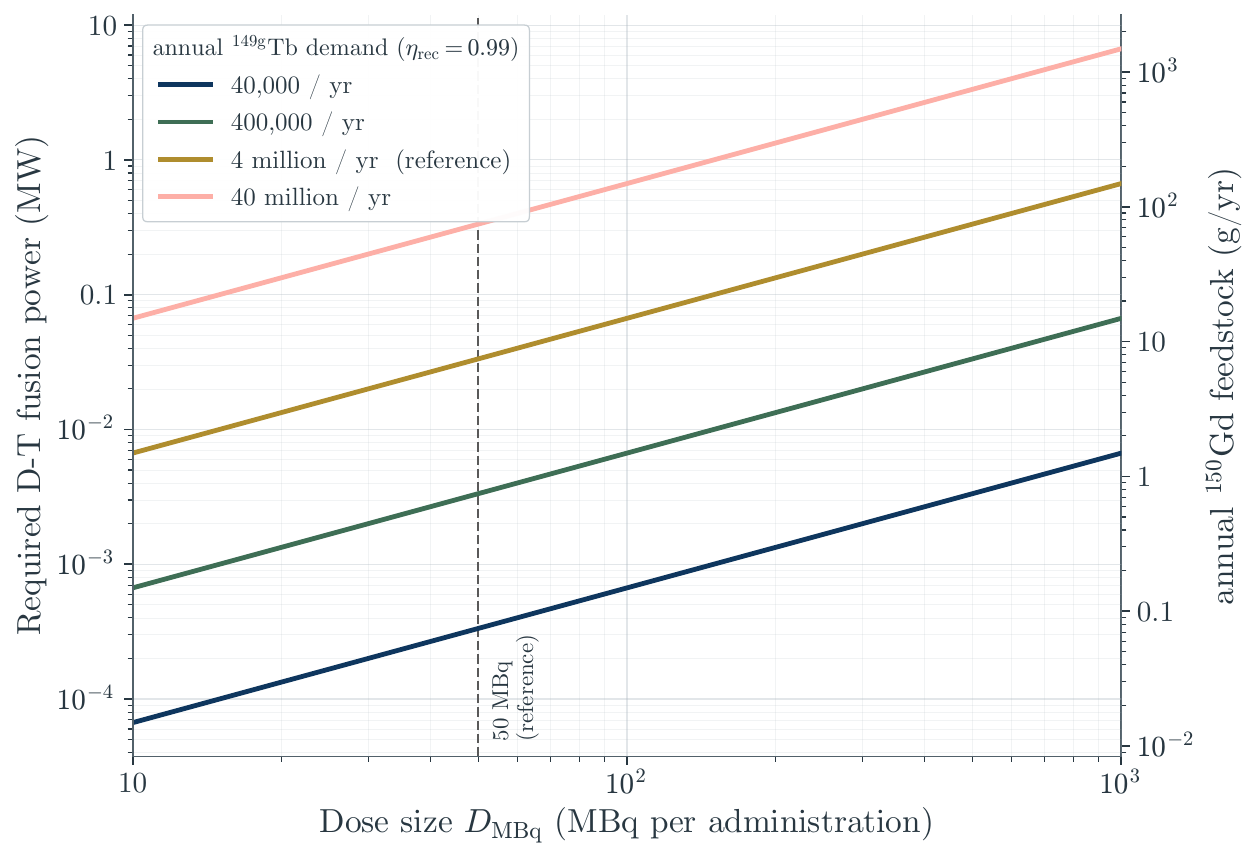}
    \caption{D-T fusion power required to breed the annual $^{150}$Gd feedstock as a function of assumed dose size $D_{\mathrm{MBq}}$, for four annual $^{149\mathrm{g}}$Tb demand levels from 40,000 to 40~million doses/year (the reference 4~million doses/year is $\sim$10\% penetration of the addressable population of \Cref{sec:production}). We use the fast-neutron breeding rate of \qty{223}{\mg} $^{150}$Gd per kW-year (\Cref{sec:step1}) and the per-dose make-up of \Cref{eq:per-dose_main} at $\eta_{\mathrm{rec}} = 0.99$, $J = \qty{100}{\uA\per\cm\squared}$, $T_{\mathrm{irr}} = \qty{4}{\hour}$. $\eta_{\mathrm{rec}} = 0.999$ shifts every curve down by $\sim$10$\times$. The right axis is the corresponding annual $^{150}$Gd mass.}
    \label{fig:dt_fusion_vs_dose}
\end{figure}

\Cref{fig:dose_size_sensitivity}(b) shows the required $^{150}$Gd per dose. At \qty{50}{MBq} and $\eta_{\mathrm{rec}} = 0.99$ the make-up is $m_{\mathrm{dose}}^{{}^{150}\mathrm{Gd}} \simeq \qty{1.9}{\ug}$ per dose, corresponding to $\sim$\qty{7.5}{\g} per year of $^{150}$Gd at $\sim$4~million doses/year (\Cref{sec:production}). A \qty{1}{GBq} dose raises both by a factor of 20, to $\sim$\qty{38}{\ug} per dose and $\sim$\qty{150}{\g} per year, within the projected annual output of a megawatt-scale fast-neutron source (\Cref{sec:step1}). The improved chemistry case $\eta_{\mathrm{rec}} = 0.999$ keeps annual ${}^{150}$Gd demand below $\sim$\qty{15}{\g} per year even at \qty{1}{GBq} per dose. The linear scaling means the paper's feedstock estimates can be rescaled directly to any adopted dose size without re-running the production model.

Only a modest capacity of D-T neutron sources is needed to supply the ${}^{150}$Gd for millions of annual doses, even at several GBq per dose. Using the fast-neutron breeding rate of \qty{223}{\mg} of $^{150}$Gd per kilowatt-year of D-T fusion power (\Cref{sec:step1}), \Cref{fig:dt_fusion_vs_dose} shows the fusion power needed to breed the annual $^{150}$Gd feedstock against dose size, for global demand levels from 40,000 to 40 million doses/year. In the pessimistic corner where each administration requires the full \qty{1}{GBq}, at $\eta_{\mathrm{rec}} = 0.99$ and 4 million doses/yr,  the world requires $\sim$\qty{152}{\g} of $^{150}$Gd per year,  which $\sim$\qty{680}{kW} of D-T fusion capacity would supply. Even serving the entire addressable population with 40 million doses/year at \qty{1}{GBq} per dose  needs approximately \qty{6.8}{MW} of D-T fusion capacity.  At the assumed neutron utilization, these demand scenarios correspond to D-T fusion powers of approximately \qty{0.68}{MW} and \qty{6.8}{MW} at continuous operation, or \qty{0.85}{MW} and \qty{8.5}{MW} at 80\% uptime, respectively. Natural Europium is also extremely abundant, and enriched $^{151}$Eu is not onerously expensive. Further optimization for quantities such as irradiation time and $^{150}$Gd target size is possible with varying dose size.

\section{Radiation emitted by the $^{150}$Tb impurity}\label{sec:tb150_photon_bound}

In this appendix, we estimate how much radiation the $^{150}$Tb impurity adds to that already emitted by the therapeutic $^{149\mathrm{g}}$Tb. Because both are terbium isotopes, chemical separation cannot remove one from the other. For the 94-98\% purity range examined here, the impurity adds a modest amount of photon energy. This is a useful check on the production method, although the acceptable impurity level will depend on the radiopharmaceutical used.

We count photons and electrons separately. Photons include gamma rays, X-rays and the radiation produced when positrons annihilate. The electron total includes the kinetic energy of electrons and positrons from beta decay, together with conversion and Auger electrons released during nuclear and atomic rearrangements. We use the complete mean energies tabulated in MIRDspecs from ICRP Publication 107~\cite{icrp107,mirdspecs2025}, given in \Cref{tab:non_alpha_dose}. For Tb, these values describe the decay of the Tb atom itself, including prompt radiation from the newly formed daughter. They exclude later radioactive decays of the daughter atoms. For the $^{212}$Pb reference, we include its short-lived daughter chain, weighting each daughter by the probability that it is produced.

\begin{table}[!b]
\centering
\caption{Mean emitted energies used in the comparison~\cite{icrp107,mirdspecs2025}. Tb rows include parent decays only. The $^{212}$Pb row includes its short-lived daughter chain per initial Pb decay. Electron energies include positrons, conversion electrons and Auger electrons.}
\label{tab:non_alpha_dose}
\small
\setlength{\tabcolsep}{7pt}
\begin{tabular}{l c c c}
\toprule
Isotope & \makecell{Half-life\\(h)} &
\makecell{Photon energy\\(MeV/decay)} &
\makecell{Electron energy\\(MeV/decay)} \\
\midrule
$^{149\mathrm{g}}$Tb & 4.118 & 1.361 & 0.0871 \\
$^{150}$Tb & 3.48 & 2.440 & 0.2890 \\
$^{212}$Pb chain & 10.64 & 1.455 & 0.9007 \\
\bottomrule
\end{tabular}
\end{table}

An activity $A_X$ in Bq represents $A_X$ decays per second at administration. If its mean life is $\tau_X=T_{1/2}/\ln 2$, in seconds, the initial sample contains $A_X\tau_X$ atoms. Multiplying by the mean energy per decay, $\bar\epsilon_k(X)$, gives the energy released as those atoms decay:
\begin{equation}
E_k(X)=A_X\tau_X\bar\epsilon_k(X),\qquad k=\gamma,e.
\label{eq:photon_total}
\end{equation}
Here $\gamma$ denotes photons and $e$ denotes electrons and positrons. This calculation counts emitted energy wherever the decays occur, including after material has left the body. It therefore does not give the energy absorbed by a patient.

The most direct comparison is between the two Tb isotopes in the same preparation. Let $r=A_{150}/A_{149\mathrm{g}}$ be their activity ratio at administration. From the activity fractions defined in the main text, $r=(P_{149+150}-P_{149})/P_{149}$, which becomes $(1-P_{149})/P_{149}$ when other Tb activities are negligible. The additional photon and electron energies, relative to those from the $^{149\mathrm{g}}$Tb parent, are
\begin{equation}
\frac{E_\gamma(^{150}\mathrm{Tb})}{E_\gamma(^{149\mathrm{g}}\mathrm{Tb})}\simeq1.52r,
\qquad
\frac{E_e(^{150}\mathrm{Tb})}{E_e(^{149\mathrm{g}}\mathrm{Tb})}\simeq2.80r.
\label{eq:combined_photon}
\end{equation}
Thus, at $P_{149}=94$-98\%, assigning the remaining activity to $^{150}$Tb gives approximately \qtyrange{1.0}{3.2}{MBq} of impurity alongside \qty{50}{MBq} of $^{149\mathrm{g}}$Tb. This adds about 3-10\% to the parent photon energy and 6-18\% to the parent electron energy. Other Tb impurities, when present, require their own emission contributions.

For clinical context, we also compare the emitted energy with an illustrative \qty{35}{MBq} administration of $^{212}$Pb, an activity of the order used in the intraperitoneal $^{212}$Pb-TCMC-trastuzumab study~\cite{meredith2014}. At the Tb activities above, the combined Tb parent photon energy is approximately 53-57\% of that emitted by the Pb chain. This comparison does not include later emissions from radioactive Tb daughters, such as $^{149}$Gd and $^{145}$Eu. Their contribution to patient dose depends on how long they remain in the body and where they accumulate. Nor are the selected Tb and Pb activities matched for therapeutic effect: $^{149\mathrm{g}}$Tb emits an alpha particle in only 16.7\% of its decays, whereas the short $^{212}$Pb chain produces one alpha particle per parent decay.

 For these 94-98\% purity preparations, the $^{150}$Tb impurity adds a relatively small photon contribution to that from $^{149\mathrm{g}}$Tb itself. Whether that contribution is acceptable depends on the dose absorbed by healthy organs. This requires knowing where the radiopharmaceutical and its daughters travel, how quickly they leave the body, and how much of their radiation reaches each organ~\cite{mird21,mird22}. A single photon absorption coefficient cannot replace that information. The emission comparison therefore supports further evaluation of these production conditions, without setting a clinical impurity limit.

\section{Cross-section comparisons and saturation yields}\label{sec:yield_validation}

In this appendix, we compare evaluated cross sections with measurements of neighboring reactions and examine the predicted saturation yields.

\subsection{Comparison with measured neighboring reactions}

\begin{figure}[!tb]
\centering
\includegraphics[width=0.8\textwidth]{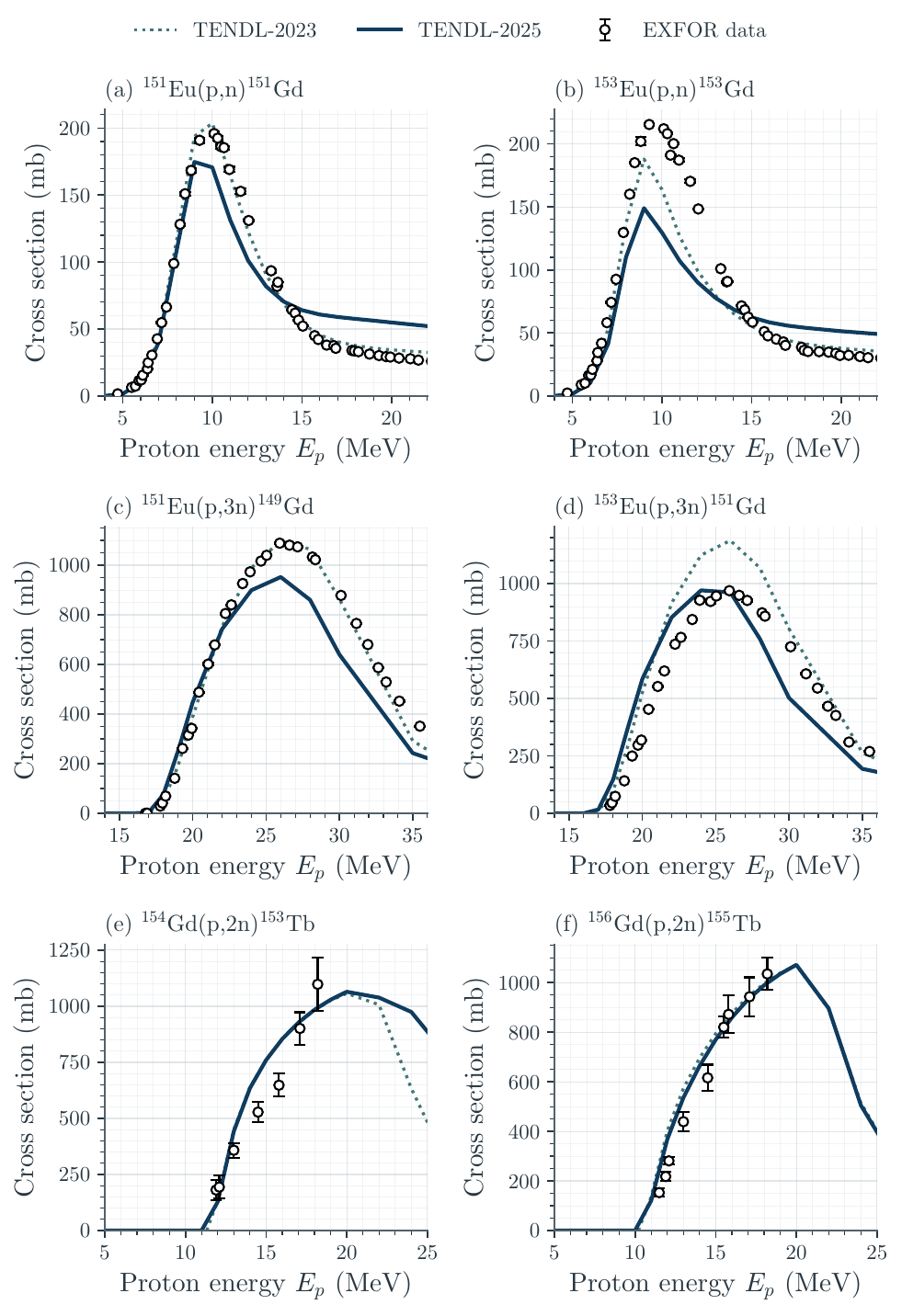}
\caption{TENDL-2023 (teal dotted) and TENDL-2025 (blue solid) vs EXFOR (open circles) for six analog (p,xn) reactions on neutron-poor lanthanide targets: (a) \(^{151}\)Eu(p,n)\(^{151}\)Gd; (b) \(^{153}\)Eu(p,n)\(^{153}\)Gd; (c) \(^{151}\)Eu(p,3n)\(^{149}\)Gd; (d) \(^{153}\)Eu(p,3n)\(^{151}\)Gd; (e) \(^{154}\)Gd(p,2n)\(^{153}\)Tb; (f) \(^{156}\)Gd(p,2n)\(^{155}\)Tb (panels e and f from Dellepiane 2023~\cite{dellepiane2023}).}
\label{fig:analog_pxn_validation}
\end{figure}

 The $^{150}$Gd(p,2n)$^{149\mathrm{g}}$Tb cross section remains unmeasured. We therefore compare the evaluations with measured neighboring channels before using them for the production estimates in Section~\ref{sec:step2}. \Cref{fig:analog_pxn_validation} compares six channels over \qtyrange{4}{36}{MeV}, covering three reaction types and three target atomic numbers. TENDL-to-data peak ratios range from \numrange{0.84}{1.22}.

For a closer cross-section prediction, we ran a TALYS-2.2 family of seven nuclear-physics ingredient choices (six level-density models plus the Jeukenne Lejeune Mahaux (JLM) optical potential) on the three measured Gd(p,2n)Tb channels of Dellepiane 2023~\cite{dellepiane2023} (\(^{154,156,157}\)Gd). Across 20 measured points in the \qtyrange{12}{20}{MeV} peak region, the mean TALYS prediction overshoots the data by an average factor of \(1.30\pm 0.23\). It is essential for the cross section to be measured - efforts are underway to do just this - but we note that a TALYS prediction failure that would push this many times smaller or higher has no precedent across the validation channels in \Cref{fig:analog_pxn_validation}, nor in other proton-driven (p,n), (p,2n), etc, reactions we have analyzed on neutron-poor nuclei.
While more detailed modeling for the \(^{150}\)Gd(p,2n)\(^{149\mathrm{g}}\)Tb cross section can be performed, we believe the fastest pathway is to measure it experimentally. The production calculations in this paper use the TENDL-2025 cross section for \(^{150}\)Gd(p,2n)\(^{149\mathrm{g}}\)Tb (TALYS-2.1), with the caveat that it has significant uncertainty. The TENDL-2025 evaluation is a modest revision of the TENDL-2023 \(^{150}\)Gd(p,2n) ground-state prediction (\qty{457}{mb} vs \qty{437}{mb} at \qty{18}{MeV}), but substantially reduces the predicted competing \(^{150}\)Gd(p,n)\(^{150\mathrm{g}}\)Tb ground-state channel (\qty{5.7}{mb} vs \qty{23}{mb} at \qty{22}{MeV}), increasing the predicted radionuclidic-purity for a range of energies.

\subsection{Saturation activity per microamp}\label{ssec:sat_per_uA}

\begin{figure}[!tb]
\centering
\includegraphics[width=0.85\textwidth]{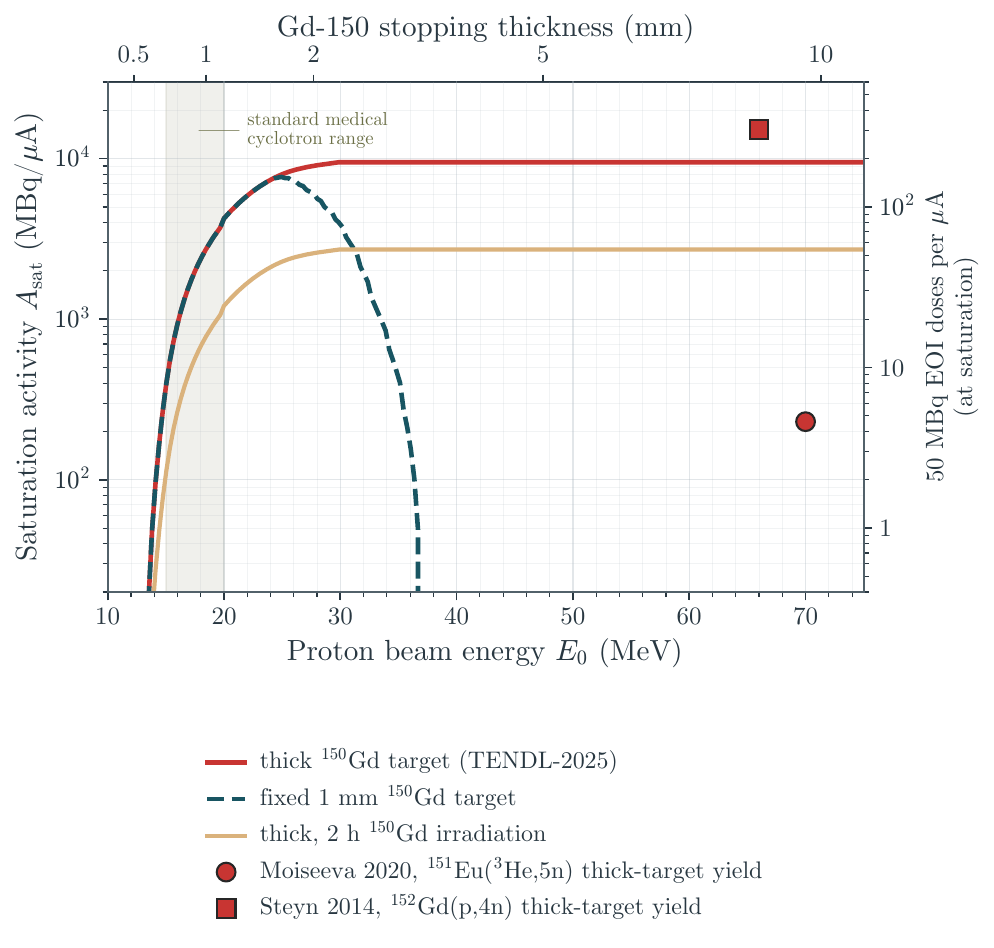}
\caption{Saturation activity per microamp for \(^{149\mathrm{g}}\)Tb production via the \(^{150}\)Gd(p,2n) route at full-stop, as a function of proton beam energy. Curves use the TENDL-2025 ground-state \(^{149\mathrm{g}}\)Tb cross section (this work). Reference points: red circle, \(^{151}\)Eu(\(^{3}\)He,5n) at \SI{70}{MeV} (Moiseeva 2020 \cite{moiseeva2020}), and red square, \(^{152}\)Gd(p,4n) at \SI{66}{MeV} (Steyn 2014 \cite{steyn2014}). The standard medical-cyclotron operating range is shaded. At \SI{20}{MeV} the \(^{150}\)Gd(p,2n) route reaches \SI{4230}{MBq/\uA}, about a factor of \(18\) above the Moiseeva (\(^{3}\)He,5n) route per unit beam current.}
\label{fig:Tb149g_sat_per_uA}
\end{figure}

A useful figure of merit for any cyclotron production route is the saturation activity per unit beam current, \(A_{\rm sat}/I\), which sets the maximum activity.
For a thick full beam stop \ce{^150Gd} metal target this is
\begin{equation}
\frac{A_{\rm sat}}{I} \;=\; \frac{1}{e}\, \frac{\,\rho_{\rm Gd}N_\mathrm{A}}{M_{\rm Gd}}\, \int_{0}^{E_0}\!\frac{\sigma(E)}{|dE/dx|(E)}\, dE,
\end{equation}
where $E_0$ is the incident proton energy, $e$ is the proton charge, $N_\mathrm{A}$ is Avogadro's number, \(M_{\rm Gd} = 150\) g\,mol\(^{-1}\) is the molar mass of the target, \(\sigma\) is the \(^{150}\)Gd(p,2n)\(^{149\rm g}\)Tb cross section, and  \(dE/dx\) is the linear proton stopping power in Gd metal (MeV\,cm$^{-1}$), and $\rho_{\rm Gd}$ is the metal density (g\,cm$^{-3}$). For $\sigma$ in cm$^2$, this expression gives Bq/A. A mass stopping power would instead absorb the density factor. \Cref{fig:Tb149g_sat_per_uA} shows the result over \qtyrange{10}{75}{MeV}, with literature reference points overlaid for the two competing direct routes: \(^{151}\)Eu(\(^{3}\)He,5n) at \qty{70}{MeV}~\cite{moiseeva2020} (\qty{230}{MBq/\uA}) and \(^{152}\)Gd(p,4n) at \qty{66}{MeV}~\cite{steyn2014} ($\sim$\qty{15}{GBq/\uA}, converted from the published thick-target yield \qty[sticky-per=true]{2556}{MBq\per\uA.\hour} integrated over the \qtyrange{30}{66}{MeV} slowing-down window).
The (p,4n) thick-target integral exceeds the present (p,2n) route at \qty{20}{MeV} by roughly a factor of  3.5, but this advantage reflects the much wider slowing-down window available at \qty{66}{MeV} (\qtyrange{30}{66}{MeV} versus \qtyrange{10}{20}{MeV}) and the lower stopping power at high proton energy, not a more favorable cross section: our peak \(^{150}\)Gd(p,2n) cross section ($\sim$\qty{457}{mb} in TENDL-2025) is almost double the \(^{152}\)Gd(p,4n) peak ($\sim$\qty{250}{mb}). The case for the (p,2n) route does not rest on achieving the highest raw activity per microamp, it rests on \qty{20}{MeV} beam access on 113 reported cyclotrons (versus 20 reaching \qty{66}{MeV})~\cite{iaea_akp_2026}, the avoidance of long-lived terbium impurities such as \(^{151}\)Tb and \(^{152}\)Tb co-produced when (p,4n) is run on the only commercially available \(^{152}\)Gd enrichment ($\sim$30\%), and the breedable \(^{150}\)Gd feedstock described in \Cref{sec:step1}.

At the standard \qty{20}{MeV} operating point the TENDL-2025 prediction is \qty{4230}{MBq/\uA}. Pushing the beam to \qty{25}{MeV} would give \qty{7960}{MBq/\uA} but at the cost of worsening the \(^{149}\)Tb m/g isomer ratio and opening the (p,3n)\(^{148}\)Tb contamination channel (not necessarily a significant issue given the \qty{1}{h} \(^{148 \mathrm{g} }\)Tb half-life).

\section{Gadolinium Self-Sufficient Targets}\label{sec:self_sufficient}\label{sec:feedstock_balance}

In this appendix, we derive the mass balance equations for a \(^{150}\)Gd breeder target. The main result is the $^{150}$Gd feedstock requirement per administered \(^{149\mathrm{g}}\)Tb dose.

\begin{figure}[!tb]
\centering
\includegraphics[width=0.85\textwidth]{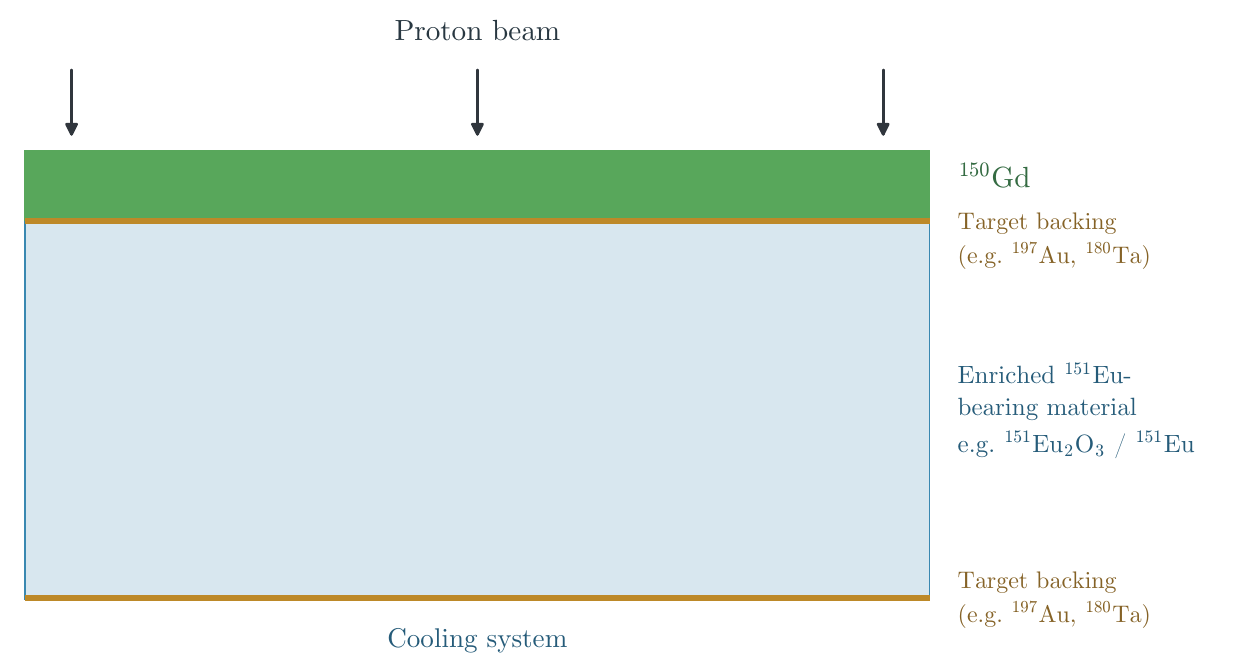}
\caption{Gadolinium breeding target. System proportions not necessarily to scale.}
\label{fig:Gd_breeder_target}
\end{figure}

While \(^{150}\)Gd can be produced in independent facilities using proton, photon, and neutron reactions, the same cyclotron used for \(^{149\mathrm{g}}\)Tb production can also co-produce \(^{150}\)Gd by backing the thin \(^{150}\)Gd front layer with a thick \(^{151}\)Eu layer (\Cref{fig:Gd_breeder_target}). Protons that pass through the thin \(^{150}\)Gd front layer drive \(^{151}\)Eu(p,2n)\(^{150}\)Gd in the back layer, partially offsetting the \(^{150}\)Gd consumed up front. We quantify this with the gadolinium breeding ratio (GBR), defined as the ratio of \(^{150}\)Gd bred in the \(^{151}\)Eu layer to \(^{150}\)Gd consumed (nuclear burn plus chemistry losses) in the front layer. At the paper's reference operating point ($I = \qty{100}{\uA}$, $J = \qty{100}{\uA\per\centi\meter\squared}$), the modest low-throughput operating points ($\sim$10 doses per irradiation)  reach GBR $\approx 2$-$3$ at 17-20 MeV with $\eta_\mathrm{rec} = 0.99$, so the inventory grows and can double in a few hundred days. Higher-yield points ($\sim$100 doses per irradiation) carry a larger $^{150}$Gd inventory whose chemistry loss outpaces the back-layer breeding, giving GBR $<1$. Here $I$ is the total proton beam current, $J$ is its current density, and $\eta_{\rm rec}$ is the fraction of Gd recovered after extraction. The following derivation defines the breeding ratio and inventory growth, including the conditions required to double the feedstock. The numerical operating points are listed in \ref{sec:target_optimization}.

\subsection{Mass balance per irradiation}
Let $N_0$ be the $^{150}$Gd atoms in the \(^{150}\)Gd target. Conservation of particles gives
\begin{equation}
N_0 = N_{{}^{150}\mathrm{Gd}}^{\mathrm{rem}} + \sum_{k\neq{}^{150}\mathrm{Gd}} N_k,
\end{equation}
where $N_{{}^{150}\mathrm{Gd}}^{\mathrm{rem}}$ is the number of remaining \(^{150}\)Gd atoms. The number transmuted away is
\begin{equation}
N_{{}^{150}\mathrm{Gd}}^{\mathrm{burn}} = N_0 - \begin{gathered}\bigl(N_{{}^{150}\mathrm{Gd}}^{\mathrm{rem}}+N_{{}^{150\mathrm{m}}\mathrm{Tb}}\bigr)\end{gathered} = \sum_{k\notin\{{}^{150}\mathrm{Gd},\, {}^{150\mathrm{m}}\mathrm{Tb}\}} N_k.
\label{eq:burn-app}
\end{equation}
 This expression credits full return of the EOI $^{150\mathrm{m}}$Tb inventory to recoverable $^{150}$Gd before chemistry. It is an approximation requiring sufficient cooling before separation. For a partial return fraction $r_m$, replace the credited term by $r_mN_{{}^{150\mathrm{m}}\mathrm{Tb}}$ and use that same convention in $f_{\rm burn}$.
The right-hand side scales linearly with $N_0$, so we write
\begin{equation}
N_{{}^{150}\mathrm{Gd}}^{\mathrm{burn}} = f_{\mathrm{burn}} N_0,
\qquad f_{\mathrm{burn}} \equiv \frac{N_{{}^{150}\mathrm{Gd}}^{\mathrm{burn}}}{N_0},
\label{eq:fburn-app}
\end{equation}
where the per-irradiation burn fraction $f_{\mathrm{burn}}$ depends on cross sections, layer thickness, and irradiation duration but not on $N_0$ in the thin-target limit.

After one cyclotron run, the remaining $^{150}$Gd is $(1-f_{\mathrm{burn}}) N_0$. Post-run extraction recovers a fraction $\eta_{\mathrm{rec}}$ of \(^{150}\)Gd, so the inventory before the next run is
\begin{equation}
N_1 = \eta_{\mathrm{rec}} (1 - f_{\mathrm{burn}}) N_0 = s N_0,
\end{equation}
with per-run $^{150}$Gd survival factor
\begin{equation}
s \equiv \eta_{\mathrm{rec}} (1 - f_{\mathrm{burn}}).
\end{equation}
We perform $n_{\mathrm{irr}}$ irradiations per year.  Iterating across one year without replenishment gives the surviving inventory fraction
\begin{equation}
\frac{N_{n_{\mathrm{irr}}}}{N_0} = s^{n_{\mathrm{irr}}} = \bigl[\eta_{\mathrm{rec}} (1-f_{\mathrm{burn}})\bigr]^{n_{\mathrm{irr}}}.
\end{equation}

\subsection{Adding back-layer breeding and external supply}
\label{app:feedstock-breed}

In the above analysis, we ignored two methods for restocking $^{150}$Gd inventory. First, the $^{151}$Eu$\mathrm{(p,2n)}^{150}$Gd back-layer breeding deposits
\begin{equation}
N_{\mathrm{breed}} = b_{\mathrm{rate}} T_{\mathrm{irr}},
\end{equation}
where $b_{\mathrm{rate}}$ depends on the Eu-151 inventory and the proton current but not on the front-Gd inventory. Second, an external make-up of $S_{\mathrm{ext}}$ atoms per run may be added directly to the feedstock pool.

The bred $^{150}$Gd sits in the back Eu-151 layer, separated from the per-run \(^{149\mathrm{g}}\)Tb chemistry that drives $\eta_{\mathrm{rec}}$. We extract the bred pool occasionally (e.g., once per campaign) via a slower, half-life-unconstrained Eu/Gd separation with efficiency $\eta_{\mathrm{recover,bred}}$, which can be much closer to unity than the per-run $\eta_{\mathrm{rec}}$ (lanthanide chromatography routinely achieves $> 99.9\%$ when run without time pressure). We treat $\eta_{\mathrm{recover,bred}}$ as the long-time-averaged transfer efficiency from the back-layer breeding pool to the front-layer feedstock pool. Including both restocking channels into a single per-run variable,
\begin{equation}
I_{\mathrm{run}} \equiv \eta_{\mathrm{recover,bred}} N_{\mathrm{breed}} + S_{\mathrm{ext}},
\end{equation}
gives the recursion relation
\begin{equation}
N_{k+1} = \eta_{\mathrm{rec}}(1 - f_{\mathrm{burn}}) N_k + \eta_{\mathrm{recover,bred}} N_{\mathrm{breed}} + S_{\mathrm{ext}} = s N_k + I_{\mathrm{run}}.
\label{eq:recursion-breed}
\end{equation}

The ratio $B_{\rm nuc}=N_{\rm breed}/(f_{\rm burn}N_0)$ in \Cref{eq:gbr-def} compares breeding with nuclear burn alone. GBR in \Cref{eq:GBR} includes chemistry loss. With post-chemistry make-up, the steady balance is $S_{\rm ext}=[(1-\eta_{\rm rec})+\eta_{\rm rec}f_{\rm burn}]N_0-\eta_{\rm recover,bred}N_{\rm breed}$. The two ratios therefore cannot be compared directly.

 The nuclear-burn breeding ratio, normalized to the reference inventory $N_0$, is
\begin{equation}
\,B_{\rm nuc} \equiv \frac{N_{\mathrm{breed}}}{N_{{}^{150}\mathrm{Gd}}^{\mathrm{burn}}} = \frac{b_{\mathrm{rate}}}{\mathrm{burn\ rate}},
\label{eq:gbr-def}
\end{equation}
the fraction of burned $^{150}$Gd replenished per run by Eu-back breeding, assuming an enriched ${}^{151}$Eu back-target. Solving the recursion in \Cref{eq:recursion-breed},
\begin{equation}
N_k = s^k N_0 + I_{\mathrm{run}} \frac{1 - s^k}{1 - s},
\end{equation}
the year end inventory fraction is
\begin{equation}
\frac{N_{\mathrm{final}}}{N_0} = s^{n_{\mathrm{irr}}} + \frac{I_{\mathrm{run}}}{N_0} \frac{1 - s^{n_{\mathrm{irr}}}}{1 - s}.
\label{eq:final-with-supply}
\end{equation}

Defining the dimensionless make-up rate
\begin{equation}
\sigma \equiv S_{\mathrm{ext}}/(f_{\mathrm{burn}} N_0),
\end{equation}
(make-up in units of per-run burn) and using $N_{\mathrm{breed}} = \,B_{\rm nuc}\, f_{\mathrm{burn}} N_0$ and $S_{\mathrm{ext}} = \sigma f_{\mathrm{burn}} N_0$,
\begin{equation}
\frac{N_{\mathrm{final}}}{N_0} = s^{n_{\mathrm{irr}}} + (\eta_{\mathrm{recover,bred}}\,\,B_{\rm nuc} + \sigma)\, f_{\mathrm{burn}} \frac{1 - s^{n_{\mathrm{irr}}}}{1 - s}.
\end{equation}
As $k \to \infty$ the iteration converges to
\begin{equation}
N_\infty = \frac{I_{\mathrm{run}}}{1 - s} = \frac{(\eta_{\mathrm{recover,bred}}\,\,B_{\rm nuc} + \sigma) f_{\mathrm{burn}}}{1 - s} N_0.
\end{equation}
The inventory is constant ($N_\infty = N_0$) when
\begin{equation}
f_{\mathrm{burn}} (\eta_{\mathrm{recover,bred}}\,\,B_{\rm nuc} + \sigma) = 1 - s = 1 - \eta_{\mathrm{rec}}(1 - f_{\mathrm{burn}}),
\end{equation}
which solved for the make-up rate gives
\begin{equation}
\sigma_{\mathrm{sustain}} = \frac{1 - \eta_{\mathrm{rec}}(1 - f_{\mathrm{burn}})}{f_{\mathrm{burn}}} - \eta_{\mathrm{recover,bred}}\,\,B_{\rm nuc}.
\label{eq:sigma-sustain}
\end{equation}
A pure breeder ($\sigma = 0$) gives the GBR self-sustaining condition\[\eta_{\mathrm{recover,bred}}\,\,B_{\rm nuc}\,f_{\mathrm{burn}} = 1 - s.\] With no breeding ($\,B_{\rm nuc} = 0$) the make-up alone must supply $\sigma_{\mathrm{sustain}} = (1 - \eta_{\mathrm{rec}}(1 - f_{\mathrm{burn}}))/f_{\mathrm{burn}}$,  which approaches unity only when $\eta_{\rm rec}\to1$ and $(1-\eta_{\rm rec})/f_{\rm burn}\ll1$.

\subsection{Steady-stream limit (no breeding, constant make-up)}
With $\,B_{\rm nuc} = 0$ and a constant external top-up $S_{\mathrm{ext}}$ per run, the recursion is
\begin{equation}
N_{k+1} = s N_k + S_{\mathrm{ext}},
\end{equation}
with closed form
\begin{equation}
N_k = s^k N_0 + S_{\mathrm{ext}} \frac{1 - s^k}{1 - s}
\;\xrightarrow[k\to\infty]{}\; N_\infty = \frac{S_{\mathrm{ext}}}{1 - s}.
\end{equation}
The make-up rate that holds the inventory at $N_0$ is
\begin{equation}
S_{\mathrm{ext}}^{\mathrm{steady}} = (1 - s) N_0 = \bigl[(1 - \eta_{\mathrm{rec}}) + \eta_{\mathrm{rec}} f_{\mathrm{burn}}\bigr] N_0,
\label{eq:steady-makeup}
\end{equation}
which for perfect recovery efficiency ($\eta_{\mathrm{rec}} \to 1$) reduces to $f_{\mathrm{burn}} N_0$, replacing exactly what is burned.

The annual $^{150}$Gd requirement is $S_{\mathrm{ext}}^{\mathrm{year}} = n_{\mathrm{irr}} S_{\mathrm{ext}}^{\mathrm{steady}}$,  which for a constant target inventory is exactly
\begin{equation}
\begin{gathered}S_{\mathrm{ext}}^{\mathrm{year}} = n_{\mathrm{irr}}(1-s)N_0 = n_{\mathrm{irr}}\bigl[(1-\eta_{\mathrm{rec}})+\eta_{\mathrm{rec}}f_{\mathrm{burn}}\bigr]N_0,\end{gathered}
\label{eq:steady-year}
\end{equation}
 splitting into chemistry loss and burn terms. The quantity $N_0(1-s^{n_{\rm irr}})$ instead describes depletion without top-up. Approximating it by $n_{\rm irr}(1-s)N_0$ requires $n_{\rm irr}(1-s)\ll1$, not merely a small loss per run. With $^{150}$Gd breeding (\Cref{eq:final-with-supply}),
\begin{equation}
S_{\mathrm{ext}}^{\mathrm{year}} \approx n_{\mathrm{irr}} \bigl[(1 - \eta_{\mathrm{rec}}) + (\eta_{\mathrm{rec}} - \eta_{\mathrm{recover,bred}}\,\,B_{\rm nuc})\, f_{\mathrm{burn}}\bigr] N_0,
\label{eq:steady-year-breed}
\end{equation}
so $\eta_{\mathrm{recover,bred}}\,\,B_{\rm nuc} \to \eta_{\mathrm{rec}}$ cancels the burn term, leaving only chemistry loss for external make-up.

\subsection{Effective $^{150}$Gd required per administrable dose} \label{sec:gd_150_per_dose}

One figure of merit is the $^{150}$Gd consumed per administered $^{149\mathrm{g}}$Tb dose at steady state. It is the per-run external make-up $S_{\mathrm{ext}}^{\mathrm{steady}}$ from \Cref{eq:steady-makeup} divided by the dose count per run.

We define the following quantities. $D_{\mathrm{MBq}}$ is the $^{149\mathrm{g}}$Tb  activity per administered dose, expressed in Bq in the equations below. $T_{\mathrm{irr}}$ is the irradiation duration of one run, $T_{\mathrm{cool}}$ is the elapsed time between end of irradiation and patient administration (covering separation chemistry, labeling, transport), and $\lambda_g = \ln 2 / T_{1/2}^{(149\mathrm{g})}$ is the $^{149\mathrm{g}}$Tb decay rate ($T_{1/2}^{(149\mathrm{g})} = 4.1$~h). The $^{149\mathrm{g}}$Tb production rate per unit time during irradiation is $R_\mathrm{p,2n} \equiv \sigma_\mathrm{p,2n}\,\phi\,N_0$, where $\sigma_\mathrm{p,2n}$ is the path-averaged $^{150}$Gd$\mathrm{(p,2n)}^{149\mathrm{g}}$Tb cross section, $\phi$ is the proton flux (cm$^{-2}$\,s$^{-1}$), and $N_0$ is the $^{150}$Gd inventory in the spot. The corresponding total $^{150}$Gd destruction rate (summed over all proton-driven channels) is $R_{\mathrm{destr}} \equiv \sigma_{\mathrm{destr}}\,\phi\,N_0$, with $\sigma_{\mathrm{destr}} \geq \sigma_\mathrm{p,2n}$ since other channels (e.g. $\mathrm{(p,n)}$, $\mathrm{(p,3n)}$) also consume $^{150}$Gd. The saturation activity of $^{149\mathrm{g}}$Tb is $A_{\mathrm{sat}}^{(149\mathrm{g})} = R_\mathrm{p,2n}$ (production matches decay at $T_{\mathrm{irr}}\to\infty$).

$D_{\mathrm{run}}$ is the number of $D_{\mathrm{MBq}}$-sized doses that one irradiation run yields after harvest. We obtain it by tracking the in-target $^{149\mathrm{g}}$Tb activity through three phases. During irradiation, $^{149\mathrm{g}}$Tb is produced at rate $R_\mathrm{p,2n}$ and decays with rate constant $\lambda_g$, so the activity at EOI is $A_{\mathrm{eoi}} = A_{\mathrm{sat}}^{(149\mathrm{g})}\,f_{\mathrm{sat}}$ with $f_{\mathrm{sat}} = 1 - e^{-\lambda_g T_{\mathrm{irr}}}$ the buildup fraction. Between EOI and administration the harvested $^{149\mathrm{g}}$Tb decays by $f_{\mathrm{cool}} = e^{-\lambda_g T_{\mathrm{cool}}}$. The remaining activity is divided into doses of size $D_{\mathrm{MBq}}$,
\begin{equation}
D_{\mathrm{run}} = \frac{\eta_{\rm Tb}\,A_{\mathrm{sat}}^{(149\mathrm{g})}\,f_{\mathrm{sat}}\,f_{\mathrm{cool}}}{D_{\mathrm{MBq}}},
\qquad
f_{\mathrm{sat}} = 1 - e^{-\lambda_g T_{\mathrm{irr}}},
\quad
f_{\mathrm{cool}} = e^{-\lambda_g T_{\mathrm{cool}}}.
\end{equation}
and the per-dose make-up is
\begin{equation}
\frac{S_{\mathrm{ext}}^{\mathrm{steady}}}{D_{\mathrm{run}}} = \bigl[(1-\eta_{\mathrm{rec}}) + \eta_{\mathrm{rec}} f_{\mathrm{burn}}\bigr] \cdot \frac{N_0}{D_{\mathrm{run}}},
\end{equation}
where $\eta_{\mathrm{rec}}$ is the per-run chemistry recovery (\Cref{eq:steady-makeup}), $f_{\mathrm{burn}}$ is the per-run nuclear burn-up fraction (\Cref{eq:fburn-app}), and $N_0$ is the $^{150}$Gd inventory at the start of the run.

Both $f_{\mathrm{burn}}$ and $D_{\mathrm{run}}/N_0$ scale linearly with proton fluence and target cross sections, so the combination $f_{\mathrm{burn}} N_0 / D_{\mathrm{run}}$ depends only on the cross-section ratio. Each (p,2n) event burns one $^{150}$Gd, so $N_{{}^{150}\mathrm{Gd}}^{\mathrm{burn}} = R_{\mathrm{destr}}\,T_{\mathrm{irr}}$  while $D_{\mathrm{run}} = \eta_{\rm Tb}R_\mathrm{p,2n}\,f_{\mathrm{sat}}\,f_{\mathrm{cool}}/D_{\mathrm{MBq}}$, giving the minimum number of $^{150}$Gd atoms required per dose from nuclear burn and decay alone,
\begin{equation}
\frac{N_{{}^{150}\mathrm{Gd}}^{\mathrm{burn}}}{D_{\mathrm{run}}} = \frac{\sigma_{\mathrm{destr}}}{\sigma_\mathrm{p,2n}} \cdot \frac{T_{\mathrm{irr}}\,D_{\mathrm{MBq}}}{\eta_{\rm Tb}\,f_{\mathrm{sat}} f_{\mathrm{cool}}}.
\label{eq:burn-per-dose}
\end{equation}
 With $T_{\rm irr}$ in seconds and $D_{\rm MBq}$ in Bq (e.g.\ \qty{50}{MBq}$=\qty{5e7}{Bq}$), the ratio $T_{\rm irr}D_{\rm MBq}/(\eta_{\rm Tb}f_{\rm sat}f_{\rm cool})$ counts the reactions producing $^{149\mathrm{g}}$Tb per administered dose.  Adding recovered breeding to the post-chemistry balance gives the full per-dose make-up requirement, using $B_{\rm nuc}$ from \Cref{eq:gbr-def},
\begin{equation}
\begin{aligned}
&
m_{\mathrm{dose}}^{{}^{150}\mathrm{Gd}}
\equiv
\frac{S_{\mathrm{ext}}^{\mathrm{steady}}}{D_{\mathrm{run}}}\cdot\frac{M_{{}^{150}\mathrm{Gd}}}{N_{\mathrm{A}}} \\
& = \begin{gathered}\left[\frac{1-\eta_{\mathrm{rec}}}{f_{\mathrm{burn}}} + \eta_{\mathrm{rec}} - \eta_{\mathrm{recover,bred}}\,B_{\rm nuc}\right]\end{gathered} \frac{\sigma_{\mathrm{destr}}}{\sigma_\mathrm{p,2n}} \cdot \frac{T_{\mathrm{irr}}\,D_{\mathrm{MBq}}}{\eta_{\rm Tb}\,f_{\mathrm{sat}} f_{\mathrm{cool}}}\cdot \frac{M_{{}^{150}\mathrm{Gd}}}{N_{\mathrm{A}}}.
\label{eq:per-dose}
\end{aligned}
\end{equation}
The atom count $T_{\mathrm{irr}}\,D_{\mathrm{MBq}}$ (units of Bq$\cdot$s = atoms) is converted to grams of $^{150}$Gd via the molar mass $M_{{}^{150}\mathrm{Gd}} = \qty{149.918}{\g/\mol}$ and Avogadro's number $N_{\mathrm{A}}$. We denote this $^{150}$Gd-per-administered-dose make-up by $m_{\mathrm{dose}}^{{}^{150}\mathrm{Gd}}$ throughout.

\paragraph{Numerical estimate at 18~MeV} TENDL-2025 path-averaged through \qty{200}{\micro\meter} of $^{150}$Gd gives $\sigma_\mathrm{p,2n}^{149\rm g} \approx \qty{457}{\milli\barn}$.
Summing all proton-induced residual-production channels except $^{150}$Gd itself (the target) and $^{150}$Tb (which $\beta^+$/EC-decays back to $^{150}$Gd within hours and rejoins the feedstock) gives the true total destruction $\sigma_{\mathrm{destr}} \approx 937$~mb, dominated by the two $^{149}$Tb isomers (g: \qty{457}{\milli\barn}, m: \qty{474}{\milli\barn}). The (p,2n) reaction destroys a $^{150}$Gd atom whether it lands in the deliverable g state or in the lost m state (which $\alpha$/EC-decays to $^{149}$Gd\,/\,$^{145}$Eu rather than back to the feedstock).
Thus $\sigma_{\mathrm{destr}}/\sigma_\mathrm{p,2n}^{149\rm g} \approx 2.05$.
With $T_{\mathrm{irr}} = \qty{2}{\hour}$ ($=\qty{7200}{\s}$), $D_{\mathrm{MBq}} = \qty{50}{MBq}$ ($= \qty{5e7}{Bq}$), $f_{\mathrm{sat}} \approx 0.286$, and $f_{\mathrm{cool}} \approx 0.5$ at one-half-life cooldown, $T_{\mathrm{irr}}\,D_{\mathrm{MBq}} \approx \num{3.6e11}$~atoms and
\begin{equation}
\frac{N_{{}^{150}\mathrm{Gd}}^{\mathrm{burn}}}{D_{\mathrm{run}}} \approx 2.05 \cdot \frac{3.6 \cdot 10^{11}}{0.286 \cdot 0.5} \approx \num{5.2e12}\;\mathrm{atoms/dose} \begin{gathered}(1.29\;\mathrm{ng/dose}).\end{gathered}
\end{equation}
This is the nuclear burn cost. With perfect extraction chemistry and no breeding, each 50~MBq dose consumes $\sim 1.3$~ng of $^{150}$Gd. The per-run burnup fraction $f_{\mathrm{burn}} = \sigma_{\mathrm{destr}}\,\phi\,T_{\mathrm{irr}}$ depends on the proton flux on the front $^{150}$Gd layer.
At a  current density of $J = \qty{100}{\uA/\cm\squared}$ ($\phi = 6.24\cdot 10^{14}\ \mathrm{p/cm^2/s}$, equivalent to \qty{1.8}{\kW/\cm\squared} deposited heat at \qty{18}{MeV}), $f_{\mathrm{burn}}(T_{\mathrm{irr}}=2\text{ h}) \approx 4.2\cdot 10^{-6}$ and
\begin{equation}
\;m_{\mathrm{dose}}^{{}^{150}\mathrm{Gd}} \approx \begin{gathered}\left[2.4 \cdot 10^5 (1 - \eta_{\mathrm{rec}}) + \eta_{\mathrm{rec}} - \eta_{\mathrm{recover,bred}}\,B_{\rm nuc}\right]\end{gathered} \cdot 1.29\;\mathrm{ng/dose}.
\end{equation}
At $\,B_{\rm nuc} = 0$, $\eta_{\mathrm{rec}} = 0.99$ requires $\sim \qty{3.1}{\micro\gram}$/dose, whereas $\eta_{\mathrm{rec}} = 0.9999$ requires $\sim 32$~ng/dose. $\eta_{\mathrm{rec}} \to 1$ and $\,B_{\rm nuc} \to 1$ drives the make-up to zero. Sub-ng/dose make-up needs both $\eta_{\mathrm{rec}} \gtrsim 1 - f_{\mathrm{burn}}$ and $\,B_{\rm nuc}$ near unity. This estimate is evaluated at $T_{\mathrm{irr}} = \qty{2}{\hour}$. The paper's \qty{4}{\hour} operating point raises the buildup fraction to $f_{\mathrm{sat}} \approx 0.49$ and lowers the chemistry-dominated $m_{\mathrm{dose}}^{{}^{150}\mathrm{Gd}}$ to $\sim\qty{1.9}{\micro\gram}$/dose at $\eta_{\mathrm{rec}} = 0.99$ (\Cref{tab:practical,tab:operating_points}).

\subsection{Recovery and current-density sensitivity}\label{sec:recovery_sensitivity}

The analytic recovery scans here and in \ref{sec:dose_sensitivity} use the TENDL-2023 inputs $\sigma_{\rm p,2n}=\qty{436.6}{mb}$ and $\sigma_{\rm destr}=\qty{936.5}{mb}$ at \qty{18}{MeV}. The target-yield scans use TENDL-2025.

We apply the no-breeding form of the mass balance, \Cref{eq:per-dose_main}, using the reference parameters in \Cref{tab:practical}. Here $J$ is proton current density and $\eta_{\rm rec}$ is Gd recovery per extraction. Tb product recovery $\eta_{\rm Tb}$ excludes decay already counted during cooling and is distinct from Gd recovery.

\begin{figure}[!tb]
    \centering
    \begin{subfigure}[t]{0.85\textwidth}
    \centering
    \includegraphics[width=\textwidth]{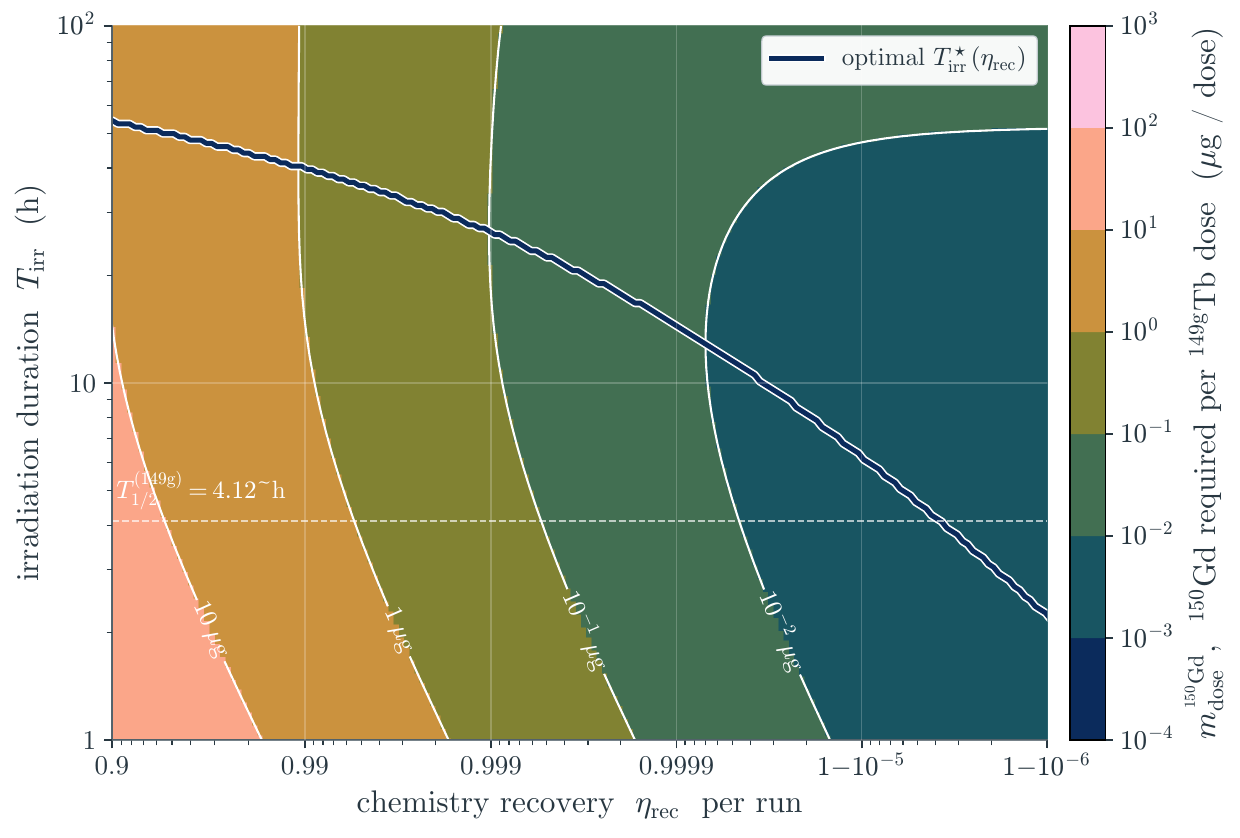}
    \caption{$m_{\mathrm{dose}}^{{}^{150}\mathrm{Gd}}$ versus irradiation duration and chemistry recovery per run. Curve indicates the optimal irradiation time.}
    \end{subfigure}
    \hfill
    \begin{subfigure}[t]{0.85\textwidth}
    \centering
    \includegraphics[width=\textwidth]{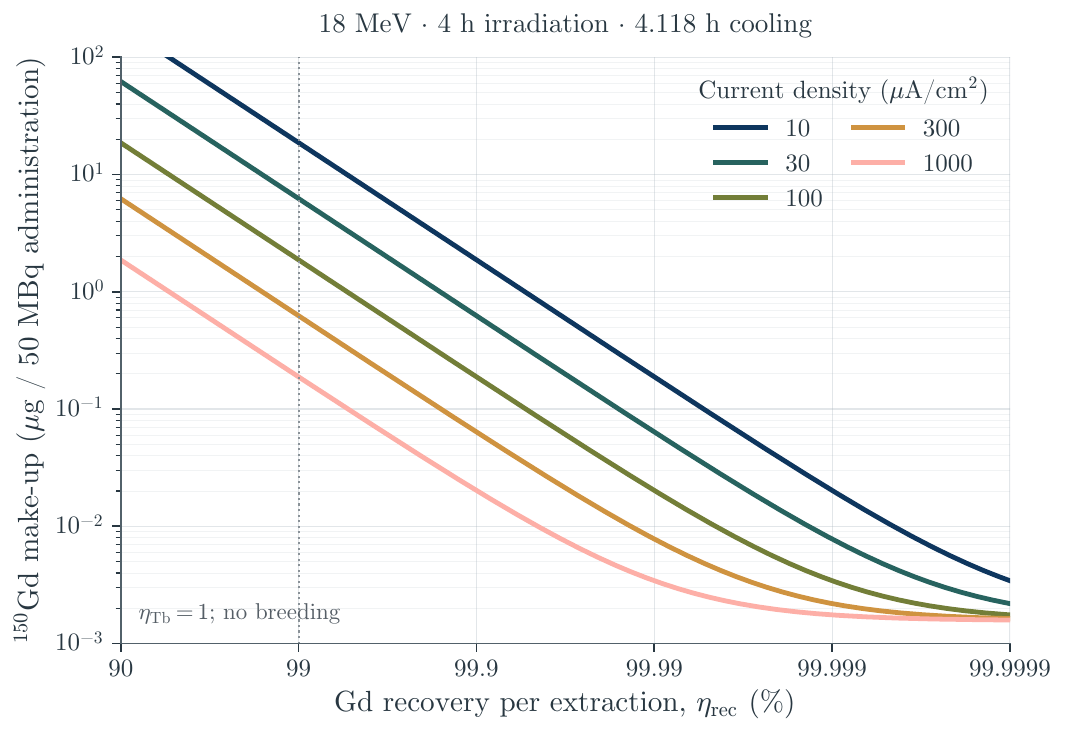}
    \caption{}
    \end{subfigure}
    \caption{Required \ce{^150Gd} per administered dose $m_{\mathrm{dose}}^{{}^{150}\mathrm{Gd}}$, see \Cref{tab:practical} for input parameter values.}
    \label{fig:gd150_per_dose}
\end{figure}

We plot $m_{\mathrm{dose}}^{{}^{150}\mathrm{Gd}}$ in \Cref{fig:gd150_per_dose}(a) versus $T_\mathrm{irr}$ and $\eta_{\mathrm{rec}}$, all other input parameter values are listed in \Cref{tab:practical}. For a \qty{4}{\hour} irradiation time and $\eta_{\mathrm{rec}} = 0.99$, the mass per dose is $m_{\mathrm{dose}}^{{}^{150}\mathrm{Gd}} \simeq \qty{1.9}{\micro\gram}$. There is a current density requirement in \Cref{fig:gd150_per_dose}(b) through $f_\mathrm{burn}$ - we show a current density scan in \Cref{fig:gd150_per_dose}(b). Given we are using $\eta_\mathrm{rec} =0.99$ and are in the regime $1-\eta_{\mathrm{rec}} \gg \eta_{\mathrm{rec}} f_{\mathrm{burn}}$,  the \ce{^150Gd} make-up mass per dose is approximately inversely proportional to current density. This means that increasing the current density can reduce the \ce{^150Gd} required per dose. As \Cref{fig:gd150_per_dose}(b) shows, only at $\eta_\mathrm{rec} \gtrsim 0.990$ does the dependence of $m_{\mathrm{dose}}^{{}^{150}\mathrm{Gd}}$ on $\eta_\mathrm{rec}$ become sublinear. In the limit of perfect recovery efficiency, $\eta_\mathrm{rec} \to 1$, \Cref{eq:per-dose_main} gives $m_{\mathrm{dose}}^{{}^{150}\mathrm{Gd}}$ requirements due to nuclear burn and decay alone.

\begin{table}[!htbp]
\centering
\color{black}
\caption{Sensitivity to Tb product recovery at fixed irradiation and cooling. Make-up values rescale the rounded reference \qty{1.9}{\micro\gram} per \qty{50}{MBq} administration at $\eta_{\rm rec}=0.99$. The dose multiplier applies at fixed initial target inventory.}
\label{tab:tb_recovery}
\begin{tabular}{cccc}
\toprule
$\eta_{\rm Tb}$ & \makecell{Dose\\multiplier} & \makecell{Make-up\\multiplier} & \makecell{Make-up\\($\mu$g/administration)}\\
\midrule
1.00 & 1.00 & 1.00 & 1.90\\
0.90 & 0.90 & 1.11 & 2.11\\
0.80 & 0.80 & 1.25 & 2.38\\
0.50 & 0.50 & 2.00 & 3.80\\
\bottomrule
\end{tabular}
\end{table}

\subsection{Mapping doses per mg to make-up per dose} \label{sec:dpm_to_mdose}

We define the number of therapeutic ${}^{149\mathrm{g}}\mathrm{Tb}$ doses per \qty{4}{\hour} irradiation period per $^{150}$Gd mass
\begin{equation}
\Dmtb \;\equiv\; \frac{D_{\mathrm{run}}}{m_\mathrm{Gd}\,[\mathrm{mg}]} \qquad [\,\text{doses}/\text{mg}\,],
\label{eq:dmtb_def}
\end{equation}
where $m_\mathrm{Gd}$ is the target mass of $^{150}$Gd. Our goal is to find an expression relating $\Dmtb$ to $m_{\mathrm{dose}}^{{}^{150}\mathrm{Gd}}$.

Consider a single irradiation followed by chemical recovery of $^{150}$Gd. Let $N$ be the number of $^{150}$Gd atoms in the target at the start of each cycle (steady state), $f_{\mathrm{burn}}$ the fraction of $N$ destroyed by all proton-induced reactions during the irradiation, $\eta_{\mathrm{rec}}$ the chemistry-recovery yield (the fraction of the surviving $^{150}$Gd that comes back out of the Tb/Gd separation), and $S_{\mathrm{ext}}$ the external $^{150}$Gd added as make-up per cycle. The process is as follows: irradiate, chemically separate \(^{149\mathrm{g}}\)Tb and recover $^{150}$Gd, mix in fresh make-up $^{150}$Gd to restore the inventory, and load the next target. The make-up is added after the chemistry extraction.

At end of irradiation the target carries $N(1-f_{\mathrm{burn}})$ atoms of $^{150}$Gd. The chemistry retains $\eta_{\mathrm{rec}} N (1-f_{\mathrm{burn}})$ of these. Mixing in $S_{\mathrm{ext}}$ atoms of make-up and requiring the inventory at the start of the next cycle to equal $N$,
\begin{equation}
\eta_{\mathrm{rec}} \cdot N (1 - f_{\mathrm{burn}}) \,+\, S_{\mathrm{ext}} \,=\, N.
\label{eq:steady_state}
\end{equation}
Solving for $S_{\mathrm{ext}}$,
\begin{equation}
S_{\mathrm{ext}} \,=\, N \big[\, 1 - \eta_{\mathrm{rec}} (1 - f_{\mathrm{burn}}) \,\big]
              \,=\, N \left[ (1 - \eta_{\mathrm{rec}}) \,+\, \eta_{\mathrm{rec}} f_{\mathrm{burn}} \right].
\label{eq:s_ext}
\end{equation}
$(1-\eta_{\mathrm{rec}})$ is the chemistry loss per cycle and $\eta_{\mathrm{rec}} f_{\mathrm{burn}}$ is the surviving fraction further reduced by the nuclear burn.

Per administered dose, the mass of make-up is
\begin{equation}
m_{\mathrm{dose}}^{{}^{150}\mathrm{Gd}} \,=\, \frac{S_{\mathrm{ext}}}{D_{\mathrm{run}}} \cdot \frac{M_{{}^{150}\mathrm{Gd}}}{N_{\mathrm{A}}}
\,=\, \left[ (1 - \eta_{\mathrm{rec}}) + \eta_{\mathrm{rec}} f_{\mathrm{burn}} \right] \frac{N}{D_{\mathrm{run}}} \cdot \frac{M_{{}^{150}\mathrm{Gd}}}{N_{\mathrm{A}}}.
\label{eq:m_dose_general}
\end{equation}
The ratio $N / D_{\mathrm{run}}$ is the number of $^{150}$Gd atoms loaded per produced dose. Converted to mass via $M_{{}^{150}\mathrm{Gd}}/N_{\mathrm{A}}$ it is the inverse of the throughput density $\Dmtb$ expressed in doses per gram (a factor of $10^3$ larger than $\Dmtb$ in doses per milligram). Substituting $\Dmtb$ in units of doses per milligram and reporting $m_{\mathrm{dose}}^{{}^{150}\mathrm{Gd}}$ in micrograms per dose gives
\begin{equation}
m_{\mathrm{dose}}^{{}^{150}\mathrm{Gd}} \,[\mu\mathrm{g}/\mathrm{dose}] = \left[ (1 - \eta_{\mathrm{rec}}) + \eta_{\mathrm{rec}} f_{\mathrm{burn}} \right] \cdot \frac{1000}{\Dmtb}.
\label{eq:m_dose_dpm}
\end{equation}

At the operating point used for the line plots ($\eta_{\mathrm{rec}} = 0.99$, $T_{\mathrm{irr}} = 4$~h, $J = 100\,\mu\mathrm{A}/\mathrm{cm}^2$ on the $^{150}$Gd front face, $E_p = \qty{18}{MeV}$) the chemistry-loss term is $(1-\eta_{\mathrm{rec}}) = 1.00\cdot 10^{-2}$ and the nuclear-burn term is $f_{\mathrm{burn}} = \sigma_{\mathrm{destr}} \phi T_{\mathrm{irr}} \approx 8.4\cdot 10^{-6}$, with $\sigma_{\mathrm{destr}} \approx \qty{937}{mb}$.
 The chemistry-loss term dominates by approximately 1200, so to within 0.1\% accuracy,
\begin{equation}
m_{\mathrm{dose}}^{{}^{150}\mathrm{Gd}} \,[\mu\mathrm{g}/\mathrm{dose}] \;\approx\; \frac{10}{\Dmtb\,[\mathrm{doses}/\mathrm{mg}]}.
\label{eq:m_dose_dpm_simplified}
\end{equation}

\subsection{$^{150}$Gd doubling time} \label{sec:doubling_time}

We calculate the time required to double the $^{150}$Gd inventory in the front target, assuming the $^{151}$Eu back-layer is held at constant inventory by external $^{151}$Eu inflow and the geometry of the front layer is held fixed. With $N$ atoms of $^{150}$Gd in the front target, the mass balance equation is
\begin{equation}
\frac{\mathrm{d}N}{\mathrm{d}t} \,=\, R_{\mathrm{breed}} - R_{\mathrm{burn}} - R_{\mathrm{ext}}
\,=\, K \,-\, \alpha\, N,
\label{eq:dN_dt}
\end{equation}
where,  with duty cycle $d \equiv T_{\mathrm{beam\,on}}/(24\,\mathrm{h})$ multiplying the beam-driven terms ($d = 0.5$ for $12$~h/day operation),
\begin{align*}
K &\,=\, \eta_{\rm recover,bred}\,d\,\sigma_{\mathrm{breed}}\,\phi\,N_{\mathrm{Eu}} \;\;\;[\text{constant, since}~N_{\mathrm{Eu}}~\text{is held fixed}], \\
\alpha &\,=\, d\,\sigma_{\mathrm{destr}}\,\phi \,+\, \lambda_{\mathrm{ext}} \;\;\;[\text{per-atom $^{150}$Gd loss rate}], \\
R_{\mathrm{ext}} &\,=\, \lambda_{\mathrm{ext}}\,N \;\;\;[\text{chemistry loss}].
\end{align*}
Because the breed rate is independent of $N$ while the loss rates scale with $N$, the inventory relaxes toward an equilibrium $N_{\mathrm{eq}} = K/\alpha$,
\begin{equation}
N(t) \,=\, N_{\mathrm{eq}} \,+\, (N_0 - N_{\mathrm{eq}}) \,e^{-\alpha\,t}.
\end{equation}
The gadolinium breeding ratio of \cref{eq:GBR} is
\begin{equation}
\mathrm{GBR} \,\equiv\, \frac{N_{\mathrm{breed}}^{{}^{150}\mathrm{Gd}}}{N_{\mathrm{cons}}^{{}^{150}\mathrm{Gd}}} \,=\, \frac{K}{\alpha\,N_0} \,=\, \frac{N_{\mathrm{eq}}}{N_0},
\label{eq:GBR}
\end{equation}
so doubling the inventory from $N_0$ to $2 N_0$ requires $\mathrm{GBR} > 2$, and the doubling time is
\begin{equation}
T_2 \,=\, \frac{1}{\alpha} \,\ln\!\left[\, \frac{\mathrm{GBR} - 1}{\mathrm{GBR} - 2} \,\right].
\label{eq:doubling_time}
\end{equation}
In the large-$\mathrm{GBR}$ limit this simplifies to $T_2 \approx 1 / (\mathrm{GBR}\,\alpha)$, i.e., the time to accumulate one additional $N_0$ at the (nearly constant) breed rate $\mathrm{GBR}\,\alpha\,N_0$.

With $100$ recovery cycles per year at $\eta_{\mathrm{rec}} = 0.99$, the chemistry-loss rate is $\lambda_{\mathrm{ext}} = -100\,\ln(\eta_{\mathrm{rec}})/\mathrm{yr} \approx 3.2\cdot 10^{-8}$~s$^{-1}$, about $100\times$ the beam-driven destruction $d\,\sigma_{\mathrm{destr}}\,\phi \approx 2.9\cdot 10^{-10}$~s$^{-1}$ (at $d = 1/2$, $J = \qty{100}{\uA/\cm^2}$), so chemistry sets the loss term $\alpha$. Doubling ($\mathrm{GBR} > 2$) then requires the breeding term to beat this loss, which favors a small $^{150}$Gd inventory (the loss $\lambda_{\mathrm{ext}} N$ grows with inventory) and a long beam-on fraction (breeding scales with duty cycle $d$). At the $I = \qty{100}{\uA}$, $A_\mathrm{spot} = \qty{1}{\cm\squared}$ reference point, the \qty{4}{\hour} ten-dose rows of \Cref{tab:operating_points} (17-20 MeV) satisfy both: $\mathrm{GBR} \approx 2$-$3$ ($T_2 \approx$ 260-530 days) at $\eta_{\mathrm{rec}} = 0.99$, rising to $\mathrm{GBR} \approx 17$-$27$ ($T_2 \approx$ 130-220 days) at $\eta_{\mathrm{rec}} = 0.999$. The high-throughput rows ($\sim$100 doses, large inventory) and the two-hour rows (short beam-on) both fall to $\mathrm{GBR} < 2$.

\subsection{GBR scans} \label{sec:gbr_sweeps}

\begin{figure}[!tb]
\centering
\includegraphics[width=\textwidth]{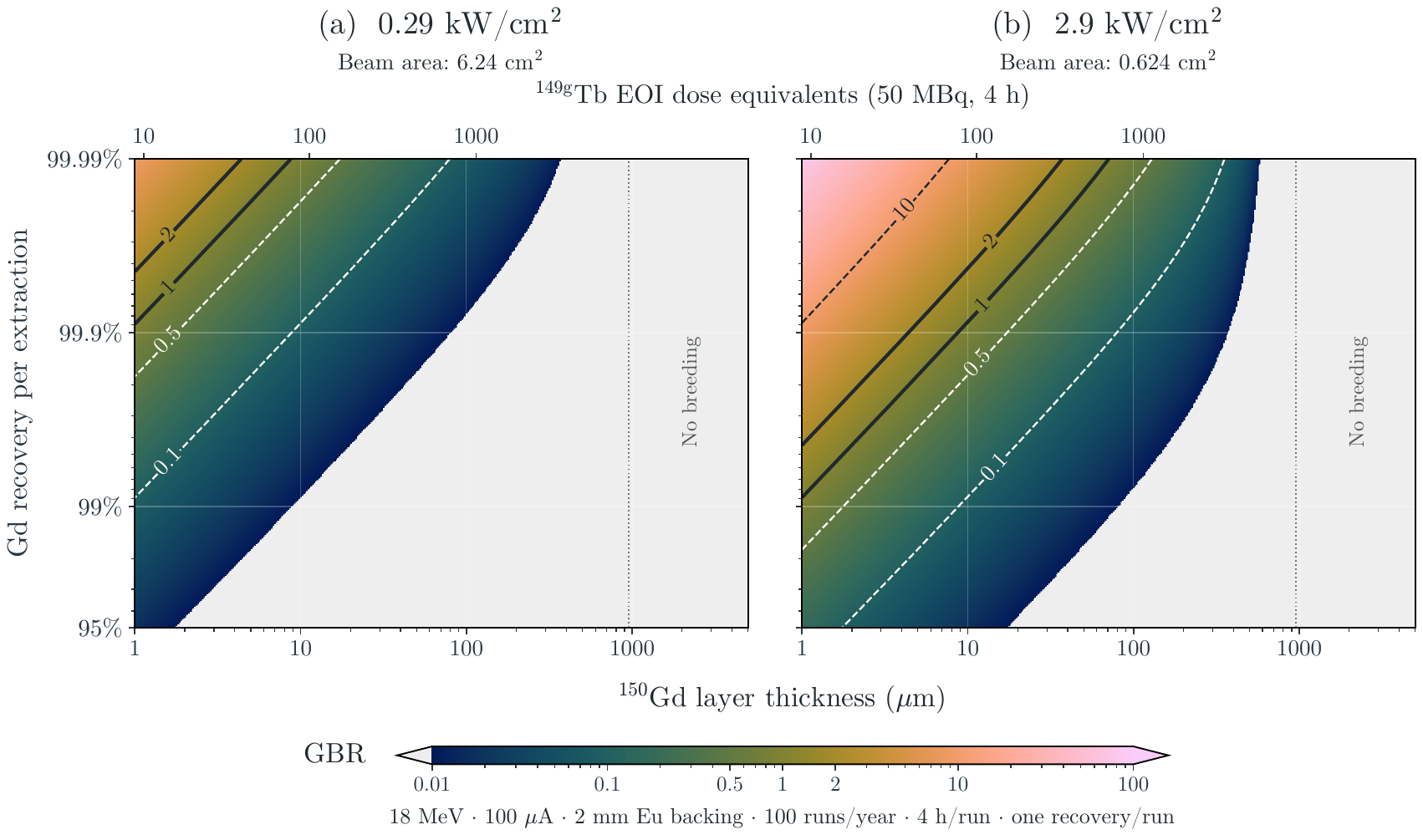}
\caption{ GBR versus Gd thickness and Gd recovery per extraction at incident beam-power densities of approximately 0.29 and \qty{2.9}{\kW/\cm^2} during irradiation. Gray means negligible breeding, and dotted vertical lines show the proton stopping distance in Gd. Conditions: \qty{18}{MeV}, \qty{100}{\uA}, \qty{2}{mm} Eu backing, beam areas of 6.24 and \qty{0.624}{\cm^2}, 100 four-hour irradiations/year ($d\simeq0.0456$), one Gd recovery per run.}
\label{fig:Gd_breeding_ratio}
\end{figure}

We now calculate plausible values for the doubling time $T_2$.  We show the recovery and thickness dependence of GBR in \Cref{fig:Gd_breeding_ratio}, with separate panels for two incident beam-power densities at \qty{18}{MeV} and \qty{100}{\uA}. The upper scale gives \qty{50}{MBq} EOI dose equivalents after \qty{4}{\hour}, before cooling or Tb recovery. We use 100 four-hour irradiations per year, with one Gd recovery after each run. The resulting beam-on fraction is $d=400/8766\simeq0.0456$ in \cref{eq:GBR}. Schemes with $\mathrm{GBR} > 2$ seed a doubling $^{150}$Gd inventory. At the paper's $I = \qty{100}{\uA}$ reference operating point this is reached by the thin, low-throughput points (GBR $\approx 2$-$3$ at $\eta_\mathrm{rec} = 0.99$, rising to $\approx 17$-$27$ at $\eta_\mathrm{rec} = 0.999$, \Cref{tab:operating_points}), while high-throughput points ($\sim$100 doses per irradiation) carry a larger inventory whose chemistry loss outpaces breeding and fall to GBR $<1$.  We show how GBR varies with Gd thickness in \Cref{fig:optimized_doses}, using \qty{4}{\hour}/day beam operation and 100 extractions/year at 99\% recovery. This duty fraction, $d=1/6$, is lower than the $d=1/2$ cases in \Cref{tab:operating_points} and therefore gives lower GBR at the same target specification.

\section{Target optimization and operating points}\label{sec:target_optimization}

In this appendix, we extend the target-choice results of Section~\ref{ssec:target_choice} to further characterize the operating space and find optimal points. Here $E_p$ is incident proton energy, $L_{\rm Gd}$ is the Gd-layer thickness, $I$ is total beam current, and $J$ is current density. The spot area is $A_{\rm spot}=I/J$. We use the dose-throughput metric $\Dmtb$ and activity fractions $P_{149}$ and $P_{149+150}$ defined in the main text. The reference Tb product recovery is $\eta_{\rm Tb}=1$. The irradiated target mass is $m_{\rm Gd}=A_{\rm spot}\rho_{\rm Gd}L_{\rm Gd}$, with $\rho_{\rm Gd}$ the Gd metal density. We set $J(E_p)=J_{\rm ref}E_{\rm ref}/E_p$, where $J_{\rm ref}=\qty{100}{\uA/\cm^2}$ at $E_{\rm ref}=\qty{18}{MeV}$.

\begin{figure}[!tb]
    \centering
    \includegraphics[width=0.97\textwidth]{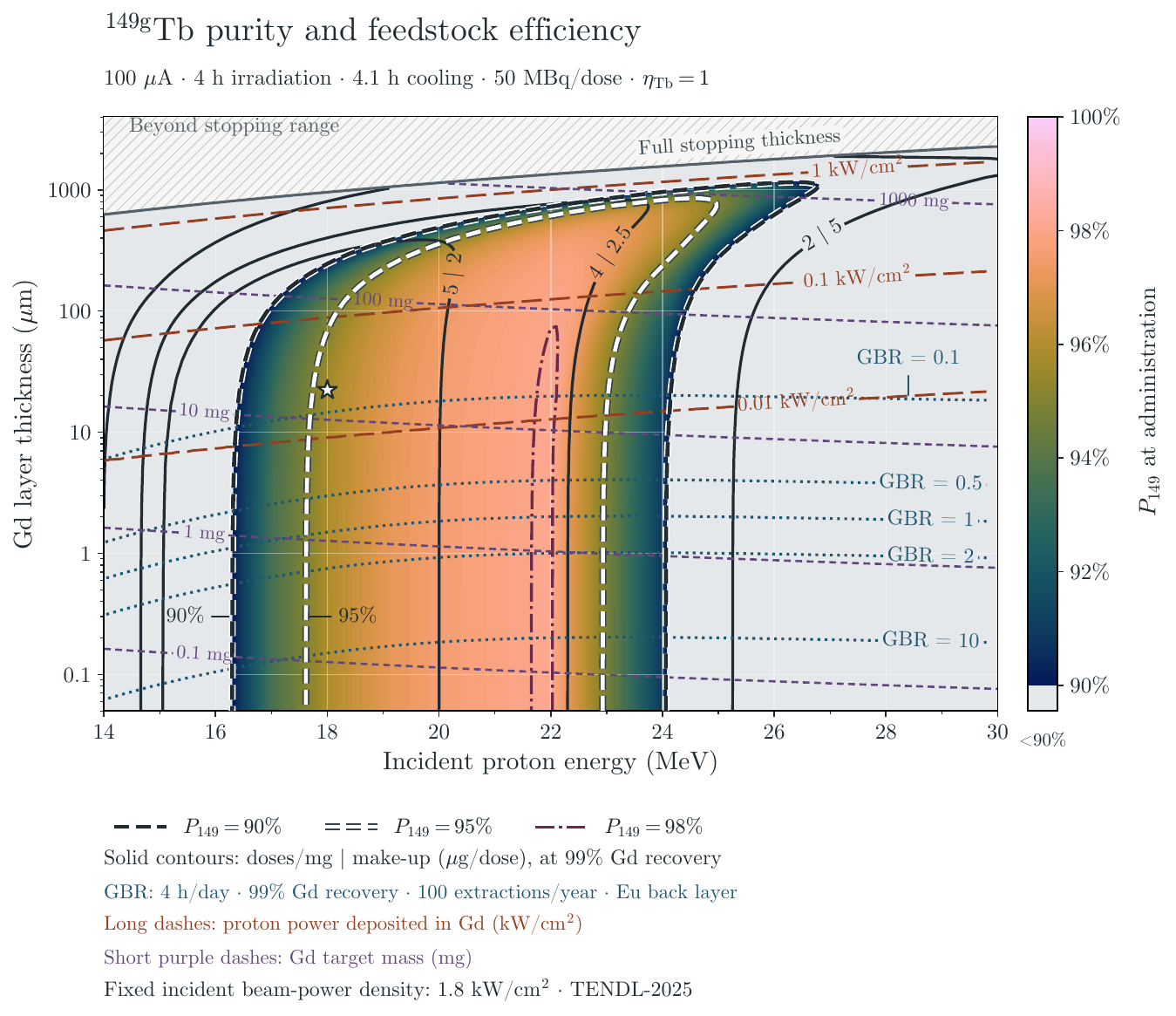}
    \caption{ Radionuclidic purity, feedstock use, breeding and heating versus incident proton energy and Gd-layer thickness. The Eu backing is 97\%-enriched $^{151}$Eu and thick enough to stop transmitted protons. Make-up per dose assumes chemical losses dominate, while GBR includes both nuclear destruction and chemical losses (\cref{eq:GBR}). The star marks the reference target in \Cref{fig:dose_throughput}.}
    \label{fig:optimized_doses}
    \label{fig:optimized_doses_makeup}
    \label{fig:optimized_doses2}
    \label{fig:optimized_doses3_gbr_app}
    \label{fig:optimized_doses3_backcool_app}
\end{figure}

We choose a nominal target specification (mass, thickness, proton energy) according to our specifications in \Cref{tab:practical}.

There are tradeoffs between \(^{149\mathrm{g}}\)Tb yield, the \(^{149\mathrm{g}}\)Tb yield per mass of \(^{150}\)Gd feedstock, and radionuclidic purity for a given tolerable target cooling rate and chemical extraction recycling efficiency. This is motivated by needing as little \(^{150}\)Gd material in the target as possible, which will be particularly useful if \(^{150}\)Gd is scarce. This involves optimizing proton energy and layer thickness.

A useful figure of merit is the number of therapeutic doses per mg of \(^{150}\)Gd feedstock after an irradiation period of \qty{4}{\hour} and decay and chemistry losses during a \qty{4.1}{\hour} cooldown, $D_\mathrm{m}^{^{149\mathrm{g}} \mathrm{Tb} } $. It is related to the $^{150}$Gd make-up per dose $m_{\mathrm{dose}}^{{}^{150}\mathrm{Gd}}$ by
\begin{equation}
D_\mathrm{m}^{^{149\mathrm{g}} \mathrm{Tb} } \approx \frac{10}{m_{\mathrm{dose}}^{{}^{150}\mathrm{Gd}}},
\end{equation}
 where $D_\mathrm{m}$ is expressed in administrations/mg, $m_{\rm dose}$ in $\mu$g/administration, and we assumed $\eta_\mathrm{rec}=0.99$ with chemistry loss dominating nuclear burn. See \Cref{eq:m_dose_dpm,eq:m_dose_dpm_simplified} for a derivation.

 In \Cref{fig:optimized_doses}, we show how purity, doses per milligram, target mass, GBR and deposited power density vary with proton energy and Gd thickness. Each efficiency contour also gives the equivalent make-up per dose. Thin Gd targets irradiated at 16-20 MeV approach approximately 5-6 doses/mg, with purity depending on energy. The GBR contours use the feedstock model of \ref{sec:feedstock_balance}. Absorbed power density increases toward the incident \qty{1.8}{\kW/\cm^2} as the Gd layer approaches full stopping thickness.

\begin{table}[!htbp]
\centering
\caption{Operating points across six irradiation modes at $T_{\mathrm{cool}} = \qty{4.1}{\hour}$, $I = \qty{100}{\uA}$ total beam current, and a fixed \qty{1.8}{\kilo\watt\per\cm^2} front-face heat flux at the $J = \qty{100}{\uA/\cm^2}$, $E_\mathrm{ref} = \qty{18}{MeV}$ reference, so that both $J(E_p)$ and $A_\mathrm{spot}(E_p) = I/J$ vary with $E_p$ (see \cref{fig:optimized_doses} caption).  For the 16-20 MeV rows below, $A_\mathrm{spot}$ ranges from $\qty{0.89}{\cm\squared}$ to $\qty{1.11}{\cm\squared}$, with throughput fixed at the stated number of administrable $^{149\mathrm{g}}$Tb doses per irradiation. Chemistry assumes 100 Gd recovery cycles per year at the stated per-extraction efficiency $\eta_{\mathrm{rec}}$ (we top up Gd losses after each irradiation).  The two activity fractions at administration are $P_{149}=A_{149\mathrm{g}}/A_{\rm Tb,total}$ and $P_{149+150}=(A_{149\mathrm{g}}+A_{150})/A_{\rm Tb,total}$. The second includes $^{150}$Tb in the numerator. It is distinct from $A_{149\mathrm{g}}/(A_{\rm Tb,total}-A_{150})$. $d$ is the beam-on fraction of calendar time. GBR is the gadolinium breeding ratio of \cref{eq:GBR} at the indicated duty cycle. $T_2$ is the \ce{^{150}Gd} doubling time (\cref{eq:doubling_time}). ``N/A'' indicates $\mathrm{GBR}\leq 2$, where the inventory cannot double.}
\label{tab:operating_points}
\scriptsize
\setlength{\tabcolsep}{3pt}
\renewcommand{\arraystretch}{0.9}
\begin{tabular}{c c c c c c c c c}
\toprule
\makecell{\textbf{$E_p$} \\ \textbf{(MeV)}} &
\makecell{\textbf{$L_{\mathrm{Gd}}$} \\ \textbf{($\mu$m)}} &
\makecell{\textbf{$m_{\mathrm{Gd}}$} \\ \textbf{(mg)}} &
\makecell{\textbf{$\Dmtb$} \\ \textbf{(doses/mg)}} &
\makecell{ \\  \\ $P_{149}$} &
\makecell{ \\  \\ $P_{149+150}$} &
\makecell{\textbf{$m_{\mathrm{dose}}^{{}^{150}\mathrm{Gd}}$} \\ \textbf{($\mu$g/dose)}} &
\makecell{\textbf{GBR}} &
\makecell{\textbf{$T_2$} \\ \textbf{(days)}} \\
\midrule
\multicolumn{9}{l}{\textit{$T_{\mathrm{irr}} = 4$~h, $d = 1/2$, 10 doses per irradiation, $\eta_\mathrm{rec} = 0.99$ per extraction}} \\
16 & 2.45 & 1.722 & 5.81 & 88.3\,\% & 99.3\,\% & 1.72 & 1.82 & N/A \\
17 & 2.27 & 1.696 & 5.90 & 93.5\,\% & 99.4\,\% & 1.70 & 2.29 & 533 \\
18 & 2.22 & 1.756 & 5.70 & 95.9\,\% & 99.4\,\% & 1.76 & 2.65 & 336 \\
19 & 2.23 & 1.860 & 5.38 & 97.0\,\% & 99.4\,\% & 1.86 & 2.87 & 274 \\
20 & 2.28 & 2.000 & 5.00 & 97.6\,\% & 99.4\,\% & 2.00 & 2.96 & 257 \\
\midrule
\multicolumn{9}{l}{\textit{$T_{\mathrm{irr}} = 2$~h, $d = 1/12$, 10 doses per irradiation, $\eta_\mathrm{rec} = 0.99$}} \\
16 & 4.16 & 2.919 & 3.43 & 88.1\,\% & 99.4\,\% & 2.92 & 0.18 & N/A \\
17 & 3.84 & 2.863 & 3.49 & 93.5\,\% & 99.4\,\% & 2.86 & 0.23 & N/A \\
18 & 3.74 & 2.955 & 3.38 & 95.9\,\% & 99.5\,\% & 2.96 & 0.26 & N/A \\
19 & 3.75 & 3.123 & 3.20 & 97.1\,\% & 99.5\,\% & 3.12 & 0.29 & N/A \\
20 & 3.82 & 3.349 & 2.99 & 97.7\,\% & 99.5\,\% & 3.35 & 0.30 & N/A \\
\midrule
\multicolumn{9}{l}{\textit{$T_{\mathrm{irr}} = 2$~h, $d = 1/12$, 5 doses per irradiation, $\eta_\mathrm{rec} = 0.99$}} \\
16 & 2.07 & 1.456 & 3.43 & 88.3\,\% & 99.4\,\% & 2.91 & 0.36 & N/A \\
17 & 1.92 & 1.430 & 3.50 & 93.5\,\% & 99.4\,\% & 2.86 & 0.46 & N/A \\
18 & 1.87 & 1.477 & 3.39 & 95.9\,\% & 99.5\,\% & 2.96 & 0.53 & N/A \\
19 & 1.87 & 1.562 & 3.20 & 97.1\,\% & 99.5\,\% & 3.12 & 0.58 & N/A \\
20 & 1.91 & 1.675 & 2.99 & 97.7\,\% & 99.5\,\% & 3.35 & 0.59 & N/A \\
\midrule
\multicolumn{9}{l}{\textit{$T_{\mathrm{irr}} = 4$~h, $d = 1/2$, 100 doses per irradiation, $\eta_\mathrm{rec} = 0.99$}} \\
16 & 25.24 & 17.721 & 5.64 & 86.5\,\% & 99.3\,\% & 1.77 & 0.17 & N/A \\
17 & 22.93 & 17.105 & 5.85 & 92.9\,\% & 99.4\,\% & 1.71 & 0.22 & N/A \\
18 & 22.28 & 17.598 & 5.68 & 95.6\,\% & 99.4\,\% & 1.76 & 0.26 & N/A \\
19 & 22.30 & 18.595 & 5.38 & 96.9\,\% & 99.4\,\% & 1.86 & 0.28 & N/A \\
20 & 22.73 & 19.951 & 5.01 & 97.5\,\% & 99.4\,\% & 2.00 & 0.29 & N/A \\
\midrule
\multicolumn{9}{l}{\textit{$T_{\mathrm{irr}} = 4$~h, $d = 1/2$, 10 doses per irradiation, $\eta_\mathrm{rec} = 0.999$ (optimistic chemistry)}} \\
16 & 2.45 & 1.722 & 5.81 & 88.3\,\% & 99.3\,\% & 0.17 & 16.86 & 217 \\
17 & 2.27 & 1.696 & 5.90 & 93.5\,\% & 99.4\,\% & 0.17 & 21.15 & 169 \\
18 & 2.22 & 1.756 & 5.70 & 95.9\,\% & 99.4\,\% & 0.18 & 24.33 & 145 \\
19 & 2.23 & 1.860 & 5.38 & 97.0\,\% & 99.4\,\% & 0.19 & 26.33 & 133 \\
20 & 2.28 & 2.000 & 5.00 & 97.6\,\% & 99.4\,\% & 0.20 & 27.07 & 129 \\
\midrule
\multicolumn{9}{l}{\textit{$T_{\mathrm{irr}} = 4$~h, $d = 1/2$, 100 doses per irradiation, $\eta_\mathrm{rec} = 0.999$ (optimistic chemistry)}} \\
16 & 25.24 & 17.721 & 5.64 & 86.5\,\% & 99.3\,\% & 0.18 & 1.56 & N/A \\
17 & 22.93 & 17.105 & 5.85 & 92.9\,\% & 99.4\,\% & 0.17 & 2.03 & 12012 \\
18 & 22.28 & 17.598 & 5.68 & 95.6\,\% & 99.4\,\% & 0.18 & 2.37 & 4329 \\
19 & 22.30 & 18.595 & 5.38 & 96.9\,\% & 99.4\,\% & 0.19 & 2.60 & 3249 \\
20 & 22.73 & 19.951 & 5.01 & 97.5\,\% & 99.4\,\% & 0.20 & 2.69 & 2939 \\
\bottomrule
\end{tabular}
\end{table}

 We select nominal target parameters from these operating maps by fixing the number of administered therapeutic doses per irradiation. We show several key parameters in \Cref{tab:operating_points} for 16-20 MeV proton beams. We also include shorter scenarios with $T_\mathrm{irr} = 2$~hr (5 and 10 doses per irradiation) and a higher-throughput scenario with $100$ doses per irradiation. The last two blocks in \Cref{tab:operating_points} show an optimistic chemistry assumption of $\eta_\mathrm{rec} = 0.999$ per extraction, which reduces $m_\mathrm{dose}^{{}^{150}\mathrm{Gd}}$ by an order of magnitude in both cases. The breeding-ratio and doubling-time columns (GBR and $T_2$) are discussed in \ref{sec:feedstock_balance}.

\section{Repeated cook-and-harvest production}\label{ssec:cook_harvest}

\begin{figure}[!tb]
\centering
\includegraphics[width=\textwidth]{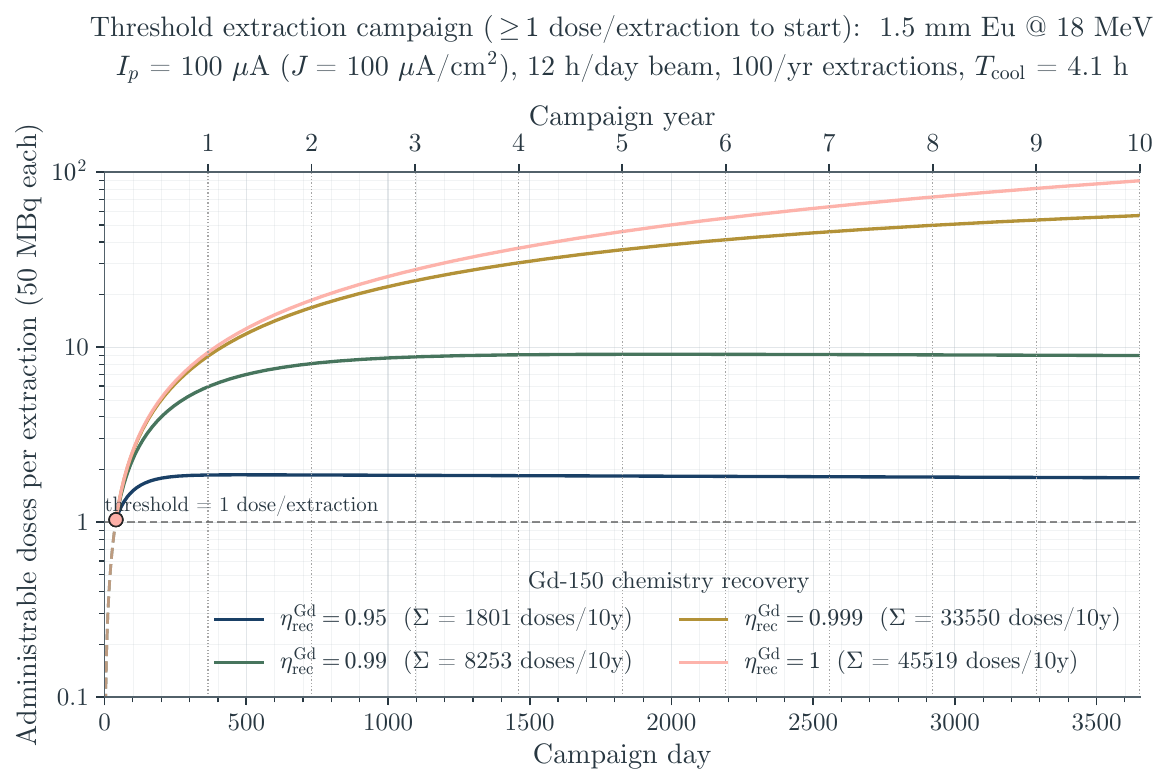}
\caption{Administrable \(^{149\mathrm{g}}\)Tb doses per extraction with `cook and harvest' of a $^{151}$Eu target.}
\label{fig:cook_harvest}
\end{figure}

In this appendix, we consider repeated extraction during extended irradiation, extending the single-extraction estimates in Section~\ref{ssec:preclinical_ramp}. This strategy for \(^{149\mathrm{g}}\)Tb production is continuous irradiation of an enriched \(^{151}\)Eu target, followed by periodic \(^{149\mathrm{g}}\)Tb extraction as needed for doses. We call this technique `cook and harvest.' We plot the number of doses available per day as a function of extraction recovery efficiency and irradiation time in \Cref{fig:cook_harvest}. Here $\eta_{\rm recovery}$ denotes Gd recovery per extraction, the same efficiency denoted by $\eta_{\rm rec}$ elsewhere. With a \qty{100}{\uA} \qty{18}{MeV} proton beam and $\eta_\mathrm{recovery}=0.99$, two administrable doses (accounting for a 4.1 hour delay after production) of  \(^{149\mathrm{g}}\)Tb are available per extraction after roughly 200 days. If the ${}^{150}$Gd recovery efficiency increases to $\eta_\mathrm{recovery}=0.999$, roughly 10 administrable doses are available each extraction. Because `cook and harvest' combines two sequential reactions to produce \(^{149\mathrm{g}}\)Tb starting with ${}^{151}$Eu,  the number of doses scales roughly quadratically with current density and irradiation time (at fixed beam current).

\section{Cyclotron Deployments} \label{sec:cyclotron_dep}

\begin{figure}[!tb]
\centering
\includegraphics[width=\columnwidth]{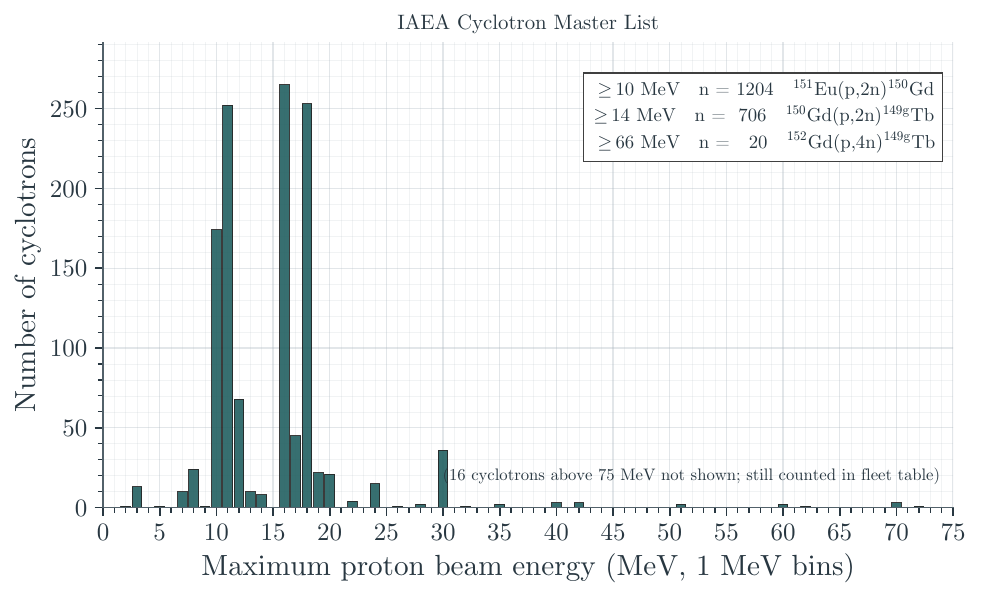}
\caption{Number of cyclotrons binned by maximum proton beam energy, source IAEA Accelerator Knowledge Portal \cite{iaea_akp_2026}.}
\label{fig:IAEA_cyclotrons}
\end{figure}

In this appendix, we plot the number of proton cyclotrons versus their energy (\Cref{fig:IAEA_cyclotrons}), showing there are over 1200 proton machines capable of \(^{151}\)Eu(p,2n)\(^{150}\)Gd, and over 700 proton machines capable of \(^{150}\)Gd(p,2n)\(^{149\mathrm{g}}\)Tb. This data was obtained from the IAEA Accelerator Knowledge Portal \cite{iaea_akp_2026}.

\section{Alternative \({}^{150}\)Gd Pathways} \label{sec:alternative_gd150}

In this appendix, we describe alternate methods for producing or sourcing \({}^{150}\)Gd. We judge them to be much less efficient than the methods outlined earlier that start from europium, but we add for completeness.

\paragraph{Multi-stage neutron route via Eu-152} Neutron capture or (n,2n) on natural Eu produces \(^{152}\)Eu (\SI{13.5}{yr}), which \(\beta^-\) decays to \(^{152}\)Gd with \SI{27.92}{\percent} branching,
\begin{equation*}
{}^{151}\mathrm{Eu}(\mathrm{n},\gamma){}^{152}\mathrm{Eu}\ \text{or}\ {}^{153}\mathrm{Eu}(\mathrm{n,2n}){}^{152}\mathrm{Eu}\ \xrightarrow{\beta^-} {}^{152}\mathrm{Gd}\ \xrightarrow{(\mathrm{n,2n})+(\mathrm{n,3n})}\ {}^{150}\mathrm{Gd}.
\end{equation*}
The \(^{152\mathrm{m}1}\)Eu isomer (\SI{9.3}{h}, \SI{73}{\percent} \(\beta^-\)) provides a much faster path than the ground state. The Gd is separated from Eu chemically (lanthanide chromatography) or physically (selective evaporation, exploiting the higher vapor pressure of Eu). This route also gives a path to \(^{152}\)Gd feedstock starting from natural Eu rather than from sourced enriched \(^{152}\)Gd.

\paragraph{Gd-target neutron route} An enriched \(^{152}\)Gd target is placed in a high-flux fast neutron environment. Two routes to \(^{150}\)Gd are available. The cascade
\begin{equation}\label{eq:cascade}
^{152}\text{Gd}\,\text{(n,2n)}\,^{151}\text{Gd}\,\text{(n,2n)}\,^{150}\text{Gd},
\end{equation}
proceeds through the unstable \(^{151}\)Gd intermediate (\SI{124}{d} EC half-life), making yields low unless neutron flux is extremely high. The alternative single-step reaction
\begin{equation}
^{152}\text{Gd}\,\text{(n,3n)}\,^{150}\text{Gd},
\end{equation}
has a threshold of $\sim$\SI{15.2}{MeV} (just above the D-T energy) but reaches \qtyrange{0.5}{1.6}{b} at \qtyrange{20}{25}{MeV}, accessible to deuteron breakup neutron sources~\cite{morrell2023}. This route bypasses the \(^{151}\)Gd bottleneck. However, we believe both of these schemes to be impractical unless far superior isotope separation becomes available, both for enriching \(^{152}\)Gd feedstock and for separating \(^{150}\)Gd from \(^{152}\)Gd.

\paragraph{Other beam-driven routes} A number of additional charged-particle reactions can produce \(^{150}\)Gd  directly or through radioactive precursors:

\noindent\emph{Protons.} \(^{149}\)Sm(p,\(\gamma\)), \(^{150}\)Sm(p,n), \(^{152}\)Sm(p,3n), \(^{153}\)Eu(p,4n), \(^{152}\)Gd(p,3n)\(^{150}\)Tb \(\to \beta^+\), \(^{152}\)Gd(p,p+2n).

\noindent\emph{Deuterons.} \(^{149}\)Sm(d,n), \(^{150}\)Sm(d,2n), \(^{152}\)Sm(d,4n), \(^{151}\)Eu(d,3n), \(^{153}\)Eu(d,5n).

\noindent\emph{Alphas.} \(^{147}\)Sm(\(\alpha\),n), \(^{148}\)Sm(\(\alpha\),2n), \(^{149}\)Sm(\(\alpha\),3n), \(^{150}\)Sm(\(\alpha\),4n).

\noindent\emph{Helium-3.} \(^{148,149,150}\)Sm(\(^{3}\)He,xn), \(^{151}\)Eu(\(^{3}\)He,p3n), and \(^{151}\)Eu(\(^{3}\)He,\(\alpha\))\(^{150}\)Eu \(\to \beta^-\). The Moiseeva 2020 measurement \cite{moiseeva2020} used \(^{151}\)Eu(\(^{3}\)He,5n) at 70 MeV.

These routes are less attractive than the three Eu-target routes in the main paper (p,2n; n,2n; \(\gamma\),n), either because they require higher-energy or more specialized beams (deuterons, \(^{3}\)He), because they pass through intermediate isotope-separation steps, and/or because they have lower cross sections. They are noted here for completeness.

\paragraph{Recovery from existing spallation targets}

Irradiated spallation targets offer a supplementary source of \(^{150}\)Gd. High-energy protons on Ta, W, Hg, and Pb produce gadolinium among the residual nuclei. Measured \(^{181}\)Ta(p,x)\(^{148}\)Gd cross sections rise from approximately \qty{15}{mb} at \qty{600}{MeV} to \qty{33}{mb} near \qty{1.6}{GeV}~\cite{titarenko2011,titarenko2008,kelley2005}. Gadolinium recovered from irradiated tantalum at PSI has a \(^{150}\)Gd/\(^{148}\)Gd atom ratio of approximately 2.4~\cite{chiera2020}. We use this ratio as a working estimate of \(\sigma_{150}/\sigma_{148}\).

For a thick target, we estimate the yield per incident proton as \(Y_{150}\approx f\sigma_{150}/\sigma_{\rm inel}\), with \(\sigma_{\rm inel}\approx\qty{1.65}{b}\) and \(f\approx1.5\) to account for secondary particles. This gives a production scale of approximately \qty{2}{g} per MW-year for Ta and W. For the facility estimates, we reduce the Ta cross section by 20\% for W, use Au as a proxy for Hg, and half the Au value for Pb~\cite{kelley2005,steinberg1968}. These substitutions provide an order-of-magnitude comparison.

\Cref{tab:gd150_spallation} summarizes our estimates, including an accumulated inventory of around \qty{40}{g} across existing facilities. The ESS estimate of \qty{5.5}{g} per full-power year corresponds to roughly \qty{30}{g} over five full-power years. In the PSI samples, \(^{150}\)Gd represents about 5\% of the recovered Gd, so isotope separation is needed in addition to chemical recovery~\cite{chiera2020}. The estimates describe material produced in targets before recovery and enrichment losses.

\begin{table}[!tb]
\centering
\caption{ Estimated \(^{150}\)Gd production and accumulated inventory in spallation targets. Production is per full-power year. Inventories are order-of-magnitude estimates before recovery and enrichment.}
\label{tab:gd150_spallation}
\begingroup
\color{black}
\small
\setlength{\tabcolsep}{7pt}
\renewcommand{\arraystretch}{1.16}
\begin{tabular*}{\textwidth}{@{\extracolsep{\fill}}l l r r@{}}
\toprule
Facility & Target & Production & Inventory \\
 & & (g/full-power yr) & (g) \\
\midrule
\multicolumn{4}{@{}l}{\textit{Existing and historical facilities}} \\
LAMPF (historic)~\cite{lisowski2006} & W, Ta & n/a & 18 \\
ISIS TS1~\cite{thomason2019} & Ta-clad W & 0.4 & 9 \\
SNS~\cite{mason2006} & Hg & 0.6 & 5 \\
LANSCE~\cite{lisowski2006} & W & 0.3 & 4 \\
J-PARC MLF~\cite{nagamiya2012} & Hg & 0.2 & 1 \\
ISIS TS2~\cite{thomason2019} & W & 0.1 & 1 \\
SINQ~\cite{chiera2020} & Pb & 0.08 & 0.7 \\
TRIUMF ISAC~\cite{dilling2014} & Ta & 0.08 & 0.3 \\
CERN ISOLDE~\cite{catherall2017} & Ta & 0.006 & 0.2 \\
\addlinespace[3pt]
\multicolumn{3}{@{}l}{\textbf{Total, existing (rounded)}} & \textbf{40} \\
\midrule
\multicolumn{4}{@{}l}{\textit{Planned facilities and upgrades}} \\
ESS~\cite{garoby2018} & W & 5.5 & \,- \\
SNS second station~\cite{sns_sts2026} & \,W & 1.0 & \,- \\
CSNS-II~\cite{wei2009} & W & 0.5 & \,- \\
\bottomrule
\end{tabular*}
\endgroup
\end{table}

As a separate comparison, our Geant4~11.3.2 calculations  combine the INCL++ nuclear cascade model with Geant4's default nuclear de-excitation model~\cite{agostinelli2003,allison2016,boudard2013,mancusi2014}. For \qtyrange{0.8}{3}{GeV} protons on \qty{40}{cm}-thick Ta and W targets, they give approximately \qtyrange{0.3}{0.6}{g} per MW-year, below the cross-section-based estimates. Given these uncertainties and the appeal of other pathways to produce $^{150}$Gd, we therefore currently consider spallation recovery as a supplementary supply option.

\section{Alternative \({}^{149\mathrm{g}}\)Tb Pathways} \label{sec:alternative_tb149g}

In this appendix, we describe alternate pathways for producing \({}^{149\mathrm{g}}\)Tb. It is important to re-iterate that we expect the \({}^{150}\)Gd(p,2n) pathway to perform best, pending measurement of the cross section.

\begin{figure}[!tb]
\centering
\includegraphics[width=0.85\textwidth]{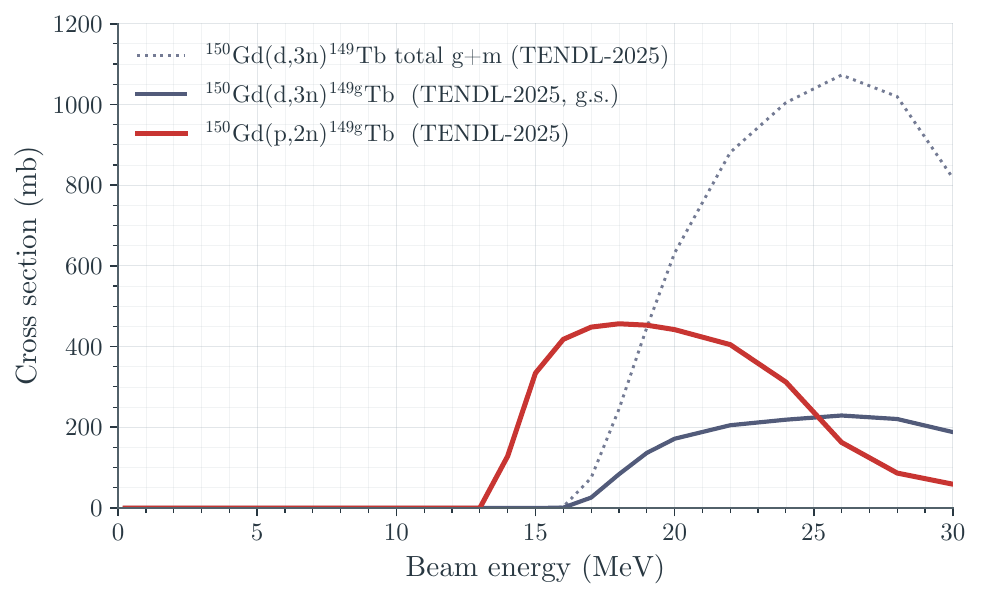}
\caption{Predicted \(^{150}\)Gd(d,3n)\(^{149}\)Tb cross sections from TENDL, alongside  \(^{150}\)Gd(p,2n)\(^{149\mathrm{g}}\)Tb predictions for reference. The total (d,3n) cross section (g + m, dotted) peaks near \qty{1}{b} at \SI{26}{MeV}, but only about \SI{25}{\percent} of that feeds the \(^{149\mathrm{g}}\)Tb ground state (solid, peak $\sim$\SI{257}{mb}). The remaining \(\sim 75\%\) populates the \SI{4.16}{min} \(^{149\mathrm{m}}\)Tb isomer, which EC-decays to \(^{149}\)Gd. After accounting for this isomer split, the (d,3n) g.s.\ peak  is below the (p,2n) g.s.\ peak ($\sim$\SI{457}{mb} in TENDL-2025).}
\label{fig:Gd150_d3n}
\end{figure}

A beam-driven pathway from \(^{150}\)Gd feedstock uses a deuteron beam,
\begin{equation}
^{150}\text{Gd}\,\text{(d,3n)}\,^{149\mathrm{g}}\text{Tb}\,.
\end{equation}
The total (d,3n) cross section (summed over Tb-149 isomers) peaks near \qty{1}{b} at \qty{26}{MeV},  about twice the (p,2n) ground-state peak.
However, TENDL predicts that only about a quarter of the (d,3n) cross section populates \(^{149\mathrm{g}}\)Tb ground state. The  remaining three quarters populate the \SI{4.16}{min} \(^{149\mathrm{m}}\)Tb isomer, which EC-decays to \(^{149}\)Gd and is lost. After this correction, the (d,3n) ground-state cross section peaks near \SI{257}{mb} at \SI{26}{MeV} (\Cref{fig:Gd150_d3n}),  below the $\sim$\SI{457}{mb} (p,2n) ground-state peak at \SI{18}{MeV}. The (d,3n) route therefore does not offer a clear yield advantage over (p,2n), but it does provide an option for facilities equipped with deuteron-capable cyclotrons that lack a suitable proton beam. However, given the scarcity of deuteron-only facilities, we expect that proton-driven production of \(^{149\mathrm{g}}\)Tb will dominate.

\section{${}^{150}$Gd(p,*) Cross Sections} \label{sec:cross_section_isomer}

\begin{figure}[!tb]
\centering
\includegraphics[width=\columnwidth]{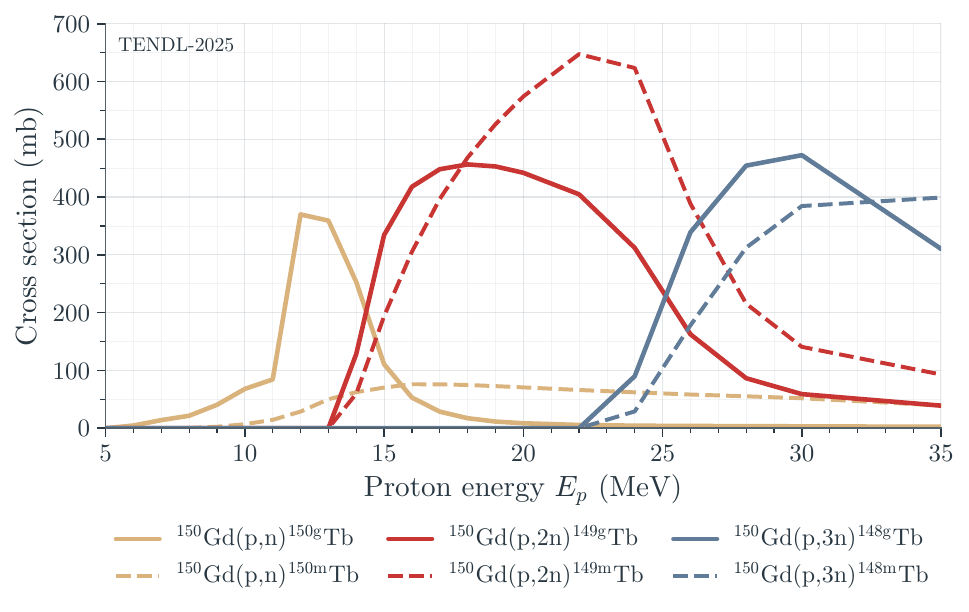}
\caption{Some ${}^{150}$Gd(p,*) cross sections, source TENDL-2025.}
\label{fig:gd150xsTENDL2025}
\end{figure}

In this appendix, we show the cross sections for ${}^{150}$Gd(p,*) that produce ${}^{148}$Tb, ${}^{149}$Tb, and ${}^{150}$Tb. Of most concern for ${}^{149\mathrm{g}}$Tb is co-producing ${}^{150\mathrm{g}}$Tb given the comparable half-lives of  these two isotopes. \Cref{fig:gd150xsTENDL2025} shows the ${}^{150}$Gd(p,n), ${}^{150}$Gd(p,2n), and ${}^{150}$Gd(p,3n) cross sections.

\FloatBarrier
\bibliographystyle{elsarticle-num}
\bibliography{references}

\end{document}